\documentstyle[12pt]{article}
\input epsf
\global\arraycolsep=2pt

\catcode`\@=11
\long\def\@makefntext#1{ %\parindent 1em
\protect\noindent \hbox to 3.2pt {\hskip-.9pt
 $^{{\ninerm\@thefnmark}}$\hfil}#1\hfill} %can be used

\def\thefootnote{\fnsymbol{footnote}}
\def\@makefnmark{\hbox to 0pt{$^{\@thefnmark}$\hss}}  %original

\def\ps@myheadings{\let\@mkboth\@gobbletwo
\def\@oddhead{\hbox{} %\sl
\rightmark\hfil\ninerm\thepage}
\def\@oddfoot{}\def\@evenhead{\ninerm\thepage\hfil %\sl
\leftmark\hbox{}}\def\@evenfoot{}
\def\sectionmark##1{}\def\subsectionmark##1{}}

\begin{document}

%----------------------------PROCSLA.STY---------------------------
\newcommand{\symbolfootnote}{\renewcommand{\thefootnote}
        {\fnsymbol{footnote}}}
\renewcommand{\thefootnote}{\fnsymbol{footnote}}
\newcommand{\alphfootnote}
        {\setcounter{footnote}{0}
         \renewcommand{\thefootnote}{\sevenrm\alph{footnote}}}

%------------------------------------------------------------------
%NEW DEFINED SECTION COMMANDS
%% FOLLOWING LINE CANNOT BE BROKEN BEFORE 70 CHAR
%% FOLLOWING LINE CANNOT BE BROKEN BEFORE 70 CHAR
%% FOLLOWING LINE CANNOT BE BROKEN BEFORE 70 CHAR
\newcounter{sectionc}\newcounter{subsectionc}\newcounter{subsubsectionc}

\renewcommand{\section}[1] {\vspace{0.6cm}\addtocounter{sectionc}{1}
\setcounter{subsectionc}{0}\setcounter{subsubsectionc}{0}\noindent
        {\bf\thesectionc. #1}\par\vspace{0.4cm}}
\renewcommand{\subsection}[1]
{\vspace{0.6cm}\addtocounter{subsectionc}{1}
        \setcounter{subsubsectionc}{0}\noindent
        {\it\thesectionc.\thesubsectionc. #1}\par\vspace{0.4cm}}
\renewcommand{\subsubsection}[1]
{\vspace{0.6cm}\addtocounter{subsubsectionc}{1}
        \noindent
{\rm\thesectionc.\thesubsectionc.\thesubsubsectionc.
        #1}\par\vspace{0.4cm}}
\newcommand{\nonumsection}[1] {\vspace{0.6cm}\noindent{\bf #1}
        \par\vspace{0.4cm}}

%NEW MACRO TO HANDLE APPENDICES
\newcounter{appendixc}
\newcounter{subappendixc}[appendixc]
\newcounter{subsubappendixc}[subappendixc]
%% FOLLOWING LINE CANNOT BE BROKEN BEFORE 70 CHAR
%% FOLLOWING LINE CANNOT BE BROKEN BEFORE 70 CHAR
%% FOLLOWING LINE CANNOT BE BROKEN BEFORE 70 CHAR
\renewcommand{\thesubappendixc}{\Alph{appendixc}.\arabic{subappendixc}}
\renewcommand{\thesubsubappendixc}

{\Alph{appendixc}.\arabic{subappendixc}.\arabic{subsubappendixc}}

\renewcommand{\appendix}[1] {\vspace{0.6cm}
        \refstepcounter{appendixc}
        \setcounter{figure}{0}
        \setcounter{table}{0}
        \setcounter{equation}{0}
        \renewcommand{\thefigure}{\Alph{appendixc}.\arabic{figure}}
        \renewcommand{\thetable}{\Alph{appendixc}.\arabic{table}}
        \renewcommand{\theappendixc}{\Alph{appendixc}}

\renewcommand{\theequation}{\Alph{appendixc}.\arabic{equation}}
%       \noindent{\bf Appendix \theappendixc. #1}\par\vspace{0.4cm}}
        \noindent{\bf Appendix \theappendixc #1}\par\vspace{0.4cm}}
\newcommand{\subappendix}[1] {\vspace{0.6cm}
        \refstepcounter{subappendixc}
        \noindent{\bf Appendix \thesubappendixc.
#1}\par\vspace{0.4cm}}
\newcommand{\subsubappendix}[1] {\vspace{0.6cm}
        \refstepcounter{subsubappendixc}
        \noindent{\it Appendix \thesubsubappendixc. #1}
        \par\vspace{0.4cm}}

%-------------------------------------------------------------------
%MARCO FOR ABSTRACT BLOCK
\def\abstracts#1{{

\centering{\begin{minipage}{30pc}\tenrm\baselineskip=12pt\noindent
        \centerline{\tenrm ABSTRACT}\vspace{0.3cm}
        \parindent=0pt #1
        \end{minipage} }\par}}

%-------------------------------------------------------------------
%NEW MACRO FOR BIBLIOGRAPHY
\newcommand{\bibit}{\it}
\newcommand{\bibbf}{\bf}
\renewenvironment{thebibliography}[1]
        {\begin{list}{\arabic{enumi}.}
        {\usecounter{enumi}\setlength{\parsep}{0pt}
%1.25cm IS STRICTLY FOR PROCSLA.TEX ONLY
\setlength{\leftmargin 1.25cm}{\rightmargin 0pt}
%0.52cm IS FOR NEW DATA FILES
%\setlength{\leftmargin 0.52cm}{\rightmargin 0pt}
         \setlength{\itemsep}{0pt} \settowidth
        {\labelwidth}{#1.}\sloppy}}{\end{list}}

%-------------------------------------------------------------------
%FOLLOWING THREE COMMANDS ARE FOR 'LIST' COMMAND.
\topsep=0in\parsep=0in\itemsep=0in
\parindent=1.5pc

%LIST ENVIRONMENTS
\newcounter{itemlistc}
\newcounter{romanlistc}
\newcounter{alphlistc}
\newcounter{arabiclistc}
\newenvironment{itemlist}
        {\setcounter{itemlistc}{0}
         \begin{list}{$\bullet$}
        {\usecounter{itemlistc}
         \setlength{\parsep}{0pt}
         \setlength{\itemsep}{0pt}}}{\end{list}}

\newenvironment{romanlist}
        {\setcounter{romanlistc}{0}
         \begin{list}{$($\roman{romanlistc}$)$}
        {\usecounter{romanlistc}
         \setlength{\parsep}{0pt}
         \setlength{\itemsep}{0pt}}}{\end{list}}

\newenvironment{alphlist}
        {\setcounter{alphlistc}{0}
         \begin{list}{$($\alph{alphlistc}$)$}
        {\usecounter{alphlistc}
         \setlength{\parsep}{0pt}
         \setlength{\itemsep}{0pt}}}{\end{list}}

\newenvironment{arabiclist}
        {\setcounter{arabiclistc}{0}
         \begin{list}{\arabic{arabiclistc}}
        {\usecounter{arabiclistc}
         \setlength{\parsep}{0pt}
         \setlength{\itemsep}{0pt}}}{\end{list}}

%-------------------------------------------------------------------
%FIGURE CAPTION
\newcommand{\fcaption}[1]{
        \refstepcounter{figure}
        \setbox\@tempboxa = \hbox{\tenrm Fig.~\thefigure. #1}
        \ifdim \wd\@tempboxa > 6in
           {\begin{center}
        \parbox{6in}{\tenrm\baselineskip=12pt Fig.~\thefigure. #1 }
            \end{center}}
        \else
             {\begin{center}
             {\tenrm Fig.~\thefigure. #1}
              \end{center}}
        \fi}

%TABLE CAPTION
\newcommand{\tcaption}[1]{
        \refstepcounter{table}
        \setbox\@tempboxa = \hbox{\tenrm Table~\thetable. #1}
        \ifdim \wd\@tempboxa > 6in
           {\begin{center}
        \parbox{6in}{\tenrm\baselineskip=12pt Table~\thetable. #1 }
            \end{center}}
        \else
             {\begin{center}
             {\tenrm Table~\thetable. #1}
              \end{center}}
        \fi}

%-------------------------------------------------------------------
%ACKNOWLEDGEMENT: this portion is from John Hershberger
\def\@citex[#1]#2{\if@filesw\immediate\write\@auxout
        {\string\citation{#2}}\fi
\def\@citea{}\@cite{\@for\@citeb:=#2\do
        {\@citea\def\@citea{,}\@ifundefined
        {b@\@citeb}{{\bf ?}\@warning
        {Citation `\@citeb' on page \thepage \space undefined}}
        {\csname b@\@citeb\endcsname}}}{#1}}

\newif\if@cghi
\def\cite{\@cghitrue\@ifnextchar [{\@tempswatrue
        \@citex}{\@tempswafalse\@citex[]}}
\def\citelow{\@cghifalse\@ifnextchar [{\@tempswatrue
        \@citex}{\@tempswafalse\@citex[]}}
\def\@cite#1#2{{$\null^{#1}$\if@tempswa\typeout
        {IJCGA warning: optional citation argument
        ignored: `#2'} \fi}}
\newcommand{\citeup}{\cite}

%-------------------------------------------------------------------
%FOR FNSYMBOL FOOTNOTE AND ALPH{FOOTNOTE}
\def\fnm#1{$^{\mbox{\scriptsize #1}}$}
\def\fnt#1#2{\footnotetext{\kern-.3em
        {$^{\mbox{\sevenrm #1}}$}{#2}}}

%-------------------------------------------------------------------
\font\twelvebf=cmbx10 scaled\magstep 1
\font\twelverm=cmr10 scaled\magstep 1
\font\twelveit=cmti10 scaled\magstep 1
\font\elevenbfit=cmbxti10 scaled\magstephalf
\font\elevenbf=cmbx10 scaled\magstephalf
\font\elevenrm=cmr10 scaled\magstephalf
\font\elevenit=cmti10 scaled\magstephalf
\font\bfit=cmbxti10
\font\tenbf=cmbx10
\font\tenrm=cmr10
\font\tenit=cmti10
\font\ninebf=cmbx9
\font\ninerm=cmr9
\font\nineit=cmti9
\font\eightbf=cmbx8
\font\eightrm=cmr8
\font\eightit=cmti8

\thispagestyle{empty}
\begin{flushright}
CERN-TH.7225/94
\end{flushright}
\vspace{2cm}

\centerline{\twelvebf HEAVY QUARK MASSES, MIXING ANGLES,}
\baselineskip=16pt
\centerline{\twelvebf AND SPIN-FLAVOUR SYMMETRY}
\vspace{1.0cm}
\centerline{\tenrm MATTHIAS NEUBERT\footnote{Address before
Oct.~1993: Stanford Linear Accelerator Center, Stanford University,
Stanford, California 94309}}
\baselineskip=15pt
\centerline{\tenit Theory Division, CERN}
\baselineskip=12pt
\centerline{\tenit CH-1211 Geneva 23, Switzerland}
\vspace{1.1cm}
\abstracts{In a series of three lectures, I review the theory and
phenomenology of heavy quark masses and mixing angles and the status
of their determination. In addition, I give an introduction to heavy
quark symmetry, with the main emphasis on the development of heavy
quark effective field theory and its application to obtain a
model-independent determination of $|\,V_{cb}|$.}

\twelverm
\baselineskip=14pt
\vspace{3.5cm}

\centerline{(Lectures presented at TASI-93, Boulder, Colorado, 1993)}
\vfill
\noindent
CERN-TH.7225/94\\
April 1994
\newpage

\setcounter{page}{1}
\noindent
{\bf Introduction}
\vspace{0.4cm}

The rich phenomenology of weak decays has always been a source of
information about the structure of elementary particle interactions.
A long time ago, $\beta$- and $\mu$-decay experiments revealed the
nature of the effective flavour-changing interactions at low momentum
transfer. Today, we are in a similar situation: Weak decays of
hadrons containing heavy quarks are employed for tests of the
standard model and measurements of its parameters. In particular,
they offer the most direct way to determine the weak mixing angles
and to test the unitarity of the Kobayashi--Maskawa matrix. On the
other hand, hadronic weak decays also serve as a probe of that part
of strong-interaction phenomenology which is least understood: the
confinement of quarks and gluons into hadrons. In fact, it is this
intricate interplay between weak and strong interactions that makes
this field challenging and attractive to many theorists.

Over the last decade or so, a lot of information has been collected
about heavy quark decays from experiments on the $\Upsilon(4s)$
resonance, and more recently at $e^+ e^-$ and hadron colliders. This
has led to a rather detailed knowledge of the flavour sector of the
standard model and many of the parameters associated with it. There
have been several great discoveries in this field, such as
$B_d$--$\bar B_d$ mixing\cite{BBbar1,BBbar2}, $b\to u$
transitions\cite{btou1}$^-$\cite{btou4}, and rare decays induced by
penguin operators\cite{BKstar}. Yet there is much more to come.
Hopefully, the approval of the first $B$-meson factory at SLAC has
opened the way for a bright future for $B$-physics. At the same time,
upgrades of the existing facilities at Cornell, Fermilab, and LEP
will provide a wealth of data within the coming years.

However, my main purpose here is to talk about theory, and
fortunately there has been a lot of progress and enthusiasm in this
field in recent years. This is related to the discovery of heavy
quark symmetry\cite{Shu1}$^-$\cite{Isgu} and the development of the
heavy quark effective theory\cite{EiFe}$^-$\cite{Luke}, which is a
low-energy effective theory that describes the strong interactions of
a heavy quark with light quarks and gluons. The excitement about
these developments is caused by the fact that they allow (some)
model-independent predictions in an area in which ``progress'' in
theory often meant nothing more than the construction of a new model,
which could be used to estimate some strong-interaction hadronic
matrix elements. Therefore, I hope that you do not mind it if the
part about the heavy quark effective theory constitutes the main body
of these notes; this is where the main progress has been achieved
from the theoretical point of view. Also, I have recently finished a
long review article on heavy quark symmetry\cite{review}, which
clearly simplifies the task of writing up lecture notes.

In the first lecture, I will discuss the theory of heavy quark
masses, their definition in perturbation theory, and their
determination from QCD sum rules. The second lecture provides an
introduction to the standard model description of quark mixing,
different parametrizations of the Kobayashi--Maskawa matrix, the
status of the determination of the mixing angles, and the physics of
the unitarity triangle. The third lecture, which covers more than
half of these notes, is devoted to a review of the ideas on heavy
quark symmetry and the formalism of the heavy quark effective theory.
In particular, I will discuss the theory of the model-independent
determination of $|\,V_{cb}|$ from exclusive semileptonic $B\to
D^*\ell\,\bar\nu$ decays. At the end of each lecture you will find
some suggestions for little exercises, which are typically related to
the derivation of important equations given in the notes. I have
tried to select problems that are fun to solve and contain some
interesting pieces of physics. You are invited to see if I am right.

%%%   Lecture 1   %%%%%%%%%%%%%%%%%%%%%%%%%%%%%%%%%%%%%%%%%%%%%%%%%%%

\newpage
\section{Heavy Quark Masses}

Because of confinement at large distance scales, quarks and gluons do
not appear among the physical states of QCD. There is thus no
natural, physical definition of quark masses. Rather, several
definitions are possible and have been proposed, and it is often a
matter of convenience which one to use. Most of these definitions are
tied to the framework of perturbation theory. In this lecture, I will
discuss some of the most common mass definitions and their
interrelation. Special emphasis is put on the discussion of the
running quark mass in the context of the renormalization group. I
will then discuss how values for the bottom and charm quark masses
are obtained from QCD sum rules. Very briefly, I will touch upon the
subtle issue of an infrared renomalon in the pole mass, which implies
an intrinsic uncertainty in these determinations.

\subsection{Quark Mass Definitions}

For heavy quarks, the nonrelativistic bound-state picture suggests
the notion of a pole mass $m_Q^{\rm pole}$ defined, order by order in
perturbation theory, as the position of the singularity in the
renormalized quark propagator. One can show that the pole mass is
gauge-invariant, infrared finite, and renormalization-scheme
independent\cite{Tarr}. It is thus a meaningful ``physical''
parameter as long as the heavy quark is not exactly on-shell. For
instance, the pole mass appears in the formula for the energy levels
in quarkonium systems.

An alternative gauge-invariant definition is provided by the running
quark mass in some subtraction scheme. One usually works with the
modified minimal subtraction ($\overline{\rm MS}$) scheme\cite{Bura}
and denotes the running mass by $\overline{m}_Q(\mu)$. Running quark
masses are used when there is a large momentum scale $\mu\gg m_Q$ in
the problem, so as to absorb large logarithms, which would otherwise
render perturbation theory invalid. Of course, one can use other
subtraction schemes such as minimal subtraction\cite{tHo2} (MS).
Sometimes, it may even be convenient to use a gauge-dependent
definition of a heavy quark mass such as the so-called Euclidean mass
$m_Q^{\rm Eucl}$, which is defined by a subtraction at the Euclidean
point\cite{SVZ} $p^2=-m_Q^2$. What is important is that these
perturbative definitions are related to each other in a calculable
way, order by order in perturbation theory.

One may also think of obvious nonperturbative definitions of a heavy
quark mass. For instance, one could define $m_Q$ to be one half of
the mass of the ground state in the corresponding $(Q\bar Q)$
quarkonium system, or as the mass of the lightest hadron containing
the heavy quark, or as that mass minus some fixed binding energy,
etc. The problem is, of course, that these definitions are ad hoc,
and relating them to each other or to the perturbative definitions
given above would require to solve QCD, a task that is presently
beyond our calculational skills. There is, however, a bridge between
the two classes of definitions, which is provided by the heavy quark
effective theory, which is an effective field theory appropriate to
describe the soft interactions of an almost on-shell heavy quark with
light quarks and gluons. There, one can define a parameter
$\bar\Lambda$, which corresponds to the effective mass of the light
degrees of freedom in a hadron $H_Q$ containing a single heavy quark
$Q$, in terms of a gauge-invariant, infrared finite, and
renormalization-scheme independent hadronic matrix element\cite{FNL}.
One can then define a heavy quark mass $m_Q^*$ as\cite{Neu3}
$m_Q^*=m_{H_Q}-\bar\Lambda$. In perturbation theory, and up to
corrections of order $1/m_Q$, the mass $m_Q^*$ coincides with the
pole mass. However, the above definition is more general, as it does
not rely on a perturbative expansion.

\subsection{Quark Masses in Perturbation Theory}

\begin{figure}[htb]
   \vspace{0.5cm}
   \epsfxsize=11cm
   \centerline{\epsffile{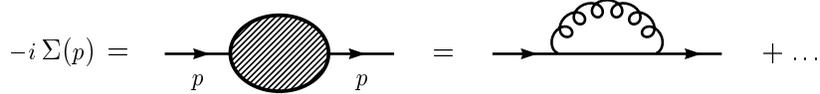}}
   \centerline{\parbox{11cm}{\caption{\label{fig:1.1}
The quark self-energy to one-loop order in QCD.
   }}}
\end{figure}

Let me now discuss the concept of heavy quark masses to one-loop
order in perturbation theory. The bare quark mass $m_Q^{\rm bare}$
appearing in the QCD Lagrangian is related to the renormalized mass
$m_Q^{\rm ren}$ by a counter term,
\begin{equation}
   m_Q^{\rm bare} = m_Q^{\rm ren} - \delta m_Q \,,
\end{equation}
where $m_Q^{\rm bare}$ and $\delta m_Q$ are divergent quantities. The
counter term is chosen such that the renormalized mass is finite. Its
value depends, however, on the subtraction prescription. In the
$\overline{\rm MS}$ scheme, the renormalized mass will depend on some
subtraction scale $\mu$, i.e.\ $m_Q^{\rm ren}=\overline{m}_Q(\mu)$.
To calculate $\delta m_Q$ in perturbation theory, one has to evaluate
the heavy quark self-energy $\Sigma(p)$ shown in Fig.~\ref{fig:1.1}.
The relation is
\begin{equation}
   \delta m_Q = \mbox{divergent part of $\Sigma(\rlap{\,/}p=m_Q)$
   + scheme-dependent finite terms.}
\end{equation}
To calculate the self-energy at one-loop order, it is convenient to
use dimensional regularization\cite{tHo1,Boll}, i.e.\ to work in
$D=4-2\epsilon$ space-time dimensions. Then the renormalized coupling
constant is related to the bare one by\cite{DGro}
\begin{equation}
   \alpha_s^{\rm bare} = Z_\alpha\,\alpha_s(\mu)\,\mu^{2\epsilon}
   = \alpha_s\,\mu^{2\epsilon} + O(\alpha_s^2) \,.
\end{equation}
{}From a straightforward calculation, one obtains at one-loop order
\begin{eqnarray}\label{Sigmap}
   \Sigma(\rlap{\,/}p=m_Q) &=& m_Q\,{\alpha_s\over 3\pi}\,
    \bigg( {m_Q^2\over 4\pi\mu^2} \bigg)^{-\epsilon}\,
    \Gamma(\epsilon)\,{3-2\epsilon\over 1-2\epsilon} \,,
    \nonumber\\
   &=& m_Q\,{\alpha_s\over\pi}\,
    \bigg( {1\over\hat\epsilon} + \ln{\mu^2\over m_Q^2}
    + {4\over 3} \bigg) + O(\epsilon) \,,
\end{eqnarray}
where $1/\hat\epsilon=1/\epsilon-\gamma+\ln 4\pi$. Notice that this
result is still $\mu$-independent, since the $\mu$ dependence of
$\alpha_s/\hat\epsilon$ cancels the $\mu$-dependent logarithm. A
scale dependence appears only when one subtracts the ultraviolet
divergence. In the $\overline{\rm MS}$ scheme, the mass counter term
is defined to subtract the $1/\hat\epsilon$-pole in the self-energy,
i.e.
\begin{eqnarray}
   \delta m_Q^{\overline{\rm MS}} &=& m_Q\,
    {\alpha_s\over\pi\hat\epsilon} \,, \nonumber\\
   \overline{m}_Q(\mu) &=& m_Q\,\bigg( 1
    + {\alpha_s\over\pi\hat\epsilon} \bigg) \,.
\end{eqnarray}
The pole mass, on the other hand, is defined so as to absorb the
entire self-energy on-shell. Hence
\begin{eqnarray}
   \delta m_Q^{\rm pole} &=& \Sigma(\rlap{\,/}p=m_Q) \,, \nonumber\\
   m_Q^{\rm pole} &=& m_Q\,\bigg\{ 1
    + {\alpha_s\over\pi}\,\bigg( {1\over\hat\epsilon}
    + \ln{\mu^2\over m_Q^2} + {4\over 3} \bigg) \bigg\} \,.
\end{eqnarray}
Comparing the two results, one obtains
\begin{eqnarray}\label{MSpolrel}
   \overline{m}_Q(\overline{m}_Q) &=& m_Q^{\rm pole}\,
    \bigg\{ 1 - {4\alpha_s(\overline{m}_Q)\over 3\pi} \bigg\}
    \,, \nonumber\\
   \overline{m}_Q(\mu) &=& \overline{m}_Q(\overline{m}_Q)\,
    \bigg\{ 1 - {\alpha_s\over\pi}\,
    \ln{\mu^2\over\overline{m}_Q^2} \bigg\} \,.
\end{eqnarray}
The first equation relates different perturbative definitions of the
heavy quark mass, namely the pole mass to the running mass in the
$\overline{\rm MS}$ scheme evaluated at the scale
$\mu=\overline{m}_Q$. I have written this relation as a perturbative
expansion in powers of $\alpha_s(\overline{m}_Q)$. This expansion is
also known to two-loop order from a calculation by Gray et
al.\cite{Gray}\/ Similar relations exist in other subtraction
schemes, e.g.
\begin{eqnarray}\label{meucl}
   m_Q^{\rm MS}(m_Q^{\rm MS}) &=& m_Q^{\rm pole}\,
    \bigg\{ 1 - {\alpha_s(m_Q^{\rm MS})\over\pi}\,
    \bigg( {4\over 3} - \gamma + \ln 4\pi \bigg) \bigg\}
    \,, \nonumber\\
   m_Q^{\rm Eucl}(m_Q^{\rm Eucl}) &=& m_Q^{\rm pole}\,
    \bigg\{ 1 - {\alpha_s(m_Q^{\rm Eucl})\over\pi}\,
    2\ln 2 \bigg\} \,,
\end{eqnarray}
where the gauge-dependent Euclidean mass is defined in the Landau
gauge\cite{SVZ}.

The second relation in (\ref{MSpolrel}) determines the running of the
heavy quark mass. In QCD, quarks become ``lighter'' at high-energy
scales. The simple one-loop calculation presented above is, however,
not sufficient to scale $\overline{m}_Q(\mu)$ up to very large
scales. For instance, using (\ref{MSpolrel}) to extrapolate the mass
of the bottom quark up to a typical grand unification scale $M_{\rm
GUT}\sim 10^{16}$ GeV, one would obtain a meaningless result:
\begin{equation}
   {\overline{m}_b(M_{\rm GUT})\over\overline{m}_b(\overline{m}_b)}
   = 1 - {\alpha_s\over\pi}\,\ln{M_{\rm GUT}^2\over\overline{m}_b^2}
   \simeq 1 - 71\,{\alpha_s\over\pi} = \,\, ?
\end{equation}
Of course, the problem of large logarithms in perturbation theory is
not specific to this case, but is encountered frequently. Large
logarithms can be controlled in a systematic way using the beautiful
machinery of the renormalization group. It is worth going through the
solution in great detail.

\subsection{Renormalization-Group Improvement}

For the purpose of this section, I will switch to a slightly more
general notation and rewrite (\ref{MSpolrel})
as
\begin{equation}\label{mumu0}
   m(\mu) = m(\mu_0)\,\bigg\{ 1 - {\alpha_s\over\pi}\,
   \ln{\mu^2\over\mu_0^2} + O(\alpha_s^2)  \bigg\} \,.
\end{equation}
If $\mu\gg\mu_0$, the logarithm can be so large that $\alpha_s\ln
\mu^2/\mu_0^2$ becomes of order unity, and ordinary perturbation
theory breaks down. In fact, one can show that at any given order in
perturbation theory there will be large logarithms of the type
\begin{equation}\label{leadlogs}
   \bigg( \alpha_s\,\ln{\mu^2\over\mu_0^2} \bigg)^n \sim
   \bigg( {\ln\mu^2/\mu_0^2\over\ln\mu^2/\Lambda_{\rm QCD}^2}
   \bigg)^n \,,
\end{equation}
where I have used the fact that the running coupling constant scales
like $\alpha_s(\mu)\sim 1/\ln(\mu^2/\Lambda_{\rm QCD}^2)$. It is
necessary to resum these ``leading logarithms'' to all orders in
perturbation theory. This is achieved by solving the
renormalization-group equation (RGE) for the running quark mass,
\begin{equation}
   \mu\,{\mbox{d}\over\mbox{d}\mu}\,m(\mu)
   = \gamma(\alpha_s)\,m(\mu) \,.
\end{equation}
The anomalous dimension $\gamma$ has a perturbative expansion in
the renormalized coupling constant:
\begin{equation}\label{g0g1def}
   \gamma(\alpha_s) = \gamma_0\,{\alpha_s\over 4\pi}
   + \gamma_1\,\bigg( {\alpha_s\over 4\pi} \bigg)^2 + \ldots \,.
\end{equation}
The coefficients $\gamma_i$ are known to three-loop
order\cite{Tara,Gori}. For our purposes, however, it is sufficient to
note that\cite{Tarr}
\begin{equation}
   \gamma_0 = -8 \,,\qquad
   \gamma_1 = -{404\over 3} + {40\over 9}\,n_f \,,
\end{equation}
where $n_f$ denotes the number of quark flavours with mass below
$\mu$. Throughout these notes I will evaluate the QCD coefficients
for $N_c=3$ colours, and only display the dependence on the number of
flavours explicitly. The value of $\gamma_0$ follows from the
one-loop result (\ref{mumu0}).

The next step is to rewrite the RGE in the form of a partial
differential equation, making explicit the scale dependence of the
renormalized coupling constant. This gives
\begin{equation}\label{RGEfin}
   \bigg( \mu\,{\partial\over\partial\mu}
   + \beta(\alpha_s)\,{\partial\over\partial\alpha_s(\mu)}
   - \gamma(\alpha_s) \bigg)\,m(\mu) = 0 \,.
\end{equation}
The $\beta$-function
\begin{equation}
   \beta\big(\alpha_s) = \mu\,
   {\partial\alpha_s(\mu)\over\partial\mu}
   = -2\alpha_s\,\bigg[\, \beta_0\,{\alpha_s\over 4\pi}
   + \beta_1\,\bigg( {\alpha_s\over 4\pi} \bigg)^2 + \ldots \bigg]
\end{equation}
describes the running of the coupling constant. The one- and two-loop
coefficients are\cite{Gros}$^-$\cite{Bela}
\begin{equation}
   \beta_0 = 11 - {2\over 3}\,n_f \,,\qquad
   \beta_1 = 102 - {38\over 3}\,n_f \,.
\end{equation}
The exact solution of the RGE can be written in the form
\begin{equation}
   m(\mu) = U(\mu,\mu_0)\,m(\mu_0) \,,
\end{equation}
with the evolution operator\cite{Bura,Flor,BJLW}
\begin{equation}\label{Uoper}
   U(\mu,\mu_0) = \exp\!\int\limits_{\displaystyle\alpha_s(\mu_0)}
   ^{\displaystyle\alpha_s(\mu)}\!\!\mbox{d}\alpha\,
   {\gamma(\alpha)\over\beta(\alpha)} \,.
\end{equation}
The trick is to obtain a perturbative expansion in the exponent of
this expression by inserting the expansions for the anomalous
dimension and $\beta$-function. After a simple calculation, one finds
\begin{equation}\label{Uevol}
   U(\mu,\mu_0) = \Bigg( {\alpha_s(\mu_0)\over\alpha_s(\mu)} \Bigg)
   ^{\displaystyle{\gamma_0/2\beta_0}} \bigg\{ 1
   + {\alpha_s(\mu_0) - \alpha_s(\mu)\over 4\pi}\,
   {\gamma_1\beta_0 - \beta_1\gamma_0\over 2\beta_0^2}
   + O(\alpha_s^2) \bigg\} \,.
\end{equation}
In this result, the running coupling constant has two-loop accuracy.
The corresponding expression is
\begin{equation}\label{asmu}
   \alpha_s(\mu) = {4\pi\over\beta_0\ln(\mu^2/\Lambda_{\rm QCD}^2)}
   \,\bigg[ 1 - {\beta_1\over\beta_0^2}\,
   {\ln\ln(\mu^2/\Lambda_{\rm QCD}^2)
    \over\ln(\mu^2/\Lambda_{\rm QCD}^2)} \bigg] \,,
\end{equation}
where $\Lambda_{\rm QCD}$ is a scheme-dependent scale
parameter\footnote{The value of $\alpha_s(\mu_0)$ at some reference
scale $\mu_0$ can be used to eliminate the dependence on
$\Lambda_{\rm QCD}$. Nowadays, it is convenient to choose
$\mu_0=m_Z$, since $\alpha_s(m_Z)$ is known with high accuracy.}.
The factor containing the ratio of the running coupling constants in
(\ref{Uevol}) sums the leading logarithms (\ref{leadlogs}) to all
orders in perturbation theory. Keeping just this factor corresponds
to the so-called leading logarithmic approximation (LLA), which is
often used to attack problems containing widely separated scales.
Note that to calculate this factor, one only needs to compute the
one-loop coefficient of the anomalous dimension. This is a rather
trivial task, as it suffices to calculate the $1/\epsilon$-pole in
the quark self-energy. Once the leading scaling behaviour has been
factored out, perturbation theory becomes well-behaved, i.e.\ the
next-to-leading corrections in the parenthesis in (\ref{Uevol}) obey
a perturbative expansion free of large logarithms. This is why the
approach is called ``renormalization-group improved perturbation
theory''. The terms of order $\alpha_s$, which I have shown
explicitly, correspond to the next-to-leading logarithmic
approximation (NLLA). Including them, one sums logarithms of the type
\begin{equation}
   \alpha_s\,\bigg( \alpha_s\,\ln{\mu^2\over\mu_0^2} \bigg)^n
\end{equation}
to all orders. To achieve such an accuracy, it is necessary to
calculate the two-loop coefficients of the anomalous dimension and
$\beta$-function.

Combining the above results and setting $\mu_0=m_Q$, one obtains the
running quark mass to next-to-leading order in renormalization-group
improved perturbation theory. The result is
\begin{equation}\label{mscal}
   m_Q(\mu) = m_Q(m_Q)\,\Bigg( {\alpha_s(\mu)\over\alpha_s(m_Q)}
   \Bigg)^{4/\beta_0} \bigg\{ 1 + S\,
   {\alpha_s(m_Q) - \alpha_s(\mu)\over\pi} + O(\alpha_s^2)
   \bigg\} \,,
\end{equation}
where
\begin{equation}
   S = -{5\over 6} + {34\over 3\beta_0} - {107\over\beta_0^2} \,.
\end{equation}
It is to be understood that the number of flavours changes as $\mu$
crosses a quark threshold\cite{ApCa}. For instance, when one uses
(\ref{mscal}) to scale the running bottom quark mass up to large
scales, $n_f$ changes from 5 to 6 when $\mu$ crosses the top quark
mass. At the same time, the QCD scale parameter $\Lambda_{\rm QCD}$
in expression (\ref{asmu}) for the running coupling constant changes,
so that $\alpha_s(\mu)$ is a continuous function. From the point of
view of convergence of perturbation theory, eq.~(\ref{mscal}) can be
used to evaluate the running mass at an arbitrarily large scale
$\mu$. However, I have derived this result assuming that there are
only QCD interactions. But at high-energy scales, other gauge and
Yukawa interactions become important as well. They are most
significant for the running of the top quark mass, as I will discuss
in detail in the next section. In the case of the bottom quark, the
result that renormalization effects lower the running mass at
high-energy scales remains true. In fact, in many extensions of the
standard model it is possible to obtain a unification of the
bottom quark and the tau lepton masses at a typical grand unification
scale:
\begin{equation}
   m_b(M_{\rm GUT}) = m_\tau(M_{\rm GUT}) \quad\mbox{for}\quad
   M_{\rm GUT}\sim 10^{16}~\mbox{GeV} \,.
\end{equation}
For more details on this, I refer to the lectures by L. Hall in this
volume.

\subsection{Running Top Quark Mass}

The interplay of gauge and Yukawa interactions leads to very
interesting effects on the evolution of the top quark
mass\cite{PeRo,CHil}. In my discussion below, I will focus on the
standard model. The analysis for extensions of the standard model
such as supersymmetry proceeds in a similar
way\cite{yyy1}$^-$\cite{yyy3}.

\begin{figure}[htb]
   \vspace{0.5cm}
   \epsfxsize=9cm
   \centerline{\epsffile{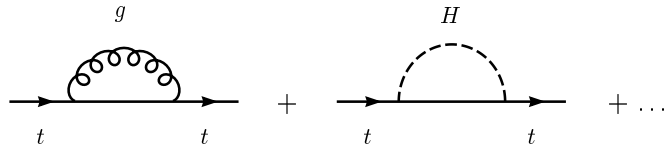}}
   \centerline{\parbox{11cm}{\caption{\label{fig:1.2}
Dominant one-loop contributions to the self-energy of the top quark
in the standard model.
   }}}
\end{figure}

\begin{figure}[htb]
   \vspace{0.5cm}
   \epsfxsize=10cm
   \centerline{\epsffile{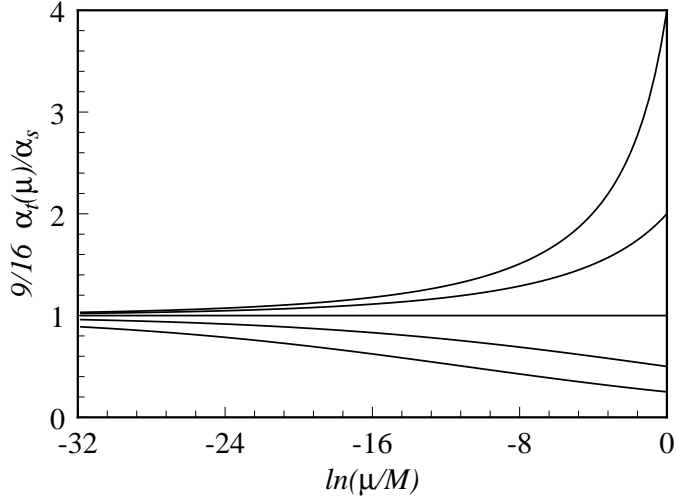}}
   \vspace{-0.5cm}
   \centerline{\parbox{11cm}{\caption{\label{fig:1.3}
The fixed-point structure of the RGE (\protect\ref{atrun}) in the
case $\alpha_s=\mbox{const}$., with $4\alpha_s/\pi=0.1$.
   }}}
\end{figure}

Let me write the running top quark mass in terms of the Yukawa
coupling $\lambda_t(\mu)$:
\begin{equation}
   m_t(\mu) = \lambda_t(\mu)\,{v(\mu)\over\sqrt{2}} \,,
\end{equation}
where $v$ denotes the vacuum expectation value of the Higgs field,
normalized so that $v(m_W)\simeq 246$ GeV. It is convenient to define
a running coupling constant $\alpha_t(\mu)$ as
\begin{equation}
   \alpha_t(\mu) = {\lambda_t^2(\mu)\over 4\pi} \,.
\end{equation}
The most important contributions to the evolution of the running mass
$m_t(\mu)$ come from the QCD interactions as well as from the large
top Yukawa coupling. The relevant one-loop diagrams are depicted in
Fig.~\ref{fig:1.2}. They lead to the RGE
\begin{equation}\label{mtrun}
   \mu\,{\mbox{d}\over\mbox{d}\mu}\,\ln m_t(\mu)
   = -8\,{\alpha_s(\mu)\over 4\pi}
   + {3\over 2}\,{\alpha_t(\mu)\over 4\pi} + \ldots \,,
\end{equation}
where the ellipses represent other contributions, which I will
neglect. Combining this with the RGE for $v(\mu)$,
\begin{equation}
   \mu\,{\mbox{d}\over\mbox{d}\mu}\,\ln v(\mu)
   = -3\,{\alpha_t(\mu)\over 4\pi} + \ldots \,,
\end{equation}
one obtains
\begin{equation}\label{atrun}
   \mu\,{\mbox{d}\over\mbox{d}\mu}\,\ln\alpha_t(\mu)
   = -16\,{\alpha_s(\mu)\over 4\pi}
   + 9\,{\alpha_t(\mu)\over 4\pi} + \ldots \,.
\end{equation}
To get an idea of the structure of this equation, suppose
first that the QCD coupling constant is not running, i.e.\ $\alpha_s=
\mbox{const}$. Then the RGE has an infrared-stable fixed point at
$\alpha_t=\frac{16}{9}\,\alpha_s$, meaning that irrespective of the
value of $\alpha_t(\mu)$ at large scales, the Yukawa coupling is
attracted into the fixed point as $\mu$ decreases. This is
illustrated in Fig.~\ref{fig:1.3}. The resulting fixed-point value of
the top quark mass, $m_t = \frac{4}{3}\sqrt{2\pi\alpha_s}\,v$, is
entirely determined by the low-energy group structure of the theory,
which is responsible for the one-loop coefficients in (\ref{atrun}).

For running $\alpha_s(\mu)$, one has to find the simultaneous
solution of (\ref{atrun}) and the evolution equation for the strong
coupling constant,
\begin{equation}
   \mu\,{\mbox{d}\over\mbox{d}\mu}\,\ln\alpha_s(\mu)
   = -14\,{\alpha_s(\mu)\over 4\pi} \,.
\end{equation}
Depending on the initial conditions, one finds that there are still
``quasi-fixed point'' solutions, where the running of $\alpha_t(\mu)$
is connected to the running of $\alpha_s(\mu)$, but independent of
the value of the Yukawa coupling at large scales. In leading
logarithmic approximation, the exact solution is not too hard to
obtain. It reads
\begin{equation}\label{exasol}
   {\alpha_t(\mu)\over\alpha_t(M)} = \Bigg(
   {\alpha_s(\mu)\over\alpha_s(M)} \Bigg)^{8/7} \Bigg\{ 1 +
   {9\over 2}\,{\alpha_t(M)\over\alpha_s(M)}\Bigg[
   \Bigg( {\alpha_s(\mu)\over\alpha_s(M)} \Bigg)^{1/7} - 1 \Bigg]
   \Bigg\}^{-1} \,,
\end{equation}
where $M\gg\mu$ denotes some large mass scale, at which the initial
conditions are imposed. In order to illuminate the structure of this
equation, let me distinguish three cases:

\bigskip
\noindent
(i) $\alpha_t(M)\ll\alpha_s(M)$:

\noindent
In this case, the top Yukawa coupling is weak, and (\ref{exasol}) can
be approximated by
\begin{equation}
   \alpha_t(\mu) \simeq \alpha_t(M)\,
   \Bigg( {\alpha_s(\mu)\over\alpha_s(M)} \Bigg)^{8/7} \,,
\end{equation}
which is nothing but the standard QCD evolution [cf.~(\ref{mscal})].

\newpage
\noindent
(ii) $\alpha_t(M)=\frac{2}{9}\,\alpha_s(M)$:

\noindent
This case corresponds to the quasi-fixed point obtained by Pendleton
and Ross\cite{PeRo}, where (\ref{exasol}) simplifies to
\begin{equation}
   \alpha_t(\mu) = {2\over 9}\,\alpha_s(\mu) \,.
\end{equation}
However, to obtain this solution requires a fine-tuning at the large
scale $M$.

\begin{figure}[htb]
   \vspace{0.5cm}
   \epsfxsize=10cm
   \centerline{\epsffile{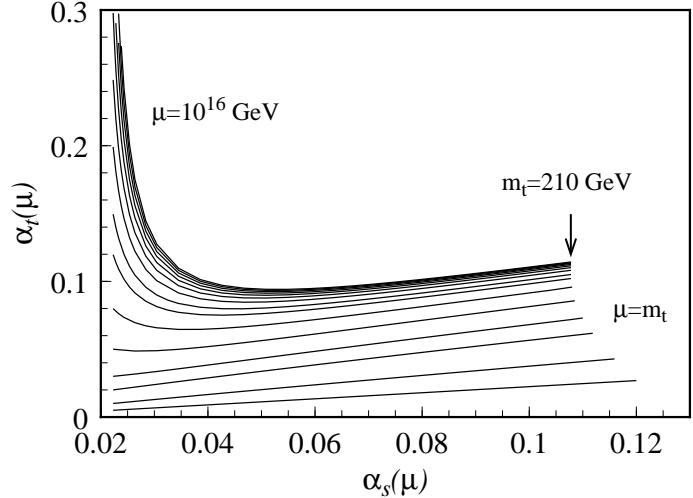}}
   \vspace{-0.5cm}
   \centerline{\parbox{11cm}{\caption{\label{fig:1.4}
Evolution of the coupling constants for different initial values of
$\alpha_t(M)$. I assume that $M=10^{16}$ GeV.
   }}}
\end{figure}

\bigskip
\noindent
(iii) $\alpha_t(M)\gg\alpha_s(M)$:

\noindent
This is the most interesting case of a large top Yukawa coupling.
Then the approximate solution of (\ref{exasol}) reads
\begin{equation}\label{Hillsol}
   \alpha_t(\mu) \simeq {2\over 9}\,\alpha_s(\mu)\,
   \Bigg[ 1 - \Bigg( {\alpha_s(M)\over\alpha_s(\mu)}
   \Bigg)^{1/7} \Bigg]^{-1} \,.
\end{equation}
This is the quasi-fixed point discovered by Hill\cite{CHil}. At low
scales $\mu\ll M$, $\alpha_t(\mu)$ follows $\alpha_s(\mu)$ and
becomes independent of the initial value $\alpha_t(M)$. The dynamics
that generates the top Yukawa coupling at some large scale $M$ cannot
be tested. An example of this mechanism is provided by the top
condensation model of Bardeen, Hill, and Lindner\cite{BHL}. Using
(\ref{Hillsol}), one obtains for the fixed-point value of the top
quark mass
\begin{equation}
   m_t(m_t) \simeq {2v(m_t)\over 3}\,\sqrt{\pi\,\alpha_s(m_t)}\,
   \Bigg[ 1 - \Bigg( {\alpha_s(M)\over\alpha_s(m_t)}
   \Bigg)^{1/7} \Bigg]^{-1/2} \simeq 210~\mbox{GeV} \,,
\end{equation}
where I have assumed $M\simeq 10^{16}$ GeV. Again, the result is
almost entirely determined by the group structure of the low-energy
theory. The dependence on $M$ is very weak.

In Fig.~\ref{fig:1.4}, I show the evolution of the coupling constants
according to (\ref{exasol}) for a set of initial values of
$\alpha_t(M)$. The quasi-fixed point behaviour is clearly visible
once $\alpha_t(M)$ is large enough. A more careful analysis including
all standard model contributions in the RGE gives\cite{CHil}
$m_t\simeq 225$ GeV (for $M\simeq 10^{16}$ GeV). Somewhat smaller
values $m_t<200$ GeV are obtained in the minimal supersymmetric
standard model\cite{yyy1}$^-$\cite{yyy3}, depending however upon the
value of $\tan\beta$.

\subsection{Determination of $m_b$ and $m_c$ from QCD Sum Rules}

In this section, I will discuss the extraction of the masses of the
bottom and charm quarks from QCD sum rules, and in particular from
the analysis of quarkonium spectra. The main idea of sum rules was
developed in the pioneering papers of Shifman, Vainshtein, and
Zakharov\cite{SVZ}. Their idea was to use quark--hadron duality to
obtain a prediction of hadron properties from calculations in the
quark--gluon theory of QCD. Consider, as an example, the vacuum
polarization induced by the electromagnetic current of a bottom
quark:
\begin{equation}
   i\int\mbox{d}^4 x\,e^{iq\cdot x}\,\langle\,0\,|\,T\,
   \big\{ j^\mu(x), j^\nu(0) \big\}\,|\,0\,\rangle
   = (q^\mu q^\nu - g^{\mu\nu} q^2)\,\Pi(Q^2) \,,
\end{equation}
where $j^\mu=\bar b\,\gamma^\mu b$, and $Q^2=-q^2$. The Lorentz
structure of the correlator follows from current conservation. The
function $\Pi(Q^2)$ satisfies a once-subtracted dispersion relation
\begin{equation}
   -{\mbox{d}\over\mbox{d}Q^2}\,\Pi(Q^2)
   = {1\over\pi} \int\mbox{d}s\,{1\over(s+Q^2)^2}\,
   \mbox{Im}\Pi(s) \,,
\end{equation}
where by the optical theorem
\begin{equation}
   \mbox{Im}\Pi(s) = {1\over 12\pi e_b^2}\,{3 s\over 4\pi\alpha^2}\,
   \sigma_s(e^+ e^-\to b\,\bar b)
\end{equation}
is related to a measurable cross section. Here, $e_b=-1/3$ is the
electric charge of the $b$-quark, and ``$b\,\bar b$'' is a short-hand
notation for ``hadrons containing $b\,\bar b$''. In the sum rule
analysis one considers the moments of the correlation function, which
are given by
\begin{equation}
   M_n = {1\over n!}\,\bigg(-{\mbox{d}\over\mbox{d}Q^2}\bigg)^n\,
   \Pi(Q^2)\Big|_{Q^2=0} = {1\over\pi} \int\mbox{d}s\,s^{-n-1}
   \,\mbox{Im}\Pi(s) \,.
\end{equation}
In principle, these moments can be extracted from experiment.

\begin{figure}[htb]
   \vspace{0.5cm}
   \epsfxsize=10cm
   \centerline{\epsffile{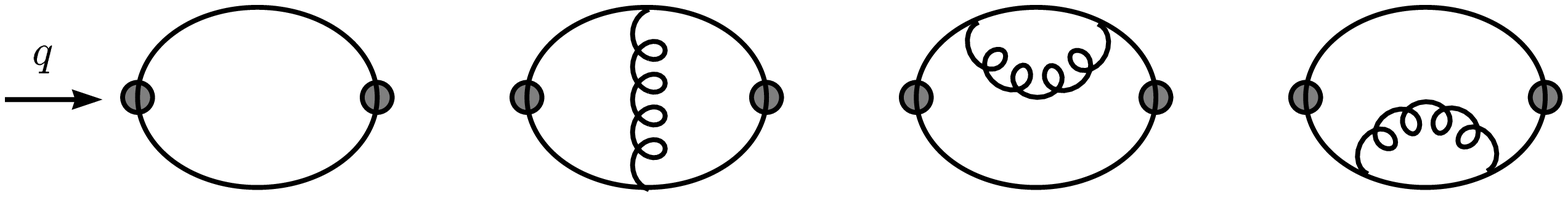}}
   \centerline{\parbox{11cm}{\caption{\label{fig:1.5}
Perturbative contributions to the correlator $\Pi(Q^2)$. The current
operators are represented by circles.
   }}}
\end{figure}

As long as $Q^2$ is not close to the resonance region, the function
$\Pi(Q^2)$ can be calculated in perturbative QCD. Since $Q^2=0$ is
far away from the physical threshold for bottomonium production, this
assumption is justified in the present case. Hence, perturbative QCD
should provide a good approximation to a calculation of the moments
$M_n$. The leading and next-to-leading perturbative contributions to
the correlator are shown in Fig.~\ref{fig:1.5}. They give rise to the
spectral density\cite{SVZ}
\begin{equation}\label{Pipert}
   \mbox{Im}\Pi_{\rm pert}(s) = {1\over 4\pi}\,{v(3-v^2)\over 2}\,
   \Theta(v^2)\,\bigg\{ 1 + {4\alpha_s\over 3\pi}\,\bigg[
   {\pi^2\over 2 v} - {3+v\over 4}\,\bigg( {\pi^2\over 2}
   - {3\over 4}\bigg) \bigg] \bigg\} \,,
\end{equation}
where $v=\sqrt{1 - 4 m_b^2/s}$ is the relative velocity of the heavy
quarks. In this expression one uses the Euclidean mass for $m_b$ in
order to minimize the effect of radiative corrections [see
(\ref{meucl})]. For $s\gg 4m_b^2$, the perturbative spectral density
leads to the well-known cross section
\begin{equation}
   \sigma_s(e^+ e^-\to b\,\bar b)_{\rm pert}
   \stackrel{s\gg 4 m_b^2}{\to} \,\,{4\pi\alpha^2\over 3 s}\,
   N_c\,e_b^2\,\bigg( 1 + {\alpha_s\over\pi} \bigg) \,.
\end{equation}
However, the above form of the spectral density includes threshold
effects, which become important when $s$ comes closer to the
threshold region.

In QCD sum rules, one usually adds to the perturbative contributions
nonperturbative corrections, which are power suppressed and appear at
higher orders in the operator product expansion (OPE) of correlation
functions such as $\Pi(Q^2)$. In the case at hand, the leading
nonperturbative corrections are proportional to the gluon condensate
$\langle\alpha_s G^2\rangle\sim 0.04$ GeV$^4$, which in sum rules is
treated as a phenomenological parameter\cite{SVZ}. In the case of the
bottomonium system, one is in the fortunate situation that the
nonperturbative contributions to the spectral function are very
small, as they are suppressed by the ratio $\langle\alpha_s G^2
\rangle/m_b^4\sim 10^{-4}$. Their effect on the mass of the bottom
quark is below 1\% and can safely be neglected.

The idea is now to construct two equivalent representations for the
moments $M_n$, a ``theoretical'' one based on expression
(\ref{Pipert}) for the spectral function, and a ``phenomenological''
one based on a measurement of the $e^+ e^-\to b\,\bar b$ cross
section:
\begin{eqnarray}\label{Mnexpr}
   M_n^{\rm th} &\simeq& {1\over\pi}
    \int\limits_{\displaystyle 4 m_b^2}^{\displaystyle\infty}
    \!\mbox{d}s\,s^{-n-1}\,\mbox{Im}\Pi_{\rm pert}(s) \,,
    \nonumber\\
   M_n^{\rm exp} &=& {1\over\pi}
    \int\limits_{\displaystyle M_{\Upsilon(1s)}^2}
    ^{\displaystyle\infty}\!\mbox{d}s\,s^{-n-1}\,
    \mbox{Im}\Pi_{\rm exp}(s) \,.
\end{eqnarray}
The moments $M_n^{\rm th}$ are very sensitive functions of $m_b$.
By comparing them to a phenomenological expression that uses
detailed experimental information in the $b\,\bar b$ vector channel,
one can extract an accurate value for the bottom quark mass.
Experimentally, six resonances have been identified in this channel.
Their masses $M_{\rm R}$ and electronic widths $\Gamma_{\rm R}(e^+
e^-)$ are known rather precisely\cite{PDG92}. To evaluate the
integrals over the experimental cross section, one then makes the
following approximation:
\begin{equation}
   \mbox{Im}\Pi_{\rm exp}(s) \simeq {3\over 4\alpha^2 e_b^2}\,
   \sum_{\rm R}\,M_{\rm R}\,\Gamma_{\rm R}(e^+ e^-)\,
   \delta(s-M_{\rm R}^2) + \mbox{Im}\Pi_{\rm pert}(s)\,
   \Theta(s-s_0) \,,
\end{equation}
where the sum is over the narrow resonances $\Upsilon(1s)$ to
$\Upsilon(6s)$, and the continuum above some threshold value $s_0$ is
approximated by perturbation theory. This can be justified by
quark--hadron duality. Inserting this ansatz into (\ref{Mnexpr}) and
equating the two representations for the moments, one obtains the sum
rules
\begin{equation}
   {1\over\pi} \int
   \limits_{\displaystyle 4 m_b^2}^{\displaystyle s_0}
   \!\mbox{d}s\,s^{-n-1}\,\mbox{Im}\Pi_{\rm pert}(s)
   \simeq {3\over 4\pi\alpha^2 e_b^2}\,
   \sum_{\rm R}\,M_{\rm R}^{-2n-1}\,\Gamma_{\rm R}(e^+ e^-) \,.
\end{equation}
The goal is to find an optimal set of parameters $(m_b,s_0)$ such
that $M_n^{\rm th}\simeq M_n^{\rm exp}$ for many values of $n$. In
practice, the perturbative calculation breaks down for large values
of $n$, since the Coulomb corrections proportional to $\alpha_s/v$ in
(\ref{Pipert}) then become too large. Typically, one is limited to
values $n<10$. As mentioned above, the nonperturbative corrections
are very small (below 1\%) in the case of the bottomonium system.
They become sizeable, however, in the charmonium system, where a
similar analysis can be performed. Therefore, $m_b$ can be extracted
with larger accuracy than $m_c$.

Let me then quote the numerical results obtained by various authors
following the strategy outlined above. Values for the Euclidean mass
of the bottom quark have been obtained by Shifman et al.\cite{SVZ}
($m_b^{\rm Eucl}=4.23\pm 0.05$ GeV), Guberina et al.\cite{Gube}
($m_b^{\rm Eucl}=4.19\pm 0.06$ GeV), and Reinders\cite{Rein}
($m_b^{\rm Eucl}=4.17\pm 0.02$ GeV). The differences in these values
are mainly due to changes in the experimental input numbers. The
result of Reinders is the most up-to-date one. Values for the
Euclidean mass of the charm quark have been obtained many years ago
by Novikov et al.\cite{Novi} ($m_c^{\rm Eucl}\simeq 1.25$~GeV) and
Shifman et al.\cite{SVZ} ($m_c^{\rm Eucl}=1.26\pm 0.02$ GeV). They
have not changed since then. Converting these results into pole
masses using (\ref{meucl}), one finds\cite{Rein}
\begin{eqnarray}
   m_b^{\rm pole} &=& 4.55\pm 0.05~\mbox{GeV} \,, \nonumber\\
   m_c^{\rm pole} &=& 1.45\pm 0.05~\mbox{GeV} \,.
\end{eqnarray}
The increase in the error is due to an additional uncertainty in the
value of $\alpha_s(m_Q)$.

Let me finally mention some alternative approaches to extract pole
masses from QCD sum rules. Voloshin has proposed to resum the large
Coulomb corrections proportional to $(\alpha_s/v)^n$ to all orders in
perturbation theory, using a nonrelativistic approach\cite{VoSR}.
This allows one to go to higher moments, thereby suppressing the
resonance contributions to the sum rule. The disadvantage is that
some relativistic corrections are not taken into account. The result
is a rather large value of the pole mass of the bottom
quark\cite{VoZa},
\begin{equation}
   m_b^{\rm pole} = 4.79\pm 0.03~\mbox{GeV} \,.
\end{equation}

Another alternative is to obtain heavy quark masses from the study of
sum rules for heavy--light bound states such as the $B$- and
$B^*$-mesons. From such an analysis, Narison obtains\cite{Nari}
\begin{equation}
   m_b^{\rm pole} = 4.56\pm 0.05~\mbox{GeV} \,.
\end{equation}
A more recent analysis using the heavy quark expansion to order
$1/m_Q$ gives\cite{review,Baga,Neu2}
\begin{eqnarray}
   m_b &=& 4.71\pm 0.07~\mbox{GeV} \,, \nonumber\\
   m_c &=& 1.37\pm 0.12~\mbox{GeV} \,.
\end{eqnarray}
It should be noticed, however, that these sum rules are more
sensitive to nonperturbative corrections than are the sum rules for
the quarkonium systems.

\subsection{Infrared Renomalons}

As emphasized at the beginning of this lecture, the fact that at low
energies quarks are always confined into hadrons prohibits a
physical, on-shell definition of quark masses. Although for heavy
quarks the notion of a pole mass is widely used, such a concept
becomes meaningless beyond perturbation theory. No precise definition
of the pole mass can be given once nonperturbative effects are taken
into account.

It is interesting that indications for an intrinsic ambiguity in the
pole mass can already be found within the context of perturbation
theory, when one studies the asymptotic behaviour of the perturbative
series for the quark self-energy\cite{renom1,renom2}. It can be shown
that the existence of so-called infrared
renomalons\cite{reno1}$^-$\cite{reno5} generates a factorial
divergence in the expansion coefficients. Roughly speaking, the
reason is that self-energy corrections to the gluon propagator in the
diagram depicted in Fig.~\ref{fig:1.1} effectively introduce the
running coupling constant $\alpha_s(\sqrt{k^2})$, where $k^2$ is the
square of the virtual gluon momentum in Euclidean space. Since
$\alpha_s(\sqrt{k^2})$ increases as $k^2$ decreases, small loop
momenta become more important. One may reorganize the perturbative
expansion as an expansion in the small coupling constant
$\alpha_s(m_Q)$ using
\begin{equation}
   \alpha_s(\sqrt{k^2}) \simeq \alpha_s(m_Q) \sum_{n=0}^\infty
   \bigg( {\beta_0\over 4\pi}\,\alpha_s(m_Q)\,\ln{m_Q^2\over k^2}
   \bigg)^n \,.
\end{equation}
When one then performs the loop integral over $k^2$, the extra
logarithms give rise to a growth of the expansion coefficients
proportional to $n!$. In such a situation, it is in principle not
possible to improve the accuracy of perturbation theory by including
more and more terms in the perturbative series. Starting from some
order $n$, the size of the corrections increases. The best that can
be achieved is to truncate the series at an optimal value of $n$.
This introduces an intrinsic error, which depends exponentially on
$1/\alpha_s(m_Q)$. In the case of the pole mass, one can show that
the irreducible uncertainty in the value of $m_Q^{\rm pole}$ is of
order\cite{renom1,renom2}
\begin{equation}\label{DmQreno}
   \Delta m_Q^{\rm pole} = {8\over 3\beta_0}\,m_Q\,
   \exp\bigg( -{2\pi\over\beta_0\,\alpha_s(m_Q)} \bigg)
   \simeq {8\Lambda_{\rm QCD}\over 3\beta_0} \,.
\end{equation}
It is thus of the order of the confinement scale $\Lambda_{\rm QCD}$.
The fact that $\Delta m_Q^{\rm pole}/m_Q^{\rm pole}$ vanishes in the
$m_Q\to\infty$ limit justifies the concept of a pole mass for heavy
quarks, at least in an approximate sense that is limited to the
context of perturbation theory.

Although one has to be careful when using (\ref{DmQreno}) to obtain
an estimate of the intrinsic uncertainty in the pole mass, I will
insert $\Lambda_{\rm QCD}\sim 150$ MeV to find $\Delta m_Q^{\rm
pole}\sim 50$ MeV. This uncertainty should be kept in mind when
considering numerical results for heavy quark masses obtained, for
instance, using QCD sum rules.

\subsection{Exercises}

\begin{itemize}
\item
Derive the one-loop expression (\ref{Sigmap}) for the on-shell quark
self-energy in QCD. The necessary loop integrals in $D$ space-time
dimensions can be found in any reasonable textbook on quantum field
theory.
\item
Show that (\ref{Uoper}) is the exact solution of the RGE
(\ref{RGEfin}), and derive the next-to-leading logarithmic
approximation (\ref{Uevol}) for the evolution operator.
\item
By calculating the $1/\epsilon$-poles of the diagrams shown in
Fig.~\ref{fig:1.2}, derive the one-loop RGE (\ref{mtrun}) for the
running top quark mass.
\item
Solve the RGE (\ref{atrun}) for the case $\alpha_s=\mbox{const.}$,
and show that there is an infrared-stable fixed point at
$\alpha_t=\frac{16}{9}\,\alpha_s$.
\item
Derive eqs.~(\ref{exasol})--(\ref{Hillsol}).
\end{itemize}

%%%   Lecture 2   %%%%%%%%%%%%%%%%%%%%%%%%%%%%%%%%%%%%%%%%%%%%%%%%%%%

\newpage
\section{Quark Mixing in the Standard Model}

In this lecture, I give an introduction to flavour-changing decays
and quark mixing in the standard model. I will discuss quark mixing
in the cases of two and three generations, introduce some useful
parametrizations of the Kobayashi--Maskawa matrix, and briefly
review the status of the direct experimental determination of the
entries in this matrix from tree-level processes. I will then turn to
the geometrical interpretation of the mixing matrix provided by the
unitarity triangle, and finish with some remarks on rare decays.
Since you probably have heard many lectures on these subjects, my
presentation will be rather short. Let me refer those of you who want
more details to two excellent review articles that I like very much:
{\it A Top Quark Story\/} by Buras and Harlander\cite{BuHa}, and {\it
CP Violation\/} by Nir\cite{Yosi}.

\subsection{Cabibbo--Kobayashi--Maskawa Matrix}

Let me briefly remind you of some facts about the flavour sector of
the standard model. Below mass scales of order $m_W\sim 80$ GeV, the
standard model gauge group $\mbox{SU}_{\rm C}(3)\times\mbox{SU}_{\rm
L}(2)\times\mbox{U}_{\rm Y}(1)$ is spontaneously broken to
$\mbox{SU}_{\rm C}(3)\times\mbox{U}_{\rm em}(1)$, since the scalar
Higgs doublet $\phi$ acquires a vacuum expectation value
\begin{equation}
   \langle\phi\rangle = \bigg\langle\bigg(
   \begin{array}{c} \phi_+ \\ \phi_0 \end{array}
   \bigg)\bigg\rangle = {v\over\sqrt{2}}\,\bigg(
   \begin{array}{c} 0 \\ 1 \end{array} \bigg) \,;\quad
   v\simeq 246~\mbox{GeV} \,.
\end{equation}
This gives masses to the $W$- and $Z$-bosons, as well as to the
quarks and leptons. The quark masses arise from the quark Yukawa
couplings to the Higgs doublet, which in the unbroken theory are
assumed to be of the most general form that is invariant under local
gauge transformations. The Yukawa interactions are written in terms
of the weak eigenstates $q'$ of the quark fields, which have definite
transformation properties under $\mbox{SU}_{\rm L}(2)\times
\mbox{U}_{\rm Y}(1)$. After the symmetry breaking, one redefines the
quark fields so as to obtain the mass terms in the canonical form.
This has an interesting effect on the form of the flavour-changing
charged-current interactions. In the weak basis, these interactions
have the form
\begin{equation}
   {\cal L}_{\rm int} = - {g\over\sqrt{2}}\,
   (\bar u'_{\rm L}, \bar c'_{\rm L}, \bar t'_{\rm L})\,\gamma^\mu
   \left( \begin{array}{c} d'_{\rm L} \\ s'_{\rm L} \\
   b'_{\rm L} \end{array} \right) W_\mu^\dagger +
   \mbox{h.c.}
\end{equation}
In terms of the mass eigenstates $q$, however, this becomes
\begin{equation}
   {\cal L}_{\rm int} = - {g\over\sqrt{2}}\,
   (\bar u_{\rm L}, \bar c_{\rm L}, \bar t_{\rm L})\,\gamma^\mu\,
   V_{\rm KM}
   \left( \begin{array}{c} d_{\rm L} \\ s_{\rm L} \\
   b_{\rm L} \end{array} \right) W_\mu^\dagger +
   \mbox{h.c.}
\end{equation}
The Kobayashi--Maskawa mixing matrix
\begin{equation}
   V_{\rm KM} \simeq \left( \begin{array}{ccc}
    V_{ud} & V_{us} & V_{ub} \\
    V_{cd} & V_{cs} & V_{cb} \\
    V_{td} & V_{ts} & V_{tb}
   \end{array} \right)
\end{equation}
is a unitary matrix in flavour space. In the general case of $n$
quark generations, $V_{\rm KM}$ would be an $n\times n$ matrix. I
will now discuss the structure of this matrix for the cases of two
and three generations.

\subsubsection{Mixing Matrix for Two Generations}

In this case, $V$ is a $2\times 2$ unitary matrix and can be
parametrized by one angle and three phases:
\begin{eqnarray}
   V &=& \left( \begin{array}{rl}
    \cos\theta_{\rm C}\,e^{i\alpha} & ~\sin\theta_{\rm C}
    \,e^{i\beta} \\
    -\sin\theta_{\rm C}\,e^{i\gamma} & ~\cos\theta_{\rm C}\,
    e^{i(\beta+\gamma-\alpha)}
   \end{array} \right) \nonumber\\
   && \nonumber\\
   &=& \left( \begin{array}{cc}
    e^{i\alpha} & 0 \\
    0 & e^{i\gamma}
   \end{array} \right)
   \left( \begin{array}{rc}
    \cos\theta_{\rm C} & ~\sin\theta_{\rm C} \\
    -\sin\theta_{\rm C} & ~\cos\theta_{\rm C}
   \end{array} \right)
   \left( \begin{array}{cc}
    1 & 0 \\
    0 & e^{i(\beta-\alpha)}
   \end{array} \right) \,.
\end{eqnarray}
The three phases are not observable, however, as they can be absorbed
into a redefinition of the phases of the quark fields $u_{\rm L}$,
$c_{\rm L}$, and $s_{\rm L}$ relative to $d_{\rm L}$. After this
redefinition, the matrix takes the standard form due to
Cabibbo\cite{Cabi}:
\begin{equation}
   V_{\rm C} = \left( \begin{array}{rc}
    \cos\theta_{\rm C} & ~\sin\theta_{\rm C} \\
    -\sin\theta_{\rm C} & ~\cos\theta_{\rm C}
   \end{array} \right) \,.
\end{equation}
The Cabibbo angle $\theta_{\rm C}$ can be extracted from
$K\to\pi\,e^+\nu_e$ decay. Experimentally, one finds\cite{PDG92}
$\sin\theta_{\rm C}\simeq 0.22$.

\subsubsection{Mixing Matrix for Three Generations}

A $3\times 3$ unitary matrix can be parametrized by three Euler
angles and six phases, five of which can be removed by adjusting the
relative phases of the left-handed quark fields. Hence, three
angles $\theta_{ij}$ and one observable phase $\delta$ remain in the
quark mixing matrix, as was first pointed out by Kobayashi and
Maskawa\cite{KoMa}. For completeness, I note that in the general case
of $n$ generations, it is easy to show that there are $\frac{1}{2}
n(n-1)$ angles and $\frac{1}{2}(n-1)(n-2)$ observable
phases\cite{Jarl}. Whereas therefore the original Cabibbo matrix was
real and had only one parameter, the Kobayashi--Maskawa matrix of the
standard model is complex and can be parametrized by four parameters.

The imaginary part of the mixing matrix is necessary to describe CP
violation in the standard model. In general, CP is violated in
flavour-changing decays if there is no degeneracy of any two quark
masses, and if the quantity $J_{\rm CP}\ne 0$, where
\begin{equation}\label{JCPdef}
   J_{\rm CP} = |\,\mbox{Im}(V_{ij} V_{kl} V_{il}^* V_{kj}^*)\,|
   \,;\quad i\ne k \,,~j\ne l
\end{equation}
is invariant under phase redefinitions of the quark fields. One can
show\cite{Jar1} that all CP-violating amplitudes in the standard
model are proportional to $J_{\rm CP}$.

I will now discuss two of the most convenient parametrizations of the
mixing matrix. The ``standard parametrization'' recommended by the
Particle Data Group\cite{PDG92} is\cite{Chau}
\begin{equation}\label{VKMstand}
   V_{\rm KM} = \left( \begin{array}{ccc}
    c_{12}\,c_{13} & s_{12}\,c_{13} & s_{13}\,e^{-i\delta} \\
    -s_{12}\,c_{23} - c_{12}\,s_{23}\,s_{13}\,e^{i\delta} &
    c_{12}\,c_{23} - s_{12}\,s_{23}\,s_{13}\,e^{i\delta} &
    s_{23}\,c_{13} \\
    s_{12}\,s_{23} - c_{12}\,c_{23}\,s_{13}\,e^{i\delta} &
    -c_{12}\,s_{23} - s_{12}\,c_{23}\,s_{13}\,e^{i\delta} &
    c_{23}\,c_{13}
   \end{array} \right) \,.
\end{equation}
Here, one uses the short-hand notation $c_{ij}=\cos\theta_{ij}$ and
$s_{ij}=\sin\theta_{ij}$. Some advantages of this parametrization are
the following:

\begin{itemize}
\item
$|\,V_{ub}|=s_{13}$ is given by a single angle, which experimentally
turns out to be very small.
\item
Because of this, several other entries are given by single angles to
an accuracy of better than four digits. They are: $V_{ud}\simeq
c_{12}$, $V_{us}\simeq s_{12}$, $V_{cb}\simeq s_{23}$, and
$V_{tb}\simeq c_{23}$.
\item
The CP-violating phase $\delta$ appears together with the small
parameter $s_{13}$, making explicit the fact that CP violation in the
standard model is a small effect. Indeed, one finds
\begin{equation}\label{JCPstan}
   J_{\rm CP} = |\,s_{13}\,s_{23}\,s_{12}\,s_\delta\,
   c_{13}^2\,c_{23}\,c_{12}\,| \,.
\end{equation}
\end{itemize}

For many purposes and applications, it is more convenient to use an
approximate parametrization of the Kobayashi--Maskawa matrix, which
makes explicit the strong hierarchy that is observed experimentally.
Setting $c_{13}=1$ (experimentally, it is known that $c_{13}>
0.99998$) and neglecting $s_{13}$ compared to terms of order unity,
one finds
\begin{equation}
   V_{\rm KM} \simeq \left( \begin{array}{ccc}
    c_{12} & s_{12} & s_{13}\,e^{-i\delta} \\
    -s_{12}\,c_{23} & c_{12}\,c_{23} & s_{23} \\
    s_{12}\,s_{23} - c_{12}\,c_{23}\,s_{13}\,e^{i\delta} &
    ~-c_{12}\,s_{23}~ & c_{23}
   \end{array} \right) \,.
\end{equation}
Now denote $s_{12}=\lambda\simeq 0.22$. Experiments indicate that
$s_{23}=O(\lambda^2)$ and $s_{13}=O(\lambda^3)$. Hence, it is natural
to define $s_{23}=A\,\lambda^2$ and $s_{13}\,e^{-i\delta}=A\,
\lambda^3 (\rho-i\eta)$, with $A$, $\rho$ and $\eta$ of order unity.
An expansion in powers of $\lambda$ then leads to the Wolfenstein
parametrization\cite{Wolf}
\begin{equation}\label{Wolpar}
   V_{\rm KM} \simeq \left( \begin{array}{ccc}
    1-{\lambda^2\over 2} & \lambda & A\,\lambda^3(\rho-i\eta) \\
    \phantom{ \bigg[ }
    -\lambda & 1-{\lambda^2\over 2} & A\,\lambda^2 \\
    A\,\lambda^3(1-\rho-i\eta) & ~-A\,\lambda^2~ & 1
   \end{array} \right) + O(\lambda^4) \,.
\end{equation}
It nicely exhibits the hierarchy of the mixing matrix: The entries in
the diagonal are close to unity, $V_{us}$ and $V_{cd}$ are of order
20\%, $V_{cb}$ and $V_{ts}$ are of order 4\%, and $V_{ub}$ and
$V_{td}$ are of order 1\% and thus the smallest entries in the
matrix. Some care has to be taken when one wants to calculate the
quantity $J_{\rm CP}$ in the Wolfenstein parametrization, since the
result is of order $\lambda^6$ and thus beyond the accuracy of the
approximation. However, taking $i=u$, $j=d$, $k=t$, and $l=b$ in
(\ref{JCPdef}), one obtains the correct answer
\begin{equation}
   J_{\rm CP} \simeq A^2\,\eta\,\lambda^6
   \simeq 1.1\times 10^{-4} A^2\,\eta \,,
\end{equation}
which shows that $J_{\rm CP}$ is generically of order $10^{-4}$ for
$\lambda\simeq 0.22$. As a consequence, CP violation in the standard
model is a small effect.

In principle, the elements in the first two rows of the mixing matrix
are accessible in so-called direct (tree-level) processes, i.e.\ in
weak decays of hadrons containing the corresponding quarks. In
practice, as I will discuss below, the entries $|\,V_{ud}|$ and
$|\,V_{us}|$ are known to an accuracy of better than 1\% from such
decays, whereas $|\,V_{cd}|$, $|\,V_{cs}|$, and $|\,V_{cb}|$ are
known to about 10--20\%. The element $|\,V_{ub}|$ has an uncertainty
of about a factor of 2. The same is true for $|\,V_{td}|$, which is
obtained from a rare process, namely $B_d$--$\bar B_d$ mixing. There
is no direct information on $|\,V_{ts}|$ and $|\,V_{tb}|$. The large
uncertainty in $|\,V_{ub}|$ and $|\,V_{td}|$ translates into an
uncertainty of the Wolfenstein parameters $\rho$ and $\eta$. A more
precise determination of these parameters will be the challenge to
experiments and theory over the next decade.

\subsection{Experimental Information on $V_{\rm KM}$ from Tree-Level
Processes}

In this section, I will briefly review what is known about entries of
the Kobayashi--Maskawa matrix from tree-level processes. Rare decays,
which are induced by loop-diagrams involving heavy particles, will be
discussed later.

\subsubsection{Determination of $|\,V_{ud}|$}

{}From a comparison of the very accurate measurements of
super-allowed nuclear $\beta$-decay with $\mu$-decay, including a
detailed analysis of radiative corrections\cite{Sirl,Jaus}, one
obtains\cite{PDG92}
\begin{equation}
   |\,V_{ud}| = 0.9744\pm 0.0010 \,.
\end{equation}

\subsubsection{Determination of $|\,V_{us}|$}

An analysis of $K_{e3}$ decays, i.e.\ $K^+\to\pi^0 e^+\nu_e$ and
$K_L^0\to\pi^- e^+\nu_e$, including isospin- and SU(3)-breaking
effects\cite{LeRo}, yields $|\,V_{us}|=0.2196\pm 0.0023$.
Alternatively, from an analysis of hyperon decays, including
SU(3)-breaking corrections\cite{DHKl}, one obtains
$|\,V_{us}|=0.222\pm 0.003$. This leads to the combined
value\cite{PDG92}
\begin{equation}\label{Vusval}
   |\,V_{us}| = 0.2205\pm 0.0018 \,.
\end{equation}

Let me add a note on $K_{e3}$ decays in this context. The fact that
the theoretical uncertainty in the description of these exclusive
hadronic decays is only about 1\% is at first sight surprising. The
reasons are the following: The approximate SU(3) flavour symmetry of
QCD implies that, at zero momentum transfer, the hadronic form factor
$f_+^{K\to\pi}$ that parametrizes $K\to\pi$ transitions induced by a
vector current equals unity, up to symmetry-breaking corrections. The
Ademollo--Gatto theorem\cite{AGTh} states that these corrections are
of second order in the symmetry-breaking parameter $(m_s-m_q)$, where
$q=u$ or $d$. Chiral perturbation theory can be employed to calculate
the leading symmetry-breaking corrections (chiral logarithms) in a
model-independent way\cite{LeRo}. In my third lecture, I will discuss
that there are analogues to all these ingredients in the case of
semileptonic $B$-decays.

\subsubsection{Determination of $|\,V_{cd}|$}

{}From the combination of data on single charm production in
deep-inelastic neutrino--nucleon scattering with the semileptonic
branching ratios of charmed me\-sons, one deduces\cite{PDG92}
\begin{equation}
   |\,V_{cd}| = 0.204\pm 0.017 \,.
\end{equation}

\subsubsection{Determination of $|\,V_{cs}|$}

Here the usefulness of deep-inelastic scattering data is limited,
since the extraction of $|\,V_{cs}|$ would depend upon an assumption
about the strange quark content of the nucleon. It is better to
proceed in analogy to the determination of $|\,V_{us}|$, i.e.\ to use
the semileptonic $D_{e3}$ decays $D\to\bar K\,e^+\nu_e$. The
experimental data are compared to various model calculations of the
decay width\cite{Wirb}$^-$\cite{ISGW}. This procedure leads, however,
to a significant amount of model dependence. The result
is\cite{PDG92}
\begin{equation}\label{Vcsval}
   |\,V_{cs}| = 1.00\pm 0.20 \,.
\end{equation}

You may wonder why the theoretical description of the decay width is
so uncertain. The reason is that, in general, our capabilities to
calculate hadronic form factors in QCD are rather limited. An
understanding of the nonperturbative confining forces would be
necessary to make the connection between quark and hadron properties,
which is a prerequisite for any calculation of hadronic matrix
elements. In the case of $K_{e3}$ decays, a symmetry of QCD helps us
eliminate most of the hadronic uncertainties. For $D_{e3}$ decays,
there is no such flavour symmetry. Hence, one has to rely on
phenomenological models. Let me briefly mention some of them:

\begin{itemize}
\item
Isgur, Scora, Grinstein, and Wise (ISGW) have proposed a
nonrelativistic constituent quark model\cite{ISGW}. They solve the
Schr\"odinger equation for a Coulomb plus linear potential to obtain
meson wave functions. Weak decay form factors are calculated at
maximum momentum transfer, where both mesons have a common rest
frame, by computing overlap integrals of the wave functions of the
initial and final meson. The nonrelativistic approximation breaks
down if $q^2\ll q^2_{\rm max}$.
\item
Bauer, Stech, and Wirbel (BSW) have constructed a relativistic
model\cite{Wirb}, which uses light-cone wave functions to calculate
weak decay form factors at zero momentum transfer. In order to
extrapolate to $q^2>0$, they assume nearest-pole dominance.
\item
K\"orner and Schuler\cite{KS} (KS) use a variation of the BSW
approach, in which the $q^2$ dependence of the form factors is
adjusted according to asymptotic QCD power-counting rules\cite{Brod}.
\item
There are a variety of other models, for instance by
Suzuki\cite{Suzu}, Altomari and Wolfenstein\cite{Alto}, and Grinstein
et al.\cite{GWI}, which are more or less closely related to one of
the above.
\end{itemize}

\noindent
Disadvantages of these models are that their relation to the
underlying theory of QCD is not obvious, that it is hard to obtain an
estimate of their intrinsic uncertainty or range of applicability,
that they rely on ad hoc assumptions (for instance, concerning the
$q^2$ dependence of form factors), and that it is difficult to
improve them in a systematic way.

Another stream of research focuses on analytical or numerical
approaches that bear a closer relation to field theory. QCD sum
rules\cite{SVZ} provide a fully relativistic approach based on QCD.
They rely, however, on the assumption of quark--hadron duality and
need some phenomenological input. Nonperturbative effects are
modelled in a simple way by introducing few universal numbers, the
so-called vacuum condensates. There have been extensive applications
of sum rules to the study of meson weak decay form
factors\cite{Ali2}$^-$\cite{Lige}. These analyses can be done for all
physical values of $q^2$. However, in spite of the many successes of
QCD sum rules it must be said that the only known approach to
low-energy QCD that truly starts from first principles is lattice
gauge theory. For more details on this, I refer to the lectures by P.
Lepage in this volume. Let me just mention that, as far as weak decay
form factors are concerned, these computations are extremely complex
and CPU-consuming. They are not yet competitive with (less rigorous)
analytical approaches.

\subsubsection{Determination of $|\,V_{ub}|$}

Since $b\to u$ transitions are strongly suppressed in nature, their
discovery in inclusive $B\to X_u\,\ell\,\bar\nu$ decays was one of
the breakthroughs in $B$-physics in recent
years\cite{btou1}$^-$\cite{btou4}. In order to account for CP
violation in the standard model, it is necessary that $V_{ub}$ be
different from zero; otherwise $J_{\rm CP}=0$, and the observed CP
violation in the kaon system\cite{Chri} could not be explained by the
Kobayashi--Maskawa mechanism.

The first direct observation of a $b\to u$ transition was provided by
two fully reconstructed events, one $B^0\to\pi^+\mu^-\bar\nu$ decay
and one $B^+\to\omega\,\mu^+\nu$ decay, reported by the ARGUS
collaboration\cite{btou3}. However, the number of exclusive charmless
$B$-decays that have been observed so far is too small to obtain a
reasonable determination of $|\,V_{ub}|$. A high luminosity
$B$-factory would improve this situation. Given enough statistics,
the idea is to obtain a model-independent measurement of $|\,V_{ub}|$
by using heavy quark symmetry to relate the decay form factors in
$B\to X_u\,\ell\,\bar\nu$ and $D\to X_d\,\ell\,\bar\nu$
transitions\cite{IsWeVub}, at the same value of the recoil energy of
the light hadron $X_q$. In the limit of infinite heavy quark masses
($m_b, m_c\to\infty$), the ratio of these form factors is fixed in a
model-independent way. In the real world, however, there are
nonperturbative power corrections to this limit. As a consequence,
even in an ideal measurement the ratio $|\,V_{ub}/V_{cd}|$ can only
be determined up to hadronic corrections of order
\begin{equation}
   1 + c\,\bigg( {\Lambda_{\rm QCD}\over m_c}
   - {\Lambda_{\rm QCD}\over m_b} \bigg) + \ldots \,,
\end{equation}
with a coefficient $c$ of order unity. Model
calculations\cite{Dib,BLNN} show that these corrections are of order
15\% for $X=\pi$ or $\rho$. I believe that it is fair to say that
even at a high-luminosity $B$-factory, the prospects for getting a
determination of $|\,V_{ub}|$ from exclusive decays, which is more
reliable than that from inclusive decays, are rather limited.
Nevertheless, it is certainly worth while to pursue this strategy as
an alternative.

\begin{figure}[htb]
   \vspace{0.5cm}
   \epsfxsize=10cm
   \centerline{\epsffile{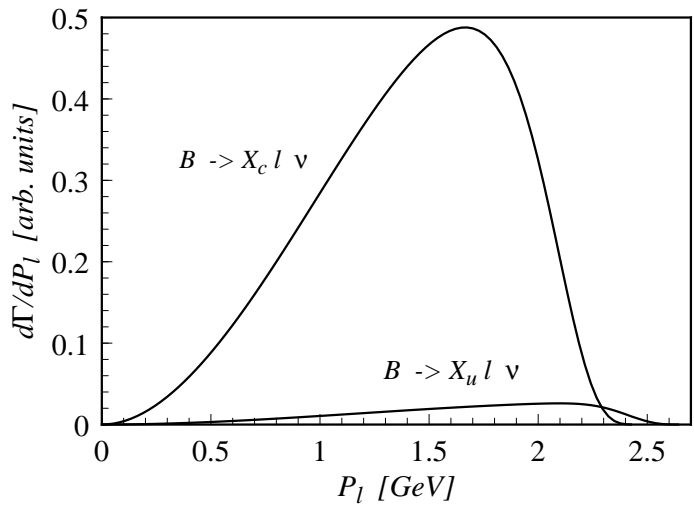}}
   \vspace{-0.5cm}
   \centerline{\parbox{11cm}{\caption{\label{fig:2.1}
Sketch of the lepton spectrum in inclusive semileptonic $B$-decays. I
assume $|\,V_{ub}/V_{cb}|\simeq 0.08$. The $b\to u$ signal has been
multiplied by a factor of 5 to be visible on this plot.
   }}}
\end{figure}

The present determinations of $|\,V_{ub}|$ are based on measurements
of the lepton momentum spectrum in inclusive $B\to X_q\,\ell\,
\bar\nu$ decays, where $X_q$ is any hadronic state containing a
$q$-quark, with $q=c$ or $u$. The expected signal is shown in
Fig.~\ref{fig:2.1}. Over most of the kinematic region, the spectrum
is dominated by $b\to c$ transitions, which are strongly enhanced
with respect to $b\to u$ decays. The only place to observe charmless
transitions is in a small window above the kinematic limit for $B\to
D\,\ell\,\bar\nu$, but below the kinematic limit for
$B\to\pi\,\ell\,\bar\nu$ decays:
\begin{equation}\label{kinlim}
   {m_B\over 2}\,\bigg( 1 - {m_D^2\over m_B^2} \bigg) \le p_\ell
   \le {m_B\over 2}\,\bigg( 1 - {m_\pi^2\over m_B^2} \bigg) \,,
\end{equation}
i.e.\ $2.31~\mbox{GeV}\le p_\ell\le 2.64~\mbox{GeV}$. Indeed, the
ARGUS and CLEO collaborations have reported signals in this
region\cite{btou1}$^-$\cite{btou4}. An extraction of $|\,V_{ub}|$
from these data is difficult, however, as the shape of the spectrum
close to the kinematic endpoint is dominated by nonperturbative
effects. This can be seen as follows. The conventional description of
inclusive decays of hadrons containing a heavy quark starts from the
free-quark decay model, in which the hadronic decay $B\to
X_q\,\ell\,\bar\nu$ is modelled by the quark decay $b\to
q\,\ell\,\bar\nu$. One usually calculates the decay distributions to
order $\alpha_s$ in perturbation theory by including the effects of
real and virtual gluon emission\cite{AlPi}$^-$\cite{JeKu}. Computing
then the lepton spectrum, one finds that the kinematic limit for
$b\to u$ decays is given by $p_\ell^{\rm max}=m_b/2\simeq 2.35$ GeV,
where I neglect the $u$-quark mass and take $m_b\simeq 4.7$ GeV for
the sake of argument. The point is that the region in which the $b\to
u$ signal is observed experimentally is almost not populated in the
free-quark decay model. Nonperturbative bound-state effects, such as
the motion of the $b$-quark inside the $B$-meson, are responsible for
the population of the spectrum beyond the parton model endpoint. The
way in which such effects are incorporated into phenomenological
approaches\cite{ACM,Pasc} is to a large extent model-dependent. Only
very recently has there been progress towards a model-independent
description of the distributions near the kinematic endpoint in the
context of QCD\cite{shape}$^-$\cite{bcshap}. It has been shown that
the leading nonperturbative effects can be parametrized in terms of a
universal structure function, which describes the momentum
distribution of the $b$-quark inside the $B$-meson. The hope is that
it will be possible to get an accurate prediction for this function
using various theoretical approaches. I believe that, within a year
or two, it should be possible to reduce the theoretical uncertainty
in the analysis of the inclusive decay spectrum by a factor of 2 or
so.

Alternatively, one can describe the lepton spectrum close to the
endpoint as the sum of contributions from semileptonic $B$-decays
into exclusive final states. These exclusive decays are then
described using one of the various bound-state
models\cite{Wirb}$^-$\cite{KS} discussed above. Of course, there is
again a strong model dependence associated with this approach.

Because of these theoretical uncertainties, the current value of
$|\,V_{ub}|$ has a model dependence of almost a factor of 2. In 1991,
the ARGUS collaboration obtained\cite{btou3} $|\,V_{ub}/V_{cb}| =
0.11\pm 0.012$, using the approach of Altarelli et al.\cite{ACM},
which is based on the parton model. Using instead bound-state models
to describe the spectrum by a sum over several exclusive decay
channels, values for $|\,V_{ub}/V_{cb}|$ between 0.10 and 0.22 are
obtained. However, the most recent data reported by the CLEO
collaboration lead to significantly smaller values\cite{btou4}:
$|\,V_{ub}/V_{cb}| = 0.076\pm 0.008$ using the approach of Altarelli
et al.\cite{ACM}, and $0.05<|\,V_{ub}/V_{cb}|<0.11$ using bound-state
models. Thus, I think a reasonable value to quote is
\begin{equation}\label{Vubval}
   \bigg|\,{V_{ub}\over V_{cb}}\,\bigg| = 0.08\pm 0.03 \,.
\end{equation}

\newpage
\subsubsection{Determination of $|\,V_{cb}|$}

As for $|\,V_{cb}|$, it can be extracted from both inclusive and
exclusive $B$-decays. I start with a discussion of inclusive decays.
The idea is to compare the total semileptonic branching ratio to the
parton model formula
\begin{equation}\label{partincl}
   \mbox{Br}(B\to X_q\,\ell\,\bar\nu)
   = {G_F^2\,m_b^5\over 192\pi^3}\,\tau_B\,\bigg\{
   \eta_c\,|\,V_{cb}|^2\,f\bigg( {m_c^2\over m_b^2} \bigg)
   + \eta_u\,|\,V_{ub}|^2 \bigg\} \,,
\end{equation}
where $\eta_c$ and $\eta_u$ contain the short-distance QCD
corrections, and
\begin{equation}
   f(x) = 1 - 8 x + 8 x^3 - x^4 - 12 x^2\ln x
\end{equation}
is a phase-space correction due to the mass of the charm quark. The
QCD correction factor for $b\to c $ decays is\cite{AlPi,CCM}
$\eta_c\simeq 0.87$. The contribution from $b\to u$ transitions is
very small and, at the present level of accuracy, can be neglected.

Let me note that it has recently become clear how to compute in a
systematic way the nonperturbative bound-state corrections to the
parton model result (\ref{partincl}). For the total inclusive decay
rates, these corrections turn out to be of
order\cite{Chay}$^-$\cite{btau3} $(\Lambda_{\rm QCD}/m_b)^2$.
Numerically, they are very small and play only a minor role in the
analysis\cite{LukSav}.

The disadvantage of using inclusive decays to extract $|\,V_{cb}|$ is
that there are substantial theoretical uncertainties due to the fact
that the total rate scales as the fifth power of the bottom quark
mass. For instance, an uncertainty of only $\pm 150$ MeV in the value
of $m_b$ translates into an uncertainty of $\pm 8\%$ in the value of
$|\,V_{cb}|$. For this reason, the last time the particle data
group\cite{PDG88} used inclusive decays to obtain a value for
$|\,V_{cb}|$ was in 1988. A compilation of more recent analyses of
inclusive decay spectra has been given by Stone\cite{Sheldon}. He
finds
\begin{equation}\label{Vcbincl}
   |\,V_{cb}|\,\bigg( {\tau_B\over 1.5~\mbox{ps}} \bigg)^{1/2}
   = 0.040\pm 0.005 \,,
\end{equation}
where I have normalized the result to a value of the $B$-meson
lifetime that is close to the world average $\tau_{B^0}=(1.48\pm
0.10)$ ps reported last year by Danilov\cite{tauBav}, as well as to
the most recent value $\tau_{B^0}=(1.52\pm 0.13)$ ps obtained from an
average of LEP results\cite{tauB}. The error quoted in
(\ref{Vcbincl}) does not take into account the theoretical
uncertainty in the value of $m_b$.

Ultimately, a more precise determination must come from the analysis
of exclusive decay modes. With the development of the heavy quark
expansion, it has become clear that certain symmetries of QCD in the
heavy quark limit can help to remove much of the model dependence
from the theoretical description of exclusive semileptonic
$B$-decays. This will be explained in detail in my third lecture. In
particular, for the decay mode $B\to D^*\ell\,\bar\nu$ the
theoretical uncertainties in the momentum distribution of the
$D^*$-meson close to the zero recoil limit can be controlled at the
level of a few per cent\cite{Neu9}. The most recent analysis of this
spectrum by the CLEO collaboration leads to the value\cite{CLEOVcb}
\begin{equation}
   |\,V_{cb}|\,\bigg( {\tau_B\over 1.5~\mbox{ps}} \bigg)^{1/2}
   = 0.039\pm 0.006 \,,
\end{equation}
which I will use in the analysis below. In contrast to
(\ref{Vcbincl}), here the uncertainty is dominated by the
experimental systematic errors and, to a lesser extent, the
statistical ones. The theoretical uncertainty is well below 10\%.
There is thus room for a significant improvement of the accuracy
within the next few years, as more data become available.

Finally, let me mention that by combining the above result with the
value of $|\,V_{us}|$ given in (\ref{Vusval}), one obtains
\begin{equation}\label{AWoval}
   A = \bigg|\,{V_{cb}\over V_{us}^2}\bigg| = 0.80\pm 0.12
\end{equation}
for the Wolfenstein parameter $A$.

\subsection{Unitarity Constraints}

The fact that the quark mixing matrix must be unitary can be employed
to put limits on the elements that are not easily accessible to
experiments. One has to assume, however, that there are only three
quark generations. Unitarity then implies that
\begin{equation}
   |\,V_{tj}|^2 = 1 - |\,V_{uj}|^2 - |\,V_{cj}|^2 \,.
\end{equation}
Inserting here the known values of $|\,V_{uj}|$ and $|\,V_{cj}|$, one
obtains
\begin{eqnarray}
   0.003 &< |\,V_{td}| &< 0.016 \,, \nonumber\\
   0.032 &< |\,V_{ts}| &< 0.048 \,, \nonumber\\
   0.9991 &< |\,V_{tb}| &<  0.9994 \,.
\end{eqnarray}
Note, in particular, that $|\,V_{tb}|$ is constrained to an accuracy
of better than $10^{-3}$.

Unitarity also imposes a rather tight constraint on $|\,V_{cs}|$.
Using that
\begin{equation}
   |\,V_{cs}|^2 = 1 - |\,V_{cd}|^2 - |\,V_{cb}|^2 \,,
\end{equation}
one finds
\begin{equation}
   0.976 < |\,V_{cs}| < 0.981 \,.
\end{equation}
This should be compared to the value (\ref{Vcsval}) extracted from a
direct measurement.

\subsection{The Unitarity Triangle}

A simple but beautiful way to visualize the implications of unitarity
is provided by the so-called unitarity triangle\cite{Bjtr}, which
uses the fact that the unitarity equation
\begin{equation}
   V_{ij}\,V_{ik}^* = 0 \qquad (j\ne k)
\end{equation}
can be represented as the equation of a closed triangle in the
complex plane. There are six such triangles, all of which have the
same area\cite{JaSt}
\begin{equation}\label{AreaJcp}
   |A_\Delta| = {1\over 2}\,J_{\rm CP} \,.
\end{equation}
Under phase reparametrizations of the quark fields, the triangles
change their orientation in the complex plane, but their shape
remains unaffected.

\begin{figure}[htb]
   \vspace{-1cm}
   \epsfysize=6cm
   \centerline{\epsffile{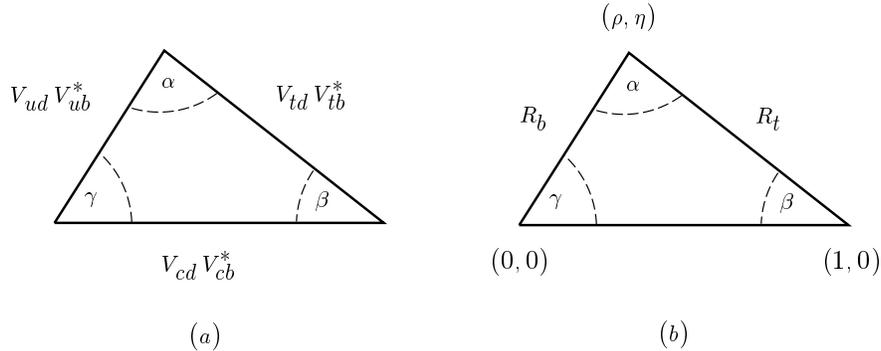}}
   \centerline{\parbox{11cm}{\caption{\label{fig:2.2}
(a) The unitarity triangle; (b) its rescaled form in the
$\rho$--$\eta$ plane. The angle $\gamma$ coincides with the phase
$\delta$ of the standard parametrization (\protect\ref{VKMstand}).
   }}}
\end{figure}

Most useful from the phenomenological point of view is the triangle
relation
\begin{equation}
   V_{ud} V_{ub}^* + V_{cd} V_{cb}^* + V_{td} V_{tb}^* = 0 \,,
\end{equation}
since it contains the most poorly known entries in the
Kobayashi--Maskawa matrix. It has been widely discussed in the
literature\cite{Bjtr}$^-$\cite{HaRo}. In the standard
parametrization, $V_{cd} V_{cb}^*$ is real, and the unitarity
triangle has the form shown in Fig.~\ref{fig:2.2}a. It is useful to
rescale the triangle by dividing all sides by $V_{cd} V_{cb}^*$. The
rescaled triangle is shown in Fig.~\ref{fig:2.2}b. It has the
coordinates $(0,0)$, $(1,0)$, and $(\rho,\eta)$, where $\rho$ and
$\eta$ appear in the Wolfenstein parametrization (\ref{Wolpar}).
Unitarity amounts to the statement that the triangle is closed, and
CP is violated when the area of the triangle does not vanish, i.e.\
when all the angles are not zero.

To determine the shape of the triangle, one can aim for measurements
of the two sides $R_b$ and $R_t$, and of the angles $\alpha$,
$\beta$, and $\gamma$. So far, experimental information is available
only on the sides of the triangle. The current value of $|\,V_{ub}|$
in (\ref{Vubval}) implies
\begin{equation}
   0.24 < R_b < 0.53 \,.
\end{equation}
As mentioned above, the uncertainty in this number is mainly a
theoretical one, so any significant improvement must originate from
new developments in the theory of inclusive $B$-decays.

Let me now discuss the determination of $|\,V_{td}|$ and of the side
$R_t$ of the unitarity triangle. The process of interest here is the
mixing between $B_d$- and $\bar B_d$-mesons, which in the standard
model is a rare process mediated through box diagrams with a virtual
top quark exchange. The ARGUS and CLEO collaborations have reported
measurements\cite{BBARG,BBCLEO} of the mixing parameter $x_d$ defined
as $x_d=\Delta m_{B_d}\,\tau_{B_d}$, where $\Delta m_B$ denotes the
mass difference between the mass eigenstates in the $B_d$--$\bar B_d$
system. The weighted average of their results is $x_d=0.68\pm 0.08$.
Combining this with the $B_d$-meson lifetime\cite{tauBav},
$\tau_{B_d}=(1.48\pm 0.10)$ ps, one obtains $\Delta m_B=(0.46\pm
0.06)~\mbox{ps}^{-1}$. Recently, direct measurements of $\Delta m_B$
have been reported by some of the LEP experiments. The average value
presented at the Moriond Conference is\cite{dmMoriond} $\Delta
m_B=(0.519_{-0.059}^{+0.063})~\mbox{ps}^{-1}$. I combine these
results to obtain
\begin{equation}
   \Delta m_B = (0.49\pm 0.04)~\mbox{ps}^{-1} \,.
\end{equation}

The theoretical expression for $\Delta m_B$ in the standard model is
\begin{equation}\label{dmBSM}
   \Delta m_B = {G_F^2 m_W^2\over 6\pi^2}\,
   \eta_{\rm QCD}\,m_B\,(B_B f_B^2)\,S(m_t/m_W)\,
   |\,V_{td} V_{tb}^*|^2 \,,
\end{equation}
where $\eta_{\rm QCD}\simeq 0.55$ contains the next-to-leading order
QCD corrections\cite{BJWe}, and $S(m_t/m_W)$ is a function of the top
quark mass\cite{InLi,Bur1}. For $100~\mbox{GeV}< m_t <
300~\mbox{GeV}$, it can be approximated by\cite{Penbox}
$S(m_t/m_W)\simeq 0.784\,(m_t/m_W)^{1.52}$. The product $(B_B f_B^2)$
parametrizes the hadronic matrix element of a local four-quark
operator between $B$-meson states. Recently, there has been a lot of
activity and improvement in theoretical calculations of the $B$-meson
decay constant $f_B$, using both lattice gauge theory and QCD sum
rule calculations. For a summary and discussion of the extensive
literature on this subject, I refer to the review articles by Buras
and Harlander\cite{BuHa} and by myself\cite{review}. Here I shall use
the range
\begin{equation}
   B_B^{1/2} f_B = (200\pm 40)~\mbox{MeV} \,,
\end{equation}
which covers most theoretical predictions. Solving (\ref{dmBSM}) for
$|\,V_{td}|$, one obtains
\begin{equation}
   |\,V_{td}| = 0.0088\,\bigg( {200~\mbox{MeV}\over B_B^{1/2} f_B}
   \bigg)\,\bigg( {170~\mbox{GeV}\over m_t} \bigg)^{0.76}\,
   \bigg( {\Delta m_B\over 0.5~\mbox{ps}^{-1}} \bigg)^{1/2} \,.
\end{equation}
Using the numbers given above, as well as $m_t=(170\pm 30)$ GeV, I
find
\begin{equation}
   |\,V_{td}| = 0.0087\pm 0.0023 \,.
\end{equation}
The corresponding range of values for $R_t$ is
\begin{equation}
   0.75 < R_t < 1.44 \,.
\end{equation}
The error in this value will hopefully be reduced soon, with a
measurement of the top quark mass at Fermilab. Further improvements
could come from a measurement of $f_B$ in $B^+\to\tau^+\nu_\tau$
decays, or from the observation of $B_s$--$\bar B_s$ mixing. In the
latter case, one could use the relation
\begin{equation}
   {x_s\over x_d} = \bigg|\,{V_{ts}\over V_{td}}\,\bigg|^2\,
   \times\Big\{ 1 + \mbox{SU(3)-breaking corrections} \Big\}
\end{equation}
to obtain a rather clean measurement of $|\,V_{td}|$. However, in the
standard model one expects large values $x_s\sim 10$--30, which will
be very hard to observe in near-future experiments.

\begin{figure}[htb]
   \vspace{0.5cm}
   \epsfxsize=10cm
   \centerline{\epsffile{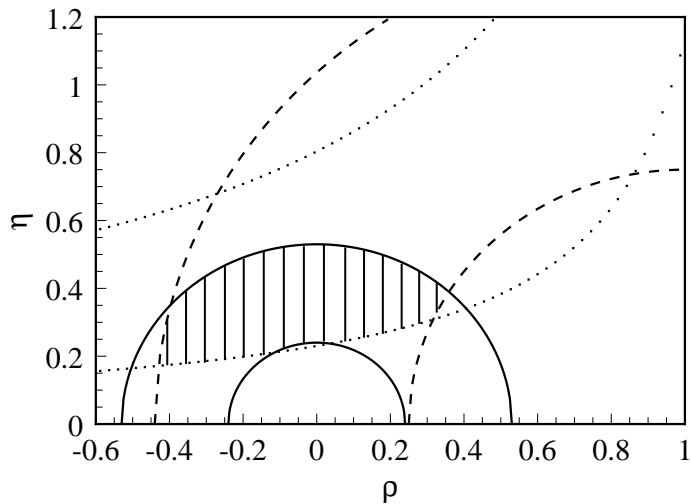}}
   \vspace{-0.5cm}
   \centerline{\parbox{11cm}{\caption{\label{fig:2.3}
Experimental constraints on the unitarity triangle in the
$\rho$--$\eta$ plane. The region between the solid (dashed) circles
is allowed by the measurement of $R_b$ ($R_t$) discussed above. The
dotted curves show the constraint following from the measurement of
the $\varepsilon$ parameter in the kaon system. The shaded region
shows the allowed range for the tip of the unitarity triangle. The
base of the triangle has the coordinates $(0,0)$ and $(1,0)$.
   }}}
\end{figure}

In Fig.~\ref{fig:2.3}, I show the constraints imposed by the above
results for $R_b$ and $R_t$ in the $\rho$--$\eta$ plane. Another
constraint can be obtained from the study of CP violation in
$K$-decays.

\subsection{CP Violation in $K$-Decays}

The experimental result on the parameter $\varepsilon$ that measures
CP violation in $K$--$\bar K$ mixing implies that the unitarity
triangle lies in the upper half plane, provided the so-called $B_K$
parameter is positive. Most theoretical
predictions\cite{DGHo}$^-$\cite{SShar} fall in the range $B_K=0.65\pm
0.20$, supporting this assertion. The arising constraint in the
$\rho$--$\eta$ plane has the form of a hyperbola approximately given
by\cite{BuHa}
\begin{equation}
   \eta\,\bigg[ (1-\rho)\,A^2\,\bigg( {m_t\over m_W} \bigg)^{1.52}
   + P_c \bigg]\,A^2\,B_K = 0.50 \,,
\end{equation}
where $P_c\simeq 0.66$ corresponds to the contributions from box
diagrams containing charm quarks, and $A=0.80\pm 0.12$ according to
(\ref{AWoval}). I have included this bound in Fig.~\ref{fig:2.3}.

In principle, the measurement of the ratio $\mbox{Re}
(\varepsilon'/\varepsilon)$ in the kaon system\cite{NA31,E731} could
provide a determination of $\eta$ independent of $\rho$. In practice,
however, the theoretical calculations\cite{Burs}$^-$\cite{Marti} of
this ratio are affected by large uncertainties, so that there
currently is no useful bound to be derived. Further information on
the parameters $\rho$ and $\eta$ could be extracted from rare
$K$-decays, such as $K_L\to\pi^0\nu\,\bar\nu$, $K^+\to\pi^+\nu\,
\bar\nu$, and $K_L\to\mu^+\mu^-$. The corresponding branching ratios
vary between $10^{-11}$ and $10^{-10}$. The experimental detection of
such decays will be very hard.

\subsection{Potential Improvements in the Determination of $\rho$ and
$\eta$}

The main conclusion to be drawn from Fig.~\ref{fig:2.3} is that,
given the present theoretical and experimental uncertainties in the
analysis of charmless $B$-decays, $B$--$\bar B$ mixing, and CP
violation in the kaon system, there is still a large region in
parameter space that is allowed for the Wolfenstein parameters $\rho$
and $\eta$. This has important implications. For instance, the
allowed region for the angle $\beta$ of the unitarity triangle (see
Fig.~\ref{fig:2.2}) is $6.9^\circ < \beta < 31.8^\circ$,
corresponding to
\begin{equation}\label{sinbe}
   0.24 < \sin 2\beta < 0.90 \,.
\end{equation}
Below I will discuss that the CP asymmetry in the decay
$B\to\psi\,K_S$, which is one of the favoured modes to search for CP
violation at a future $B$-factory, is proportional to $\sin 2\beta$.
Obviously, the prospects for discovering CP violation with such a
machine crucially depend on whether $\sin 2\beta$ is closer to the
upper or lower bound in (\ref{sinbe}). A more reliable determination
of the shape of the unitarity triangle is thus of the utmost
importance.

\begin{figure}[htb]
   \vspace{0.5cm}
   \epsfxsize=10cm
   \centerline{\epsffile{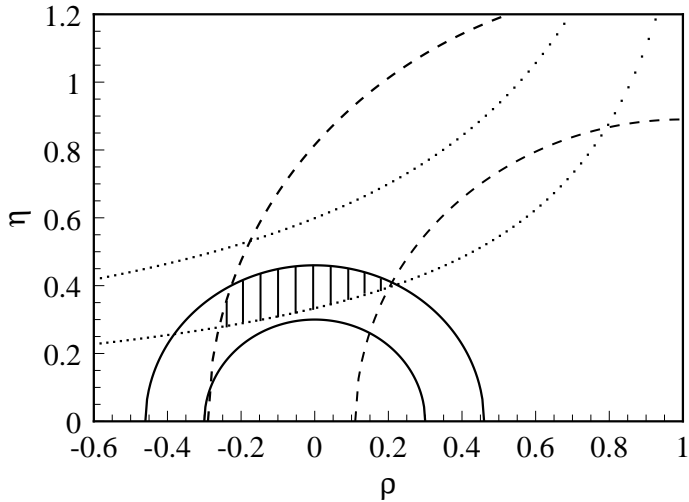}}
   \vspace{-0.5cm}
   \centerline{\parbox{11cm}{\caption{\label{fig:2.4}
Constraints on the unitarity triangle obtained assuming a more
precise determination of the various input parameters, as explained
in the text.
   }}}
\end{figure}

Let me briefly discuss what I think are the potential improvements in
this determination in the near future. It is not unrealistic to
assume the following:

\begin{itemize}
\item
The top quark will be discovered at Fermilab.
\item
The uncertainty in the value of $|\,V_{cb}|$ will be reduced to $\pm
0.04$, mainly through a better control of the experimental systematic
errors.
\item
Both theoretical and experimental improvements will reduce the
uncertainty in the determination of $|\,V_{ub}|$ by a factor of 2.
\item
Improved theoretical calculations, in particular using lattice gauge
theory, will allow the determination of the nonperturbative
parameters $B_B^{1/2} f_B$ and $B_K$ with an accuracy of 10\%.
\end{itemize}

\noindent
For the purpose of illustration, I shall assume that $m_t=170$ GeV,
$|\,V_{cb}|=0.039\pm 0.004$, $|\,V_{ub}/V_{cb}|=0.080\pm 0.015$,
$B_B^{1/2} f_B=(200\pm 20)$ MeV, and $B_K=0.65\pm 0.07$. This leads
to $R_b=0.38\pm 0.08$ and $R_t=1.09\pm 0.20$. Similarly, the
constraint derived from the measurement of $\varepsilon$ becomes more
restrictive. This fictitious scenario is illustrated in
Fig.~\ref{fig:2.4}. Obviously, the parameters of the triangle (the
value of $\eta$, in particular) would be much more constrained than
they are at present. Notice also that once a sufficient accuracy is
obtained, there is a potential for tests of the standard model. If
the three bands in Fig.~\ref{fig:2.4} did not overlap, this would be
an indication of new physics.

\subsection{Rare $B$-Decays}

Let me briefly touch upon the interesting subject of rare $B$-decays,
i.e.\ decays induced by loop diagrams. There are nice review articles
on this subject by Bertolini\cite{Berto} and Ali\cite{Ali}. Rare
$B$-decays are dominated by short-distance physics, namely the top
quark contribution in penguin and box diagrams. The decay rates are
usually rather sensitive to the mass of the top quark\cite{Penbox}.
Long-distance effects are much less important than in rare
$K$-decays. Hence, rare $B$-decays are ``clean'' from the theoretical
point of view. They are ideal to determine the parameters of the
unitarity triangle. In addition, these decays can provide important
tests of the standard model, since they are often sensitive probes of
new physics.

\begin{figure}[htb]
   \epsfxsize=9cm
   \centerline{\epsffile{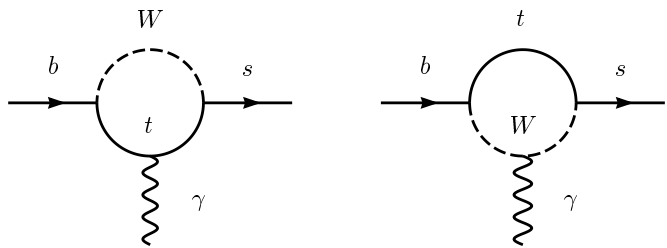}}
   \centerline{\parbox{11cm}{\caption{\label{fig:2.5}
Penguin diagrams contributing to the rare decay $B\to X_s\,\gamma$.
   }}}
\end{figure}

So far, the only rare $B$-decay that has been observed
is\cite{BKstar} $B\to K^*\gamma$. In general, decays of the type
$B\to X_s\,\gamma$ are mediated by penguin diagrams such as the ones
shown in Fig.~\ref{fig:2.5}. These decays receive very large, but
calculable, short-distance QCD corrections\cite{Bert}$^-$\cite{Ciuc},
which strongly enhance the decay rate. In addition, it is important
to take into account the effect of gluon bremsstrahlung, which
affects the total decay rate as well as the photon
spectrum\cite{AlGr}. All such effects are reasonably well under
control for the inclusive $B\to X_s\,\gamma$ decay rate. For
$m_t=170\pm 30$ GeV, the prediction is\cite{BuHa}
\begin{equation}\label{Brrare}
   \mbox{Br}(B\to X_s\,\gamma) = (2.8\mbox{--}4.2)\times 10^{-4} \,.
\end{equation}
Of more interest from the experimental point of view are exclusive
rare decay modes. Since $B\to K\,\gamma$ is forbidden by angular
momentum conservation, the lightest final state that can appear is
$K^*\gamma$. To obtain from (\ref{Brrare}) a prediction for
$\mbox{Br}(B\to K^*\gamma)$ requires an estimate of the ratio
\begin{equation}
   R_{K^*} = {\Gamma(B\to K^*\gamma)\over
   \Gamma(B\to X_s\,\gamma)} \,.
\end{equation}
This is where hadronic uncertainties enter the analysis. Estimates of
$R_{K^*}$ presented in the literature vary between 4\% and 40\%. I
believe that a reasonable value is\cite{AlGr}$^-$\cite{SNar}
$R_{K^*}\simeq 10$--15\%. This leads to
\begin{equation}
   \mbox{Br}(B\to K^*\gamma) \sim (3\mbox{--}6)\times 10^{-5} \,.
\end{equation}
This is in agreement with the result reported by the CLEO
collaboration\cite{BKstar}:
\begin{equation}
   \mbox{Br}(B\to K^*\gamma) = (4.5\pm 1.9\pm 0.9)\times 10^{-5} \,.
\end{equation}

Further rare $B$-decays, for which there are ``clean'' theoretical
predictions (which depend, however, on the top quark mass), include
$B\to X_s\,\nu\,\bar\nu$ with a branching ratio of
$10^{-5}$--$10^{-4}$, $B\to X_s\,\ell\,\bar\ell$ with a branching
ratio of about $10^{-5}$, and $B_s\to\ell\,\bar\ell$ with a branching
ratio of about $10^{-7}$.

\subsection{Measurements of the Angles of the Unitarity Triangle}

Let me finish this short introduction into the phenomenology of rare
decays with a very beautiful application, namely the direct
measurement of CP violation in decays of neutral $B$-mesons into CP
eigenstates $f$. For these processes, there are often no (or only
very small) hadronic uncertainties. The CP asymmetries arise from an
interference of mixing and decay. When such a decay is dominated by a
single weak decay amplitude, the time-integrated asymmetry obeys a
very simple relation to one of the angles $\varphi$ of the unitarity
triangle\cite{BuHa,Yosi,Khoz}:
\begin{equation}
   {\int\limits_0^\infty\!\mbox{d}t\,\Big[
    \Gamma(B^0(t)\to f) - \Gamma(\bar B^0(t)\to f) \Big]\over
    \int\limits_0^\infty\!\mbox{d}t\,\Big[
    \Gamma(B^0(t)\to f) + \Gamma(\bar B^0(t)\to f) \Big]}
   = \pm\sin(2\varphi)\,{x\over 1+x^2} \,,
\end{equation}
where $x=\Delta m_B\,\tau_B$. Examples of such decays are:
\begin{eqnarray}
   B_d &\to \psi\,K_S \qquad &(-\sin 2\beta) \,, \nonumber\\
   B_d &\to \phi\,K_S \qquad &(-\sin 2\beta) \,, \nonumber\\
   B_d &\to D^+ D^- \qquad &(-\sin 2\beta) \,, \nonumber\\
   B_d &\to \pi^+\pi^- \qquad &(\sin 2\alpha) \,, \nonumber\\
   B_s &\to \rho\,K_S \qquad &(-\sin 2\gamma) \,, \nonumber\\
   B_s &\to \phi\,K_S \qquad &(\sin 2\beta) \,.
\end{eqnarray}
The decay mode $B_d\to\psi\,K_S$ is particularly ``clean'' and is
often referred to as the ``gold-plated'' mode of a future
$B$-factory.

\subsection{Exercises}

\begin{itemize}
\item
Show that the generalization of the Kobayashi--Maskawa matrix to $n$
quark generations can be parametrized by $\frac{1}{2} n(n-1)$ angles
and $\frac{1}{2}(n-1)(n-2)$ observable phases.
\item
Show that the quantity $J_{\rm CP}$ defined in (\ref{JCPdef}) is
invariant under phase redefinitions of the quark fields. Verify that
in the standard parametrization $J_{\rm CP}$ is given by
(\ref{JCPstan}).
\item
Derive that the maximum lepton momentum in the decay $B\to X\,\ell\,
\bar\nu$ is given by $\frac{1}{2} m_B(1-m_X^2/m_B^2)$
[cf.~(\ref{kinlim})].
\item
Convince yourself that all six unitarity triangles have the same area
(\ref{AreaJcp}).
\item
Show that the decay $B\to K\,\gamma$ is forbidden by angular momentum
conservation.
\end{itemize}

%%%   Lecture 3   %%%%%%%%%%%%%%%%%%%%%%%%%%%%%%%%%%%%%%%%%%%%%%%%%%%

\newpage
\section{Heavy Quark Symmetry}

In this lecture, I give an introduction to one of the most active and
exciting recent developments in theoretical particle physics: the
discovery of heavy quark symmetry and the development of the heavy
quark effective field theory. For a more detailed description, I
refer to my recent review article in {\it Physics
Reports\/}\cite{review}. Earlier introductions into this field were
given by Georgi\cite{GeRev}, Grinstein\cite{GrRev}, Isgur and
Wise\cite{IWRev}, and Mannel\cite{MaRev}.

\subsection{The Physical Picture\/\cite{Shu1}$^-$\cite{Isgu}}

There are several reasons why the strong interactions of systems
containing heavy quarks are easier to understand than those of
systems containing only light quarks. The first is asymptotic
freedom, the fact that the effective coupling constant of QCD becomes
weak in processes with large momentum transfer, corresponding to
interactions at short-distance scales\cite{Gros,Poli}. At large
distances, on the other hand, the coupling becomes strong, leading to
nonperturbative phenomena such as the confinement of quarks and
gluons on a length scale $R_{\rm had}\sim 1/\Lambda_{\rm QCD}\sim 1$
fm, which determines the typical size of hadrons. Roughly speaking,
$\Lambda_{\rm QCD}\sim 0.2$ GeV is the energy scale that separates
the regions of large and small coupling constant. When the mass of a
quark $Q$ is much larger than this scale, $m_Q\gg\Lambda_{\rm QCD}$,
$Q$ is called a heavy quark. The quarks of the standard model fall
naturally into two classes: up, down and strange are light quarks,
whereas charm, bottom and top are heavy quarks\footnote{Ironically,
the top quark is of no relevance to my discussion here, since it is
too heavy to form hadronic bound states before it decays.}.
For heavy quarks, the effective coupling constant $\alpha_s(m_Q)$ is
small, implying that on length scales comparable to the Compton
wavelength $\lambda_Q\sim 1/m_Q$ the strong interactions are
perturbative and much like the electromagnetic interactions. In fact,
the quarkonium systems $(\bar QQ)$, whose size is of order
$\lambda_Q/\alpha_s(m_Q)\ll R_{\rm had}$, are very much
hydrogen-like. Since the discovery of asymptotic freedom, their
properties could be predicted\cite{Appe} before the observation of
charmonium, and later of bottomonium states.

Things are more complicated for systems composed of a heavy quark and
other light constituents. The size of such systems is determined by
$R_{\rm had}$, and the typical momenta exchanged between the heavy
and light constituents are of order $\Lambda_{\rm QCD}$. The heavy
quark is surrounded by a most complicated, strongly interacting cloud
of light quarks, antiquarks, and gluons. This cloud is sometimes
referred to as the ``brown muck'', a term invented by Isgur to
emphasize that the properties of such systems cannot be calculated
from first principles (at least not in a perturbative way). In this
case, it is the fact that $m_Q\gg\Lambda_{\rm QCD}$, or better
$\lambda_Q\ll R_{\rm had}$, i.e.\ the fact that the Compton
wavelength of the heavy quark is much smaller than the size of the
hadron, which leads to simplifications. To resolve the quantum
numbers of the heavy quark would require a hard probe. The soft
gluons which couple to the ``brown muck'' can only resolve distances
much larger than $\lambda_Q$. Therefore, the light degrees of freedom
are blind to the flavour (mass) and spin orientation of the heavy
quark. They only experience its colour field, which extends over
large distances because of confinement. In the rest frame of the
heavy quark, it is in fact only the electric colour field that is
important; relativistic effects such as colour magnetism vanish as
$m_Q\to\infty$. Since the heavy quark spin participates in
interactions only through such relativistic effects, it decouples.
That the heavy quark mass becomes irrelevant can be seen as follows:
As $m_Q\to\infty$, the heavy quark and the hadron that contains it
have the same velocity. In the rest frame of the hadron, the heavy
quark is at rest, too. The wave function of the ``brown muck''
follows from a solution of the field equations of QCD subject to the
boundary condition of a static triplet source of colour at the
location of the heavy quark. This boundary condition is independent
of $m_Q$, and so is the solution for the configuration of the light
degrees of freedom.

It follows that, in the $m_Q\to\infty$ limit, hadronic systems which
differ only in the flavour or spin quantum numbers of the heavy quark
have the same configuration of their light degrees of freedom.
Although this observation still does not allow us to calculate what
this configuration is, it provides relations between the properties
of such particles as the heavy mesons $B$, $D$, $B^*$ and $D^*$, or
the heavy baryons $\Lambda_b$ and $\Lambda_c$ (to the extent that
corrections to the infinite quark mass limit are small in these
systems). Isgur and Wise realized that these relations result from
new symmetries of the effective strong interactions of heavy quarks
at low energies\cite{Isgu}. The configuration of light degrees of
freedom in a hadron containing a single heavy quark $Q(v,s)$ with
velocity $v$ and spin $s$ does not change if this quark is replaced
by another heavy quark $Q'(v,s')$ with different flavour or spin, but
with the same velocity. Both heavy quarks lead to the same static
colour field. It is not necessary that the heavy quarks $Q$ and $Q'$
have similar masses. What is important is that their masses are large
compared to $\Lambda_{\rm QCD}$. For $N_h$ heavy quark flavours,
there is thus an SU$(2 N_h)$ spin-flavour symmetry group. These
symmetries are in close correspondence to familiar properties of
atoms: The flavour symmetry is analogous to the fact that different
isotopes have the same chemistry, since to good approximation the
wave function of the electrons is independent of the mass of the
nucleus. The electrons only see the total nuclear charge. The spin
symmetry is analogous to the fact that the hyperfine levels in atoms
are nearly degenerate. The nuclear spin decouples in the limit
$m_e/m_N\to 0$.

The heavy quark symmetry is an approximate symmetry, and corrections
arise since the quark masses are not infinite. These corrections are
of order $\Lambda_{\rm QCD}/m_Q$. The condition $m_Q\gg\Lambda_{\rm
QCD}$ is necessary and sufficient for a system containing a heavy
quark to be close to the symmetry limit. In many respects, heavy
quark symmetry is complementary to chiral symmetry, which arises in
the opposite limit of small quark masses, $m_q\ll\Lambda_{\rm QCD}$.
There is an important distinction, however. Whereas chiral symmetry
is a symmetry of the QCD Lagrangian in the limit of vanishing quark
masses, heavy quark symmetry is not a symmetry of the Lagrangian (not
even an approximate one), but rather a symmetry of an effective
theory which is a good approximation to QCD in a certain kinematic
region. It is realized only in systems in which a heavy quark
interacts predominantly by the exchange of soft gluons. In such
systems the heavy quark is almost on-shell; its momentum fluctuates
around the mass shell by an amount of order $\Lambda_{\rm QCD}$. The
corresponding changes in the velocity of the heavy quark vanish as
$\Lambda_{\rm QCD}/m_Q\to 0$. The velocity becomes a conserved
quantity and is no longer a dynamical degree of freedom\cite{Geor}.

Results derived based on heavy quark symmetry are model-independent
consequences of QCD in a well-defined limit. The symmetry-breaking
corrections can, at least in principle, be studied in a systematic
way. To this end, it is however necessary to recast the QCD
Lagrangian for a heavy quark,
\begin{equation}\label{QCDLag}
   {\cal L}_Q = \bar Q\,(i\,\rlap{\,/}D - m_Q)\,Q \,,
\end{equation}
into a form suitable for taking the limit $m_Q\to\infty$.

\subsection{Heavy Quark Effective Theory\/\cite{EiFe}$^-$\cite{Luke}}

In particle physics, it is often the case that the effects of a very
heavy particle become irrelevant at low energies. It is then useful
to construct a low-energy effective theory, in which this heavy
particle no longer appears. Eventually, this effective theory will be
easier to deal with than the full theory. A familiar example is
Fermi's theory of the weak interactions. For the description of weak
decays of hadrons, one can approximate the weak interactions by
point-like four-fermion couplings governed by a dimensionful coupling
constant $G_F$. Only at energies much larger than the masses of
hadrons can one resolve the structure of the intermediate vector
bosons $W$ and $Z$.

Technically, the process of removing the degrees of freedom of a
heavy particle involves the following
steps\cite{SVZ1}$^-$\cite{Polc}: One first identifies the heavy
particle fields and ``integrates them out'' in the generating
functional of the Green functions of the theory. This is possible
since at low energies the heavy particle does not appear as an
external state. However, although the action of the full theory is
usually a local one, what results after this first step is a nonlocal
effective action. The nonlocality is related to the fact that in the
full theory the heavy particle (with mass $M$) can appear in virtual
processes and propagate over a short but finite distance $\Delta
x\sim 1/M$. Thus a second step is required to get to a local
effective Lagrangian: The nonlocal effective action is rewritten as
an infinite series of local terms in an operator product
expansion (OPE)\cite{Wils,Zimm}. Roughly speaking, this corresponds
to an expansion in powers of $1/M$. It is in this step that the
short- and long-distance physics is disentangled. The long-distance
physics corresponds to interactions at low energies and is the same
in the full and the effective theory. But short-distance effects
arising from quantum corrections involving large virtual momenta (of
order $M$) are not reproduced in the effective theory, once the heavy
particle has been integrated out. In a third step, they have to be
added in a perturbative way using renormalization-group techniques.
This procedure is called ``matching''. It leads to renormalizations
of the coefficients of the local operators in the effective
Lagrangian. An example is the effective Lagrangian for nonleptonic
weak decays, in which radiative corrections from hard gluons with
virtual momenta in the range between $m_W$ and some renormalization
scale $\mu\sim 1$ GeV give rise to Wilson coefficients, which
renormalize the local four-fermion
interactions\cite{Alta}$^-$\cite{Gilm}.

The heavy quark effective theory (HQET) is constructed to provide a
simplified description of processes where a heavy quark interacts
with light degrees of freedom by the exchange of soft gluons.
Clearly, $m_Q$ is the high-energy scale in this case, and
$\Lambda_{\rm QCD}$ is the scale of the hadronic physics one is
interested in. However, a subtlety arises since one wants to describe
the properties and decays of hadrons which contain a heavy quark.
Hence, it is not possible to remove the heavy quark completely from
the effective theory. What is possible, however, is to integrate out
the ``small components'' in the full heavy quark spinor, which
describe fluctuations around the mass shell.

The starting point in the construction of the low-energy effective
theory is the observation that a very heavy quark bound inside a
hadron moves with essentially the hadron's velocity $v$ and is almost
on-shell. Its momentum can be written as
\begin{equation}\label{kresdef}
   p_Q^\mu = m_Q v^\mu + k^\mu \,,
\end{equation}
where the components of the so-called residual momentum $k^\mu$ are
much smaller than $m_Q$. Interactions of the heavy quark with light
degrees of freedom change the residual momentum by an amount of order
$\Delta k^\mu\sim\Lambda_{\rm QCD}$, but the corresponding changes in
the heavy quark velocity vanish as $\Lambda_{\rm QCD}/m_Q\to 0$. In
this situation, it is appropriate to introduce ``large'' and
``small'' component fields $h_v$ and $H_v$ by
\begin{eqnarray}
   h_v(x) &=& e^{i m_Q v\cdot x}\,P_+\,Q(x) \,, \nonumber\\
   H_v(x) &=& e^{i m_Q v\cdot x}\,P_-\,Q(x) \,,
\end{eqnarray}
where $P_+$ and $P_-$ are projection operators defined as
\begin{equation}
   P_\pm = {1\pm\rlap/v\over 2} \,.
\end{equation}
It follows that
\begin{equation}\label{redef}
   Q(x) = e^{-i m_Q v\cdot x}\,\Big[ h_v(x) + H_v(x) \Big] \,.
\end{equation}
Because of the projection operators, the new fields satisfy
$\rlap/v\,h_v=h_v$ and $\rlap/v\,H_v=-H_v$. In the rest frame, $h_v$
corresponds to the upper two components of $Q$, while $H_v$
corresponds to the lower ones. Whereas $h_v$ annihilates a heavy
quark with velocity $v$, $H_v$ creates a heavy antiquark with
velocity $v$. If the heavy quark was on-shell, the field $H_v$ would
be absent.

\begin{figure}[htb]
   \vspace{0.5cm}
   \epsfxsize=9cm
   \centerline{\epsffile{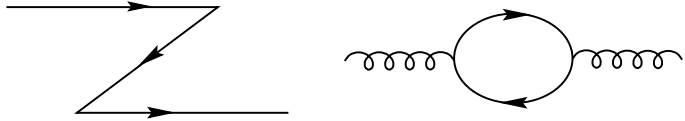}}
   \centerline{\parbox{11cm}{\caption{\label{fig:3.1}
Virtual fluctuations involving pair creation of heavy quarks. Time
flows to the right.
   }}}
\end{figure}

In terms of the new fields, the QCD Lagrangian (\ref{QCDLag}) for a
heavy quark takes the following form:
\begin{equation}\label{Lhchi}
   {\cal L}_Q = \bar h_v\,i v\!\cdot\!D\,h_v
   - \bar H_v\,(i v\!\cdot\!D + 2 m_Q)\,H_v
   + \bar h_v\,i\,\rlap{\,/}D_\perp H_v
   + \bar H_v\,i\,\rlap{\,/}D_\perp h_v \,,
\end{equation}
where $D_\perp^\mu = D^\mu - v^\mu\,v\!\cdot\!D$ is orthogonal to the
heavy quark velocity: $v\!\cdot\!D_\perp=0$. In the rest frame,
$D_\perp^\mu=(0,\vec D\,)$ contains the spatial components of the
covariant derivative. From (\ref{Lhchi}), it is apparent that $h_v$
describes massless degrees of freedom, whereas $H_v$ corresponds to
fluctuations with twice the heavy quark mass. These are the heavy
degrees of freedom, which will be eliminated in the construction of
the effective theory. The fields are mixed by the presence of the
third and fourth terms, which describe pair creation or annihilation
of heavy quarks and antiquarks. As shown in the first diagram in
Fig.~\ref{fig:3.1}, in a virtual process a heavy quark propagating
forward in time can turn into a virtual antiquark propagating
backward in time, and then turn back into a quark. The energy of the
intermediate quantum state $h h\bar H$ is larger than the energy of
the incoming heavy quark by at least $2 m_Q$. Because of this large
energy gap, the virtual quantum fluctuation can only propagate over a
short distance $\Delta x\sim 1/m_Q$. On hadronic scales set by
$R_{\rm had}=1/\Lambda_{\rm QCD}$, the process essentially looks like
a local interaction of the form
\begin{equation}
   \bar h_v\,i\,\rlap{\,/}D_\perp\,{1\over 2 m_Q}\,
   i\,\rlap{\,/}D_\perp h_v \,,
\end{equation}
where I have simply replaced the propagator for $H_v$ by $1/2 m_Q$. A
more correct treatment is to integrate out the small component field
$H_v$, thereby deriving a nonlocal effective action for the large
component field $h_v$, which can then be expanded in terms of local
operators. Before doing this, let me mention a second type of virtual
corrections that involve pair creation, namely heavy quark loops.
An example is depicted in the second diagram in Fig.~\ref{fig:3.1}.
Heavy quark loops cannot be described in terms of the effective
fields $h_v$ and $H_v$, since the quark velocities in the loop are
not conserved, and are in no way related to hadron velocities.
However, such short-distance processes are proportional to the small
coupling constant $\alpha_s(m_Q)$ and can be calculated in
perturbation theory. They lead to corrections which are added onto
the low-energy effective theory in the matching procedure.

On a classical level, the heavy degrees of freedom represented by
$H_v$ can be eliminated using the equations of motion of QCD.
Substituting (\ref{redef}) into $(i\,\rlap{\,/}D-m_Q)\,Q=0$ gives
\begin{equation}
   i\,\rlap{\,/}D\,h_v + (i\,\rlap{\,/}D - 2 m_Q)\,H_v =0 \,,
\end{equation}
and multiplying this by $P_\pm$ one derives the two equations
\begin{eqnarray}
   -i v\!\cdot\!D\,h_v &=& i\,\rlap{\,/}D_\perp H_v \,, \nonumber\\
   (i v\!\cdot\!D + 2 m_Q)\,H_v &=& i\,\rlap{\,/}D_\perp h_v \,.
\end{eqnarray}
The second can be solved to give ($\eta\to +0$)
\begin{equation}\label{Hfield}
   H_v = {1\over(2 m_Q + i v\!\cdot\!D - i\eta)}\,
   i\,\rlap{\,/}D_\perp h_v \,,
\end{equation}
which shows that the small component field $H_v$ is indeed of order
$1/m_Q$. One can now insert this solution into the first equation to
obtain the equation of motion for $h_v$. It is easy to see that this
equation follows from the nonlocal effective Lagrangian
\begin{equation}\label{Lnonloc}
   {\cal L}_{\rm eff} = \bar h_v\,i v\!\cdot\!D\,h_v
   + \bar h_v\,i\,\rlap{\,/}D_\perp\,
   {1\over(2 m_Q + i v\!\cdot\!D - i\eta)}\,
   i\,\rlap{\,/}D_\perp h_v \,.
\end{equation}
Clearly, the second term precisely corresponds to the first class of
virtual processes depicted in Fig.~\ref{fig:3.1}.

Mannel, Roberts, and Ryzak have derived this Lagrangian in a more
elegant way by manipulating the generating functional for QCD Green's
functions containing heavy quark fields\cite{Mann}. They start from
the field redefinition (\ref{redef}) and couple the large component
fields $h_v$ to external sources $\rho_v$. Green's functions with an
arbitrary number of $h_v$ fields can be constructed by taking
derivatives with respect to $\rho_v$. No sources are needed for the
heavy degrees of freedom represented by $H_v$. The functional
integral over these fields is Gaussian and can be performed
explicitly, leading to the nonlocal effective action
\begin{equation}\label{SeffMRR}
   S_{\rm eff} = \int\!{\rm d}^4 x\,{\cal L}_{\rm eff}
   - i \ln\Delta \,,
\end{equation}
with ${\cal L}_{\rm eff}$ as given in (\ref{Lnonloc}). The
appearance
of the logarithm of the determinant
\begin{equation}
   \Delta = \exp\bigg( {1\over 2}\,{\rm Tr}\,
   \ln\big[ 2 m_Q + i v\!\cdot\!D - i\eta \big] \bigg)
\end{equation}
is a quantum effect not present in the classical derivation given
above. However, in this case the determinant can be regulated in a
gauge-invariant way, and by choosing the axial gauge $v\cdot A=0$ one
shows that $\ln\Delta$ is just an irrelevant
constant\cite{Mann,Soto}.

Because of the phase factor in (\ref{redef}), the $x$ dependence of
the effective heavy quark field $h_v$ is weak. In momentum space,
derivatives acting on $h_v$ produce powers of the residual momentum
$k$, which is much smaller than $m_Q$. Hence, the nonlocal effective
Lagrangian (\ref{Lnonloc}) allows for a derivative expansion in
powers of $iD/m_Q$. Taking into account that $h_v$ contains a $P_+$
projection operator, and using the identity
\begin{equation}\label{pplusid}
   P_+\,i\,\rlap{\,/}D_\perp\,i\,\rlap{\,/}D_\perp P_+
   = P_+\,\bigg[ (i D_\perp)^2 + {g_s\over 2}\,
   \sigma_{\alpha\beta }\,G^{\alpha\beta } \bigg]\,P_+ \,,
\end{equation}
where $[i D^\alpha,i D^\beta]=i g_s\,G^{\alpha\beta}$ is the gluon
field-strength tensor, one finds that\cite{EiH1,FGL}
\begin{equation}\label{Lsubl}
   {\cal L}_{\rm eff} = \bar h_v\,i v\!\cdot\!D\,h_v
   + {1\over 2 m_Q}\,\bar h_v\,(i D_\perp)^2\,h_v
   + {g_s\over 4 m_Q}\,\bar h_v\,\sigma_{\alpha\beta}\,
   G^{\alpha\beta}\,h_v + O(1/m_Q^2) \,.
\end{equation}

\begin{figure}[htb]
   \vspace{0.5cm}
   \epsfysize=3.5cm
   \centerline{\epsffile{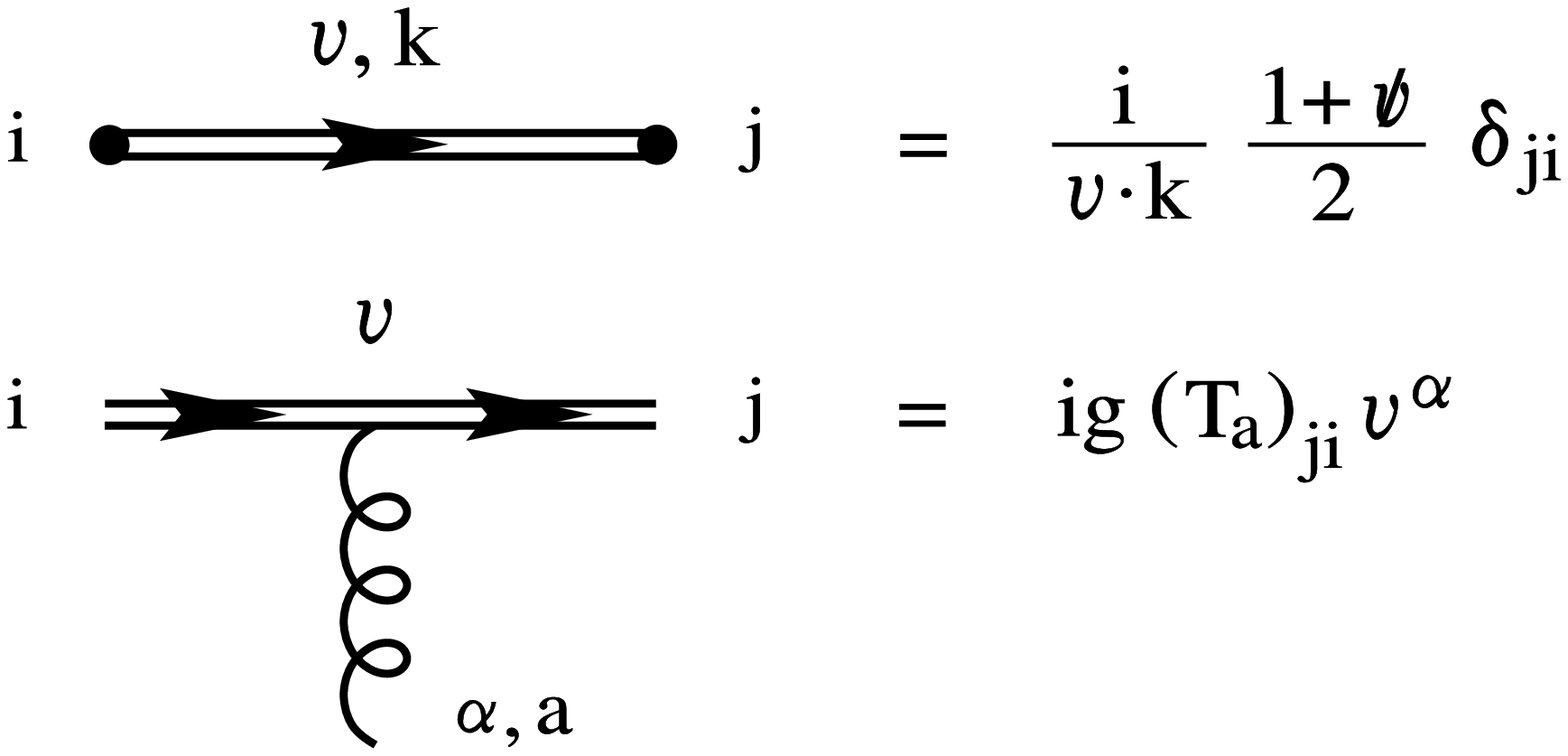}}
   \centerline{\parbox{11cm}{\caption{\label{fig:3.2}
Feynman rules of the heavy quark effective theory ($i,j$ and $a$ are
colour indices). It has become standard to represent a heavy quark by
a double line. The residual momentum $k$ is defined in
(\protect\ref{kresdef}).
   }}}
\end{figure}

In the limit $m_Q\to\infty$, only the first terms remains:
\begin{equation}\label{Leff}
   {\cal L}_\infty = \bar h_v\,i v\!\cdot\!D\,h_v \,.
\end{equation}
This is the effective Lagrangian of HQET. From it are derived the
Feynman rules depicted in Fig.~\ref{fig:3.2}. Let me take a moment to
study the symmetries of this Lagrangian\cite{Geor}. Since there
appear no Dirac matrices, interactions of the heavy quark with gluons
leave its spin unchanged. Associated with this is an SU(2) symmetry
group, under which ${\cal L}_\infty$ is invariant. The action of this
symmetry on the heavy quark fields becomes most transparent in the
rest frame, where the generators $S^i$ of spin SU(2) can be chosen as
\begin{equation}\label{Si}
   S^i = {1\over 2} \left( \begin{array}{cc}
                           \sigma^i ~&~ 0 \\
                           0 ~&~ \sigma^i \end{array} \right)
   = {1\over 2}\,\gamma_5 \gamma^0 \gamma^i \,,
\end{equation}
where $\sigma^i$ are the Pauli matrices. In a general frame, one
defines a set of three orthonormal vectors $e^i$ orthogonal to $v$,
and takes the generators of the spin symmetry as $S^i=\frac{1}{2}
\gamma_5\,\rlap/v\,\rlap/e^i$. These matrices satisfy the commutation
relations of SU(2) and commute with $\rlap/v$:
\begin{equation}
   [S^i,S^j] = i \epsilon^{ijk} S^k \,, \qquad
   [\rlap/v,S^i] = 0 \,.
\end{equation}
An infinitesimal SU(2) transformation $h_v\to (1 + i\vec\epsilon
\cdot\vec S\,)\,h_v$ leaves the Lagrangian invariant,
\begin{equation}\label{SU2tr}
   \delta{\cal L}_\infty = \bar h_v\,
   [i v\!\cdot\! D,i \vec\epsilon\cdot\vec S\,]\,h_v = 0 \,,
\end{equation}
and preserves the on-shell condition $\rlap/v\,h_v=h_v$, since
\begin{equation}
   \rlap/v\,(1+i\vec\epsilon\cdot\vec S\,)\,h_v
   = (1+i\vec\epsilon\cdot\vec S\,)\,\rlap/v\,h_v
   = (1+i\vec\epsilon\cdot\vec S\,)\,h_v \,.
\end{equation}

There is another symmetry of HQET, which arises since the mass of the
heavy quark does not appear in the effective Lagrangian. When there
are $N_h$ heavy quarks moving at the same velocity, one can simply
extend (\ref{Leff}) by writing
\begin{equation}\label{Leff2}
   {\cal L}_\infty
   = \sum_{i=1}^{N_h} \bar h_v^i\,i v\!\cdot\! D\,h_v^i \,.
\end{equation}
This is clearly invariant under rotations in flavour space. When
combined with the spin symmetry, the symmetry group becomes promoted
to SU$(2 N_h)$. This is the heavy quark spin-flavour
symmetry\cite{Isgu,Geor}. Its physical content is that, in the
$m_Q\to\infty$ limit, the strong interactions of a heavy quark
become independent of its mass and spin.

Let me now have a look at the operators appearing at order $1/m_Q$ in
(\ref{Lsubl}). They are easiest to identify in the rest frame. The
first operator,
\begin{equation}\label{Okin}
   {\cal O}_{\rm kin} = {1\over 2 m_Q}\,\bar h_v\,(i D_\perp)^2\,
   h_v \to - {1\over 2 m_Q}\,\bar h_v\,(i \vec D\,)^2\,h_v \,,
\end{equation}
is the gauge-covariant extension of the kinetic energy arising from
the off-shell residual motion of the heavy quark. The second operator
is the non-Abelian analogue of the Pauli term, which describes the
colour-magnetic interaction of the heavy quark spin with the gluon
field:
\begin{equation}\label{Omag}
   {\cal O}_{\rm mag} = {g_s\over 4 m_Q}\,\bar h_v\,
   \sigma_{\alpha\beta}\,G^{\alpha\beta}\,h_v \to
   - {g_s\over m_Q}\,\bar h_v\,\vec S\!\cdot\!\vec B_c\,h_v \,.
\end{equation}
Here $\vec S$ is the spin operator defined in (\ref{Si}), and $B_c^i
= -\frac{1}{2}\epsilon^{ijk} G^{jk}$ are the components of the
colour-magnetic field. This chromo-magnetic interaction is a
relativistic effect, which scales like $1/m_Q$. This is the origin of
the heavy quark spin symmetry.

\subsection{Spectroscopic Implications}

The spin-flavour symmetry leads to many interesting relations between
the properties of hadrons containing a heavy quark. The most direct
consequences concern the spectroscopy of such states\cite{IsWi}. In
the $m_Q\to\infty$ limit, the spin of the heavy quark and the total
angular momentum $j$ of the light degrees of freedom are separately
conserved by the strong interactions. Because of heavy quark
symmetry, the dynamics is independent of the spin and mass of the
heavy quark. Hadronic states can thus be classified by the quantum
numbers (flavour, spin, parity, etc.) of the light degrees of
freedom. The spin symmetry predicts that, for fixed $j\neq 0$, there
is a doublet of degenerate states with total spin
$J=j\pm\frac{1}{2}$. The flavour symmetry relates the properties of
states with different heavy quark flavour.

Consider, as an example, the ground-state mesons containing a heavy
quark. In this case the light degrees of freedom have the quantum
numbers of an antiquark, and the degenerate states are the
pseudoscalar ($J=0$) and vector ($J=1$) mesons. In the charm and
bottom systems, one knows experimentally
\begin{eqnarray}
   m_{B^*} - m_B &\simeq& 46~{\rm MeV} \,, \nonumber\\
   m_{D^*} - m_D &\simeq& 142~{\rm MeV} \,.
\end{eqnarray}
These mass splittings are in fact reasonably small. To be more
specific, at order $1/m_Q$ one expects hyperfine corrections to
resolve the degeneracy, for instance $m_{B^*}-m_B\propto 1/m_b$. This
leads to the refined prediction $m_{B^*}^2 - m_B^2 \simeq m_{D^*}^2
- m_D^2 \simeq {\rm const}$. The data are compatible with this:
\begin{eqnarray}\label{VPexp}
   m_{B^*}^2 - m_B^2 &\simeq& 0.49~{\rm GeV}^2 \,, \nonumber\\
   m_{D^*}^2 - m_D^2 &\simeq& 0.55~{\rm GeV}^2 \,.
\end{eqnarray}

One can also study excited meson states, in which the light
constituents carry orbital angular momentum. It is tempting to
interpret $D_1(2420)$ with $J^P=1^+$ and $D_2(2460)$ with $J^P=2^+$
as the spin doublet corresponding to $j=\frac{3}{2}$. The small mass
difference $m_{D_2^*} - m_{D_1} \simeq 35$ MeV supports this
assertion. One then expects
\begin{equation}
   m_{B_2^*}^2 - m_{B_1}^2 \simeq m_{D_2^*}^2 - m_{D_1}^2
   \simeq 0.17~{\rm GeV}^2
\end{equation}
for the corresponding states in the bottom system. The fact that this
mass splitting is smaller than for the ground-state mesons is not
unexpected. For instance, in the nonrelativistic constituent quark
model the light antiquark in these excited mesons is in a $p$-wave
state, and its wave function at the location of the heavy quark
vanishes. Hence, in this model hyperfine corrections are strongly
suppressed.

A typical prediction of the flavour symmetry is that the ``excitation
energies'' for states with different quantum numbers of the light
degrees of freedom are approximately the same in the charm and bottom
systems. For instance, one expects
\begin{eqnarray}
   m_{B_S} - m_B &\simeq m_{D_S} - m_D &\simeq 100~{\rm MeV} \,,
    \nonumber\\
   m_{B_1} - m_B &\simeq m_{D_1} - m_D &\simeq 557~{\rm MeV} \,,
    \nonumber\\
   m_{B_2^*} - m_B &\simeq m_{D_2^*} - m_D
   &\simeq 593~{\rm MeV} \,.
\end{eqnarray}
The first relation has been very nicely confirmed by the discovery of
the $B_S$ meson at LEP\cite{LEPBs}. The observed mass\cite{tauB}
$m_{B_S}=5.368\pm 0.005$ GeV corresponds to an excitation energy
$m_{B_S}-m_B=89\pm 5$ MeV.

\subsection{Weak Decay Form Factors}

Of particular interest are the relations between the weak decay form
factors of heavy mesons, which parametrize hadronic matrix elements
of currents between two meson states containing a heavy quark. These
relations have been derived by Isgur and Wise\cite{Isgu},
generalizing ideas developed by Nussinov and Wetzel\cite{Nuss} and by
Voloshin and Shifman\cite{Vol1,Vol2}. For the purpose of this
discussion, it is convenient to work with a mass-independent
normalization of meson states,
\begin{equation}\label{nonrelnorm}
   \langle M(p')|M(p)\rangle = {2 p^0\over m_M}\,(2\pi)^3\,
   \delta^3(\vec p-\vec p\,') \,,
\end{equation}
instead of the more conventional, relativistic normalization
\begin{equation}\label{relnorm}
   \langle\widetilde{M}(p')|\widetilde{M}(p)\rangle
   = 2 p^0\,(2\pi)^3\,\delta^3(\vec p-\vec p\,') \,.
\end{equation}
In the first case, $p^0/m_M=v^0$ depends only on the meson velocity.
In fact, it is more natural for heavy quark systems to use velocity
rather than momentum variables. I will thus write $|M(v)\rangle$
instead of $|M(p)\rangle$. The relation to the conventionally
normalized states is $|M(v)\rangle = m_M^{-1/2}\,\widetilde{M}(p)
\rangle$.

\begin{figure}[htb]
   \vspace{0.5cm}
   \epsfysize=3.4cm
   \centerline{\epsffile{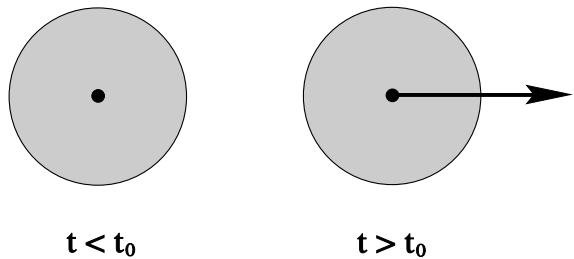}}
   \centerline{\parbox{11cm}{\caption{\label{fig:3.3}
The action of an external heavy quark current, as seen by the light
degrees of freedom in the initial state.
   }}}
\end{figure}

Consider now the elastic scattering of a pseudoscalar meson, $P(v)\to
P(v')$, induced by an external vector current coupled to the heavy
quark contained in $P$. Before the action of the current, the light
degrees of freedom in the initial state orbit around the heavy quark,
which acts as a source of colour moving with the meson's velocity
$v$. On average, this is also the velocity of the ``brown muck''. The
action of the current is to replace instantaneously (at $t=t_0$) the
colour source by one moving at velocity $v'$, as indicated in
Fig.~\ref{fig:3.3}. If $v=v'$, nothing really happens; the light
degrees of freedom do not realize that there was a current acting on
the heavy quark. If the velocities are different, however, the
``brown muck'' suddenly finds itself interacting with a moving
(relative to its rest frame) colour source. Soft gluons have to be
exchanged in order to rearrange the light degrees of freedom and
build the final state meson moving at velocity $v'$. This
rearrangement leads to a form factor suppression, which reflects the
fact that as the velocities become more and more different, the
probability for an elastic transition decreases. The important
observation is that, in the $m_Q\to\infty$ limit, the form factor can
only depend on the Lorentz boost $\gamma = v\cdot v'$ that connects
the rest frames of the initial- and final-state mesons. That the form
factor is independent of the heavy quark mass can also be seen from
the following intuitive argument: The light constituents in the
initial and final state carry momenta that are typically of order
$\Lambda_{\rm QCD} v$ and $\Lambda_{\rm QCD} v'$, respectively. Thus,
their momentum transfer is $q^2\sim\Lambda_{\rm QCD}^2 (v\cdot
v'-1)$, which is in fact independent of $m_Q$.

The result of this discussion is that in the effective theory, which
provides the appropriate framework to consider the limit
$m_Q\to\infty$ with the quark velocities kept fixed, the hadronic
matrix element describing the scattering process can be written
as\cite{Isgu}
\begin{equation}\label{elast}
   \langle P(v')|\,\bar h_{v'}\gamma^\mu h_v\,|P(v)\rangle
   = \xi(v\cdot v')\,(v+v')^\mu \,,
\end{equation}
with a form factor $\xi(v\cdot v')$ that does not depend on $m_Q$.
Since the matrix element is invariant under complex conjugation
combined with an interchange of $v$ and $v'$, the function
$\xi(v\cdot v')$ must be real. That there is no term proportional to
$(v-v')^\mu$ can be seen by contracting the matrix element with
$(v-v')_\mu$, and using $\rlap/v h_v = h_v$ and $\bar h_{v'}\rlap/v'
= \bar h_{v'}$.

One can now use the flavour symmetry to replace the heavy quark $Q$
in one of the meson states by a heavy quark $Q'$ of a different
flavour, thereby turning $P$ into another pseudoscalar meson $P'$. At
the same time, the current becomes a flavour-changing vector current.
In the infinite mass limit, this is a symmetry transformation, under
which the effective Lagrangian is invariant. Hence, the matrix
element
\begin{equation}\label{inelast}
   \langle P'(v')|\,\bar h'_{v'}\gamma^\mu h_v\,|P(v)\rangle
   = \xi(v\cdot v')\,(v+v')^\mu
\end{equation}
is still determined by the same function $\xi(v\cdot v')$. This
universal form factor is called the Isgur--Wise function, after the
discoverers of this relation\cite{Isgu}.

For equal velocities, the vector current $J^\mu=\bar h'_v\gamma^\mu
h_v = \bar h'_v v^\mu h_v$ is conserved in the effective theory,
irrespective of the flavour of the heavy quarks, since
\begin{equation}\label{Jcons1}
   i\partial_\mu J^\mu = \bar h'_v\,(i v\!\cdot\!D\,h_v)
   + (i v\!\cdot\!D\,\bar h'_v)\,h_v = 0
\end{equation}
by the equation of motion, $i v\!\cdot\!D\,h_v=0$, which follows from
the effective Lagrangian (\ref{Leff}). The conserved charges
\begin{equation}
   N_{Q' Q} = \int\!{\rm d}^3 x\,J^0(x)
   = \int\!{\rm d}^3 x\,h'^{\dagger}_v(x)\,h_v(x)
\end{equation}
are generators of the flavour symmetry. The diagonal generators
simply count the number of heavy quarks, whereas the off-diagonal
ones replace a heavy quark by another: $N_{QQ}|P(v)\rangle=
|P(v)\rangle$ and $N_{Q' Q}|P(v)\rangle=|P'(v)\rangle$. It follows
that
\begin{equation}
   \langle P'(v)|\,N_{Q' Q}\,|P(v)\rangle = \langle P(v)|P(v)\rangle
   = 2 v^0\,(2\pi)^3\,\delta^3(\vec 0\,) \,,
\end{equation}
and from a comparison with (\ref{inelast}) one concludes that the
Isgur--Wise function is normalized at the point of equal velocities:
\begin{equation}\label{Jcons2}
   \xi(1) = 1 \,.
\end{equation}
This can easily be understood in terms of the physical picture
discussed above: When there is no velocity change, the light degrees
of freedom see the same colour field and are in an identical
configuration before and after the action of the current. There is no
form factor suppression. Since $E_{\rm {recoil}}= m_{P'}\,(v\cdot
v'-1)$ is the recoil energy of the daughter meson $P'$ in the rest
frame of the parent meson $P$, the point $v\cdot v'=1$ is referred to
as the zero recoil limit.

The heavy quark spin symmetry leads to additional relations among
weak decay form factors. It can be used to relate matrix elements
involving vector mesons to those involving pseudoscalar mesons. In
the effective theory, a vector meson with longitudinal polarization
$\epsilon_3$ is related to a pseudoscalar meson by
\begin{equation}\label{VSPrel}
   |V(v,\epsilon_3)\rangle = 2\,{\bf S}_Q^3\,|P(v)\rangle \,,
\end{equation}
where ${\bf S}_Q^3$ is a Hermitian operator that acts on the spin of
the heavy quark $Q$. It follows that
\begin{eqnarray}\label{VPrel}
   \langle V'(v',\epsilon_3)|\,\bar h'_{v'}\Gamma\,h_v\,|P(v)\rangle
   &=& \langle P'(v')|\,2\,
    \big[ {\bf S}_{Q'}^3,\bar h'_{v'}\Gamma\,h_v \big]\,|P(v)\rangle
    \nonumber\\
   &=& \langle P'(v')|\,\bar h'_{v'} (2 S^3\Gamma)\,h_v\,|P(v)\rangle
    \,,
\end{eqnarray}
where $\Gamma$ can be an arbitrary combination of Dirac matrices, and
$S^3$ is a matrix representation of the operator ${\bf S}_{Q'}^3$. It
is easiest to evaluate this expression in the rest frame of the
final-state meson, where [cf.~(\ref{Si})]
\begin{equation}
   v'^\mu = (1,0,0,0) \,,\qquad
   \epsilon_3^\mu = (0,0,0,1) \,,\qquad
   S^3 = {1\over 2}\,\gamma_5 \gamma^0 \gamma^3 \,.
\end{equation}
It is then straightforward to obtain the following commutation
relations for the components of the weak current $(V-A)^\mu =
\bar h'_{v'}\gamma^\mu(1-\gamma_5)\,h_v$:
\begin{eqnarray}\label{commut}
   2\,{[ {\bf S}_{Q'}^3, V^0-A^0 ]} &=& A^3-V^3 \,, \nonumber\\
   2\,{[ {\bf S}_{Q'}^3, V^3-A^3 ]} &=& A^0-V^0 \,, \nonumber\\
   2\,{[ {\bf S}_{Q'}^3, V^1-A^1 ]} &=& \phantom{ - } i(A^2-V^2) \,,
    \nonumber\\
   2\,{[ {\bf S}_{Q'}^3, V^2-A^2 ]} &=& -i(A^1-V^1) \,.
\end{eqnarray}
Using (\ref{VPrel}) and (\ref{commut}), one can relate the matrix
element of the weak current between a pseudoscalar and a vector meson
to the matrix element of the vector current between two pseudoscalar
mesons given in (\ref{inelast}). The result can be written in the
covariant form\cite{Isgu}:
\begin{eqnarray}\label{PVff}
   \langle V'(v',\epsilon)|\,\bar h'_{v'}\gamma^\mu
   (1-\gamma_5)\,h_v\,
   |P(v)\rangle &=& i\epsilon^{\mu\nu\alpha\beta}\,\epsilon_\nu^*\,
    v'_\alpha v_\beta\,\,\xi(v\cdot v') \nonumber\\
   &&\mbox{}- \!\Big[ \epsilon^{*\mu}\,(v\cdot v'+1)
    - v'^\mu\,\epsilon^*\!\cdot v \Big]\,\xi(v\cdot v') \,, \quad
\end{eqnarray}
where $\epsilon^{0123}=-\epsilon_{0123}=-1$. Once again, the matrix
element is completely described in terms of the universal Isgur--Wise
form factor.

Equations (\ref{inelast}) and (\ref{PVff}) summarize the relations
imposed by heavy quark symmetry on the weak decay form factors
describing the semileptonic decay processes $B\to D\,\ell\, \bar\nu$
and $B\to D^*\ell\,\bar\nu$. These relations are model-independent
consequences of QCD in the limit where $m_b, m_c\gg\Lambda_{\rm
QCD}$. They play a crucial role in the determination of $|\,V_{cb}|$,
as I will discuss later in this lecture.

\subsection{The $1/m_Q$ Expansion}

At tree level, eq.~(\ref{Lsubl}) defines the Lagrangian of HQET as a
series of local, higher-dimension operators multiplied by powers of
$1/m_Q$. The expression for $H_v$ in (\ref{Hfield}) can be used to
derive a similar expansion for the full heavy quark field $Q(x)$:
\begin{eqnarray}
   Q (x) &=& e^{-i m_Q v\cdot x}\,\bigg[ 1
    + {1\over(2 m_Q + i v\!\cdot\!D - i\eta)}\,
    i\,\rlap{\,/}D_\perp \bigg]\,h_v(x) \nonumber\\
   &=& e^{-i m_Q v\cdot x}\,\bigg( 1
    + {i\,\rlap{\,/}D_\perp\over 2 m_Q} + \ldots \bigg)\,h_v(x) \,.
\end{eqnarray}
This relation contains the recipe for the construction (at tree
level) of any operator in HQET that contains one or more heavy quark
fields. For instance, the vector current $V^\mu=\bar q\,\gamma^\mu Q$
composed of a heavy and a light quark is represented as
\begin{equation}\label{Jmux}
   V^\mu(x) = e^{-i m_Q v\cdot x}\,\bar q(x)\,\gamma^\mu
   \bigg( 1 + {i\,\rlap{\,/}D_\perp\over 2 m_Q} + \ldots
\bigg)\,h_v(x)
   \,.
\end{equation}
Matrix elements of this current can be parametrized by hadronic form
factors, and the purpose of using an effective theory is to make the
$m_Q$ dependence of these form factors explicit.

Consider, as an example, the matrix element of $V^\mu(0)$ between a
heavy meson $M(v)$ and some light final state $\ell$:
\begin{equation}\label{example}
   \langle\,\ell\,|\,V^\mu\,|M(v)\rangle
   = \langle\,\ell\,|\,\bar q\,\gamma^\mu h_v\,|M(v)\rangle
   + {1\over 2 m_Q}\,\langle\,\ell\,|\,\bar q\,\gamma^\mu\,
   i\,\rlap{\,/}D_\perp h_v\,|M(v)\rangle + \ldots \,.
\end{equation}
It would be nice if the matrix elements on the right-hand side of
this equation were independent of $m_Q$. Then the second term would
give the leading power correction to the first one. However, the
equation of motion for $h_v$ derived from (\ref{Lsubl}) contains
$1/m_Q$ corrections, too. This leads to an $m_Q$ dependence of any
hadronic matrix element of operators containing such fields. Another
way to say this is that the eigenstates of the effective Lagrangian
${\cal L}_{\rm eff}$ (supplemented by the standard QCD Lagrangian for
the light quarks and gluons) depend, at higher order in $1/m_Q$, on
the heavy quark mass. This is no surprise, since the effective
Lagrangian is equivalent to the Lagrangian of the full theory.

It is better to organize things in a slightly different way by
working with the eigenstates of only the leading-order effective
Lagrangian ${\cal L}_\infty$ in (\ref{Leff}), treating the
higher-dimension operators perturbatively as power
corrections\cite{Luke,GGW,FaNe}. Then the equation of motion
satisfied by $h_v$ is simply
\begin{equation}\label{EoM}
   i v\!\cdot\!D\,h_v = 0 \,,
\end{equation}
and the states of the effective theory are truly independent of
$m_Q$. However, these states are now different from the states of the
full theory. For instance, instead of (\ref{example}) one should
write
\begin{eqnarray}\label{toypower}
   \langle\,\ell\,|\,V^\mu\,|M(v)\rangle_{\rm QCD}
   &=& \langle\,\ell\,|\,\bar q\,\gamma^\mu h_v\,|M(v)
    \rangle_\infty \nonumber\\
   &&\mbox{}+ {1\over 2 m_Q}\,\langle\,\ell\,|\,\bar q\,\gamma^\mu\,
    i\,\rlap{\,/}D_\perp h_v\,|M(v)\rangle_\infty \nonumber\\
   &&\mbox{}+ {1\over 2 m_Q}\,\langle\,\ell\,|\,i\!\int\!{\rm d}y\,
    \,T\,\big\{\,\bar q\,\gamma^\mu h_v(0),
    \bar h_v\,(i D_\perp)^2\,h_v(y)\,\big\}\,
    |M(v)\rangle_\infty \nonumber\\
   &&\mbox{}+ {g_s\over 4 m_Q}\,\langle\,\ell\,|\,i\!\int\!{\rm d}y\,
    \,T\,\big\{\,\bar q\,\gamma^\mu h_v(0),
    \bar h_v\,\sigma_{\alpha\beta}\,G^{\alpha\beta}\,h_v(y)\,
    \big\}\,|M(v)\rangle_\infty \nonumber\\
   &&\mbox{}+ O(1/m_Q^2) \,.
\end{eqnarray}
The matrix elements on the right-hand side are independent of the
heavy quark mass. The mass dependence of the states $|M(v)
\rangle_{\rm QCD}$ is reflected in HQET by the appearance of the
third and fourth term, which arise from an insertion of the
first-order power corrections in the Lagrangian into the
leading-order matrix element of the current. This insertion can be
thought of as being a correction to the wave function of the heavy
meson. From now on I will omit the subscripts on the states. I will
instead use the symbol ``$\cong$'' and write
\begin{eqnarray}\label{Jexptree}
   V^\mu(0) \cong \bar q\,\gamma^\mu h_v(0)
   &+& {1\over 2 m_Q}\,\bar q\,\gamma^\mu\,i\,\rlap{\,/}D_\perp
    h_v(0) \nonumber\\
   &+& {1\over 2 m_Q}\,i\!\int\!{\rm d}y\,\,T\,\big\{\,
    \bar q\,\gamma^\mu h_v(0),
    \bar h_v\,(i D_\perp)^2\,h_v(y)\,\big\} \nonumber\\
   &+& {g_s\over 4 m_Q}\,i\!\int\!{\rm d}y\,\,T\,\big\{\,
    \bar q\,\gamma^\mu h_v(0),
    \bar h_v\,\sigma_{\alpha\beta}\,G^{\alpha\beta}\,h_v(y)\,\big\}
    + \ldots
\end{eqnarray}
to indicate that the operators on the two sides of this equation have
to be evaluated between different states.

\subsection{Matching and Running}

In Section 3.2, I have discussed the first two steps in the
construction of HQET. Integrating out the small components in the
heavy quark spinor fields, a nonlocal effective action was derived.
This was then expanded in a series of local operators of increasing
dimension to obtain an effective Lagrangian. A similar expansion
could be written down for any external current. The effective
Lagrangian and the effective currents derived that way correctly
reproduce the long-distance physics of the full theory. They cannot
describe the short-distance physics correctly, however. The reason is
obvious: The heavy quark participates in strong interactions through
its coupling to gluons. These gluons can be soft or hard, i.e.\ their
virtual momenta can be small, of the order of the confinement scale,
or
large, of the order of the heavy quark mass. But hard gluons can
resolve
the nonlocality of the propagator of the small component fields
$H_v$. Their effects are not taken into account in the na\"\i ve
version of the OPE, which was used in the derivation of the effective
Lagrangian in (\ref{Lsubl}) and the effective vector current in
(\ref{Jexptree}). So far, the effective theory provides an
appropriate description only at scales $\mu\ll m_Q$.

In this section, I will discuss the systematic treatment of
short-distance corrections. A new feature of such corrections is that
through the running coupling constant they induce a logarithmic
dependence on the heavy quark mass\cite{Vol1}. The important
observation is that $\alpha_s(m_Q)$ is small, so that these effects
can be calculated in perturbation theory. Consider, as an example,
matrix elements of the vector current $V^\mu = \bar q\,\gamma^\mu Q$.
In QCD this current is partially conserved and needs no
renormalization\cite{Prep}. Its matrix elements are free of
ultraviolet divergences. Still, these matrix elements can have
logarithmic dependence on $m_Q$ from the exchange of hard gluons with
virtual momenta of the order of the heavy quark mass. If one goes
over to the effective theory by taking the limit $m_Q\to\infty$,
these logarithms diverge. Consequently, the vector current in the
effective theory does require a renormalization\cite{PoWi}. Its
matrix elements depend on an arbitrary renormalization scale $\mu$,
which separates the regions of short- and long-distance physics. If
$\mu$ is chosen such that $\Lambda_{\rm QCD}\ll\mu\ll m_Q$, the
effective coupling constant in the region between $\mu$ and $m_Q$ is
small, and perturbation theory can be used to compute the
short-distance corrections. These corrections have to be added to the
matrix elements of the effective theory, which only contain the
long-distance physics below the scale $\mu$. Schematically, then, the
relation between matrix elements in the full and in the effective
theory is
\begin{equation}\label{OPEex}
   \langle\,V^\mu(m_Q)\,\rangle_{\rm QCD}
   = C_0(m_Q,\mu)\,\langle V_0(\mu)\rangle_\infty
   + {C_1(m_Q,\mu)\over 2 m_Q}\,\langle V_1(\mu)\rangle_\infty
   + \ldots \,,
\end{equation}
where I have indicated that matrix elements of $V^\mu$ in the full
theory depend on $m_Q$, whereas matrix elements of operators in the
effective theory are mass-independent, but do depend on the
renormalization scale. The Wilson coefficients $C_i(m_Q,\mu)$ are
defined by this relation. Order by order in perturbation theory, they
can be computed from a comparison of the matrix elements in both
theories. Since the effective theory is constructed to reproduce
correctly the low-energy behaviour of the full theory, this
``matching'' procedure is independent of any long-distance physics,
such as infrared singularities, nonperturbative effects, the nature
of the external states used in the matrix elements, physical cuts,
etc. Only at high energies do the two theories differ, and these
differences are corrected for by the short-distance coefficients.

The calculation of the coefficient functions in perturbation theory
uses the powerful methods of the renormalization group, which I have
discussed in my first lecture. It is in principle straightforward,
yet in practice rather tedious. Much of the recent work on heavy
quark symmetry has been devoted to this subject. In this section, I
will discuss as an illustration the wave-function renormalization of
the heavy quark field in HQET, and the short-distance expansion of a
heavy--light vector current to leading order in $1/m_Q$. For a more
comprehensive presentation of short-distance corrections, I refer to
my review article\cite{review}.

In quantum field theory, the parameters and fields of the Lagrangian
have no direct physical significance. They have to be renormalized
before they can be related to observable quantities. In an
intermediate step the theory has to be regularized. The most
convenient regularization scheme in QCD is dimensional
regularization\cite{tHo2,tHo1,Boll}, in which the dimension of
space-time is analytically continued to $D=4-2\epsilon$, with
$\epsilon$ being infinitesimal. Loop integrals that are
logarithmically divergent in four dimensions become finite for
$\epsilon>0$. From the fact that the action $S=\int{\rm d}^Dx\, {\cal
L}(x)$ is dimensionless, one can derive the mass dimensions of the
fields and parameters of the theory. For instance, one finds that the
``bare'' coupling constant $\alpha_s^{\rm bare}$ is no longer
dimensionless if $D\ne 4$: ${\rm dim}[\,\alpha_s^{\rm bare}\,] =
2\epsilon$. In a renormalizable theory, it is possible to rewrite the
Lagrangian in terms of renormalized quantities, in such a way that
Green's functions of the renormalized fields remain finite as
$\epsilon\to 0$. For QCD, one introduces renormalized quantities by
$Q^{\rm bare} = Z_Q^{1/2}\,Q^{\rm ren}$, $A^{\rm bare} = Z_A^{1/2}
A^{\rm ren}$, $\alpha_s^{\rm bare} = \mu^{2\epsilon}
Z_\alpha\,\alpha_s^{\rm ren}$, etc., where $\mu$ is an arbitrary mass
scale introduced to render the renormalized coupling constant
dimensionless. Similarly, in HQET one defines a renormalized heavy
quark field by $h_v^{\rm bare} = Z_h^{1/2}\,h_v^{\rm ren}$. From now
on, the superscript ``ren'' will be omitted.

\begin{figure}[htb]
   \vspace{0.5cm}
   \epsfxsize=4cm
   \centerline{\epsffile{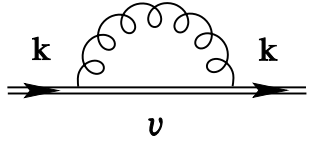}}
   \centerline{\parbox{11cm}{\caption{\label{fig:3.4}
Self-energy $-i\Sigma(v\cdot k)$ of a heavy quark in HQET. The
velocity $v$ is conserved by the strong interactions.
   }}}
\end{figure}

As a warm-up, let me sketch the calculation of the wave-function
renormalization for a heavy quark in HQET. In the $\overline{\rm MS}$
scheme, $Z_h$ can be computed from the $1/\hat\epsilon$-pole in the
quark self-energy shown in Fig.~\ref{fig:3.4}:
\begin{equation}
   1 - Z_h^{-1} = {1\over\hat\epsilon}\mbox{-pole of }\,
   {\partial\Sigma(v\cdot k)\over\partial v\cdot k} \,.
\end{equation}
As long as $v\cdot k<0$, the self-energy is infrared finite and real.
The result is gauge-dependent, however. In Feynman gauge, one obtains
at one-loop order
\begin{eqnarray}
   \Sigma(v\cdot k) &=& - {4i g_s^2\over 3} \int
    {\mbox{d}^D t\over(2\pi)^D}\,{1\over (t^2+i\eta)
     \big[ v\cdot(t+k)+i\eta \big]} \nonumber\\
   &=& - {8i g_s^2\over 3} \int\limits_0^\infty\!\mbox{d}\rho
    \int{\mbox{d}^D t\over(2\pi)^D}\,{1\over\big[ t^2 +
     2\rho\,v\cdot(t+k) + i\eta \big]^2} \nonumber\\
   &=& {2\alpha_s\over 3\pi}\,\Gamma(\epsilon)
    \int\limits_0^\infty\!\mbox{d}\rho\,\bigg(
    {\rho^2 + \rho\omega\over 4\pi\mu^2} \bigg)^{-\epsilon} \,,
\end{eqnarray}
where $\rho$ is a dimensionful Feynman parameter, and $\omega=-2
v\cdot k>0$ acts as an infrared cutoff. A straightforward calculation
leads to
\begin{eqnarray}
   {\partial\Sigma(v\cdot k)\over\partial v\cdot k}
   &=& {4\alpha_s\over 3\pi}\,\Gamma(1+\epsilon)\,
    \bigg( {\omega^2\over 4\pi\mu^2} \bigg)^{-\epsilon}
    \int\limits_0^1\!\mbox{d}z\,z^{-1+2\epsilon}\,
    (1-z)^{-\epsilon} \nonumber\\
   &=& {4\alpha_s\over 3\pi}\,\Gamma(2\epsilon)\,\Gamma(1-\epsilon)\,
    \bigg( {\omega^2\over 4\pi\mu^2} \bigg)^{-\epsilon} \,,
\end{eqnarray}
where I have substituted $\rho=\omega\,(1-z)/z$. From an expansion
around $\epsilon=0$, one obtains
\begin{equation}\label{ZfacMS}
   Z_h = 1 + {2\alpha_s\over 3\pi\hat\epsilon} \,.
\end{equation}
This result was first derived by Politzer and Wise\cite{PoWi}. In the
meantime, the calculation was also done at the two-loop
order\cite{JiMu}$^-$\cite{BGSc}.

Similar to the fields and coupling constants, any composite operator
built from quark and gluon fields may require a renormalization
beyond that of its component fields. Such operators can be divided
into three classes: gauge-invariant operators that do not vanish by
the equations of motion (class-I), gauge-invariant operators that
vanish by the equations of motion (class-II), and operators which are
not gauge-invariant (class-III). In general, operators with the same
dimension and quantum numbers mix under renormalization. However,
things simplify if one works with the background field
technique\cite{DeWi}$^-$\cite{Abbo}, which serves for quantizing
gauge theories, preserving explicit gauge invariance. This offers the
advantage that a class-I operator cannot mix with class-III
operators, so that only gauge-invariant operators need be
considered\cite{Klug}. Furthermore, class-II operators are irrelevant
since their matrix elements vanish by the equations of motion. It it
thus sufficient to consider class-I operators only. For a set
$\{{\cal O}_i\}$ of $n$ class-I operators that mix under
renormalization, one defines an $n\times n$ matrix of renormalization
factors $Z_{ij}$ by (summation over $j$ is understood) ${\cal
O}_i^{\rm bare} = Z_{ij}\,{\cal O}_j$, such that the matrix elements
of the renormalized operators ${\cal O}_j$ remain finite as
$\epsilon\to 0$. In contrast to the bare operators, the renormalized
ones depend on the subtraction scale via the $\mu$ dependence of
$Z_{ij}$:
\begin{equation}
   \mu {{\rm d}\over{\rm d}\mu}\,{\cal O}_i
   = \bigg(\mu {{\rm d}\over{\rm d}\mu} Z_{ij}^{-1}\bigg)\,
   {\cal O}_j^{\rm bare} = - \gamma_{ik}\,{\cal O}_k \,,
\end{equation}
where
\begin{equation}
   \gamma_{ik} = Z_{ij}^{-1}\,\mu {{\rm d}\over{\rm d}\mu}\,Z_{jk}
\end{equation}
are called the anomalous dimensions. It is sometimes convenient to
introduce a compact matrix notation, in which $\vec{\cal O}$ is the
vector of renormalized operators, $\hat Z$ is the matrix of
renormalization factors, and $\hat\gamma$ denotes the anomalous
dimension matrix. Then the running of the renormalized operators is
controlled by the renormalization-group equation (RGE)
\begin{equation}\label{RGEops}
   \bigg( \mu {{\rm d}\over{\rm d}\mu} + \hat\gamma \bigg)\,
   \vec{\cal O} = 0 \,.
\end{equation}

As an example, consider the short-distance expansion of the
heavy--light vector current $V^\mu=\bar q\,\gamma^\mu Q$ to leading
order in $1/m_Q$. In the full theory, the vector current is not
renormalized. But the effective current operators in HQET do require
renormalization. The general form of the short-distance expansion
reads
\begin{eqnarray}\label{OPEcor}
   V^\mu(m_Q) &\cong& C_i(m_Q,\mu)\,{\cal O}_i(\mu) + O(1/m_Q)
    \nonumber\\
   \phantom{ \bigg[ }
   &=& C_i(m_Q,\mu)\,Z_{ij}^{-1}(\mu)\,{\cal O}_j^{\rm bare}
    + O(1/m_Q) \,,
\end{eqnarray}
where I have indicated the $\mu$ dependence of the renormalized
operators. This is the correct generalization of (\ref{OPEex}) in the
case of operator mixing. In general, a complete set of operators with
the same quantum numbers appears on the right-hand side. In the case
at hand, there are two such operators, namely
\begin{equation}
   {\cal O}_1 = \bar q\,\gamma^\mu h_v \,, \qquad
   {\cal O}_2 = \bar q\,v^\mu h_v \,.
\end{equation}
{}From (\ref{RGEops}) and the fact that the product $C_i(m_Q,\mu)\,
{\cal O}_i(\mu)$ must be $\mu$-independent, one can derive the RGE
satisfied by the coefficient functions. It reads
\begin{equation}
   \bigg(\mu {{\rm d}\over{\rm d}\mu} - \hat\gamma^t \bigg)\,
   \vec C(m_Q,\mu) = 0 \,,
\end{equation}
where $\hat\gamma^t$ denotes the transposed anomalous dimension
matrix, and I have collected the coefficients into a vector.
The solution of the RGE proceeds as I described in
my first lecture; however, it is technically more complicated in
the case when there is operator mixing. One obtains
\begin{equation}\label{RGEsol}
   \vec C(m_Q,\mu) = \hat U(\mu,m_Q)\,\vec C(m_Q,m_Q) \,,
\end{equation}
with the evolution matrix\cite{Bura,Flor,BJLW}
\begin{equation}
   \hat U(\mu,m_Q) = T_\alpha\,\exp\!\int
   \limits_{\displaystyle\alpha_s(m_Q)}^{\displaystyle\alpha_s(\mu)}
   \!{\rm d}\alpha\,{\hat\gamma^t(\alpha)\over\beta(\alpha)} \,.
\end{equation}
Here ``$T_\alpha$'' means an ordering in the coupling constant such
that the couplings increase from right to left (for $\mu<m_Q$). This
is necessary since, in general, the anomalous dimension matrices at
different values of $\alpha$ do not commute: $[\hat\gamma(\alpha_1),
\hat\gamma(\alpha_2)]\ne 0$. In the present case, however, these
complications are absent, since the anomalous dimension matrix turns
out to be proportional to the unit matrix.

\begin{figure}[htb]
   \vspace{0.5cm}
   \epsfxsize=6.5cm
   \centerline{\epsffile{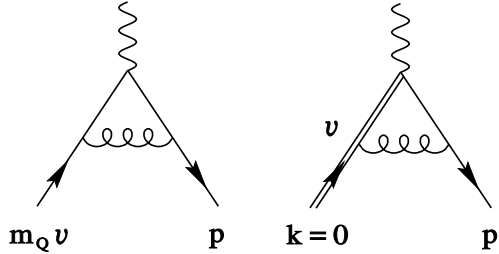}}
   \centerline{\parbox{11cm}{\caption{\label{fig:3.5}
One-loop corrections to the matrix elements of the vector current in
QCD and in HQET. The wavy line represents the current. The external
momenta are on-shell.
   }}}
\end{figure}

I will now illustrate the solution of (\ref{RGEsol}) to
next-to-leading order in renormaliza\-tion-group improved
perturbation theory. There are two ingredients to this calculation.
First, the Wilson coefficients at the high-energy scale $\mu=m_Q$
must be calculated to one-loop order,
\begin{equation}\label{Cinit}
   \vec C(m_Q,m_Q) = \vec C_{(0)} + \vec C_{(1)}\,
   {\alpha_s(m_Q)\over 4\pi} + \ldots \,,
\end{equation}
by matching QCD onto the effective theory. Then, in order to control
the running of the HQET operators, the anomalous dimension matrix
$\hat\gamma$ must be calculated to two-loop accuracy. Let me start
with a comparison of the one-loop matrix elements of the vector
current in the full and in the effective theory. It is legitimate to
perform the matching calculation with on-shell quark states. Then the
matrix elements can be written as $\langle V^\mu\rangle = \bar u_q\,
\Gamma^\mu\, u_Q$, where the heavy quark spinor satisfies
$\rlap/v\,u_Q=u_Q$. There will be infrared divergences in the
calculation, which I regulate by introducing a fictitious gluon mass
$\lambda$. At tree level, the vertex function in both theories is
simply $\Gamma^\mu = \gamma^\mu$. The one-particle irreducible
one-loop diagrams are shown in Fig.~\ref{fig:3.5}. They have to be
supplemented by an on-shell wave-function renormalization of the
external lines. The complete one-loop vertex function in QCD
is\cite{review}
\begin{equation}\label{GamQCD}
   \Gamma_{\rm QCD}^\mu = \bigg\{ 1 + {\alpha_s\over 2\pi}\,
   \bigg( \ln{m_Q^2\over\lambda^2} - {11\over 6}\bigg) \bigg\}\,
   \gamma^\mu + {2\alpha_s\over 3\pi}\,v^\mu \,.
\end{equation}
As required by the nonrenormalization theorem for partially conserved
currents, this result is gauge-independent and ultraviolet
finite\cite{Prep}. In the limit of degenerate quark masses, the
vector is conserved. The space integral over its time component is
the generator of a flavour symmetry. It cannot be renormalized. When
the symmetry is softly broken by the presence of mass splittings,
this is only relevant for small loop momenta but does not affect the
ultraviolet region. Thus, there can only be a finite renormalization.
Notice, however, that (\ref{GamQCD}) does contain an infrared
singularity, because the matrix element was calculated using
unphysical states.

In the effective theory, the one-loop vertex function for any
heavy--light current $\bar q\,\Gamma\,h_v$ is found to
be\cite{review}
\begin{equation}\label{Gamfac}
   \bigg\{ 1 + {\alpha_s\over 2\pi}\,\bigg( {1\over\hat\epsilon}
   + \ln{\mu^2\over\lambda^2} + {5\over 6} \bigg) \bigg\}\,
   \Gamma \,,
\end{equation}
where the Dirac structure of $\Gamma$ is completely arbitrary. It is
generally true that these operators are renormalized multiplicatively
and irrespectively of their Dirac structure\cite{Vol1}. Since there
is no approximate flavour symmetry relating light and heavy quarks in
the effective theory, it is not unexpected that the matrix elements
of the bare currents are ultraviolet divergent. In the $\overline{\rm
MS}$ scheme, the currents are renormalized by a diagonal matrix $\hat
Z$ with components
\begin{equation}\label{ZijHL}
   Z_{11} = Z_{22} = 1 + {\alpha_s\over 2\pi\hat\epsilon} \,, \qquad
   Z_{12} = Z_{21} = 0 \,.
\end{equation}
{}From (\ref{OPEcor}), it then follows that the renormalized one-loop
vertex function is
\begin{equation}
   \Gamma_{\rm HQET}^\mu = \bigg\{ 1 + {\alpha_s\over 2\pi}\,
   \bigg( \ln{\mu^2\over\lambda^2} + {5\over 6}\bigg) \bigg\}\,
   \Big[ C_1(m_Q,\mu)\,\gamma^\mu + C_2(m_Q,\mu)\,v^\mu \Big] \,.
\end{equation}
Notice that the $\lambda$ dependence is the same as in
(\ref{GamQCD}). The short-distance coefficients, which follow from a
comparison of the two results, are independent of the infrared
regulator:
\begin{eqnarray}\label{CHLmu}
   C_1(m_Q,\mu) &=& 1 + {\alpha_s\over\pi}\,\bigg(
    \ln{m_Q\over\mu} - {4\over 3} \bigg) \,, \nonumber\\
   C_2(m_Q,\mu) &=& {2\alpha_s\over 3\pi} \,.
\end{eqnarray}
In particular, one obtains for the matching conditions at $\mu=m_Q$:
\begin{eqnarray}
   C_1(m_Q,m_Q) &=& 1 - {4\alpha_s(m_Q)\over 3\pi} \,, \nonumber\\
   C_2(m_Q,m_Q) &=& {2\alpha_s(m_Q)\over 3\pi} \,.
\end{eqnarray}
This completes the first step of the calculation.

The next step is to calculate the one- and two-loop coefficients of
the anomalous dimension, as defined in (\ref{g0g1def}). From
(\ref{Gamfac}), it follows that the anomalous dimension matrix is
proportional to the unit matrix, with the one-loop coefficient
\begin{equation}
   \gamma_0 = - 4 \,.
\end{equation}
This is the so-called hybrid anomalous dimension of heavy--light
currents derived by Voloshin and Shifman\cite{Vol1}. In the context
of the effective theory, it has been calculated by Politzer and
Wise\cite{PoWi}. The two-loop coefficient $\gamma_1$ has been
calculated by Ji and Musolf\cite{JiMu}, and by Broadhurst and
Grozin\cite{BrGr}. They obtain
\begin{equation}
   \gamma_1 = - {254\over 9} - {56\over 27}\,\pi^2
   + {20\over 9}\,n_f \,.
\end{equation}
The final result for the Wilson coefficients can be derived using
(\ref{Uevol}). It is possible to write the answer in the factorized
form $C_i(m_Q,\mu) = \widehat{C}_i(m_Q)\,K_{\rm hl}(\mu)$, where the
dependence on the two scales has been completely separated. One finds
\begin{eqnarray}\label{C1C2hl}
   \widehat{C}_1(m_Q) &=& \Big[ \alpha_s(m_Q) \Big]^{-2/\beta_0}\,
    \bigg\{ 1 + {\alpha_s(m_Q)\over\pi}\,\bigg( S - {4\over 3}
    \bigg) \bigg\} \,, \nonumber\\
   \widehat{C}_2(m_Q) &=& \Big[ \alpha_s(m_Q) \Big]^{-2/\beta_0}\,
    \,{2\alpha_s(m_Q)\over3\pi} \,, \nonumber\\
   K_{\rm hl}(\mu) &=& \Big[ \alpha_s(\mu) \Big]^{2/\beta_0}\,
    \bigg\{ 1 - {\alpha_s(\mu)\over\pi}\,S \bigg\} \,,
\end{eqnarray}
where
\begin{equation}\label{Zcoef}
   S = {\gamma_1\beta_0 - \beta_1\gamma_0\over 8\beta_0^2}
   = 3\,{153-19 n_f\over(33-2 n_f)^2}
   - {381+28\pi^2-30 n_f\over 36\,(33-2 n_f)} \,.
\end{equation}

\subsection{Flavour-Changing Heavy Quark Currents}

The OPE of currents in the effective theory becomes considerably more
complicated in the case of flavour-changing currents. The weak
current $\bar c\,\gamma^\mu (1-\gamma_5)\,b$ is of this form,
however, so it is necessary to consider this case in detail. The
complications arise from the fact that there are two different heavy
quark masses, $m_b$ and $m_c$. Thus the calculation of the Wilson
coefficients becomes a two-scale problem. In addition, the
coefficients will depend in a nontrivial way on the velocity transfer
$w=v\cdot v'$.

There are two obvious ways of performing the transition from QCD to
an effective low-energy theory in which both the bottom and the charm
quarks are treated as heavy quarks (in the sense of HQET): The
transition can either be done in a single step, or by considering
first an intermediate theory with a static bottom quark, but a
dynamical charm quark. The latter becomes heavy in a second step. If
one could solve perturbation theory to all orders, both treatments
would lead to the same results for the Wilson coefficients. The
calculation differs in both cases, however, and the results also
differ if the perturbation series is truncated.

To see what the differences are, suppose first the two heavy quarks
have similar masses, i.e.\ $m_b\sim m_c\sim m$, with $m$ being some
average mass. It is then natural to remove at the same time the
dynamical degrees of freedom of both heavy quarks. Let me consider
the cases of the vector current $V^\mu = \bar c\,\gamma^\mu b$ and of
the axial vector current $A^\mu = \bar c\,\gamma^\mu\gamma_5\,b$
explicitly and denote them collectively by $J^\mu$. Each of these
currents obeys an expansion of the form
\begin{equation}\label{CiJi}
  J^\mu(m_b,m_c) \cong \sum_{j=1}^3 C_j(m_b,m_c,m,w,\mu)\,J_j(\mu)
  + O(1/m) \,,
\end{equation}
where $J_j=\bar h_{v'}^c\,\Gamma_j h_v^b$ are local current
operators in the effective theory with matrices
\begin{equation}\label{vecmatr}
   \Gamma_1 = \gamma^\mu \,,\qquad
   \Gamma_2 = v^\mu \,,\qquad
   \Gamma_3 = v'^\mu
\end{equation}
for the vector current, and
\begin{equation}\label{aximatr}
   \Gamma_1 = \gamma^\mu\gamma_5 \,,\qquad
   \Gamma_2 = v^\mu\gamma_5 \,,\qquad
   \Gamma_3 = v'^\mu\gamma_5
\end{equation}
for the axial vector current. In the above expression, I have
indicated that matrix elements of the original current $J^\mu$
between states of the full theory depend on $m_b$ and $m_c$, whereas
matrix elements of the operators $J_j$ between states of the
effective theory are mass-independent, but do depend on the
renormalization scale $\mu$. The short-distance coefficients are
functions of the heavy quark masses, the renormalization scale, and
the matching scale $m$. The dependence on $m$ would disappear if one
could sum the perturbative series to all orders. The advantage of
this first approach is its simplicity. At leading order in the $1/m$
expansion, only three current operators contribute. Matrix elements
of higher-dimension operators are suppressed by powers of
$\Lambda_{\rm QCD}/m$. The short-distance coefficients contain the
dependence on $m_b$ and $m_c$ correctly to a given order in
$\alpha_s$, via matching at $\mu=m$. The only disadvantage is the
residual dependence on the matching scale $m$, which arises when one
calculates to finite order in perturbation theory. Although in a
next-to-leading order calculation the scale in the leading anomalous
scaling factor is determined, one has no control over the scale in
the next-to-leading corrections proportional to $\alpha_s(m)$. Since
$m_b>m>m_c$, this introduces an uncertainty of order
$\alpha_s^2\ln(m_b/m_c)$.

The alternative approach is to consider first, as an intermediate
step for $m_b>\mu>m_c$, an effective theory with only a heavy bottom
quark. Denoting the effective current operators of dimension $(3+k)$
in this theory by $\widetilde{J}_i^{(k)}(m_c,\mu)$, indicating that
their matrix elements depend both on the charm quark mass and the
renormalization scale, I write the short-distance expansion as an
expansion in $1/m_b$:
\begin{equation}\label{interm}
  J^\mu(m_b,m_c) \cong \sum_{k=0}^\infty\,{1\over m_b^k}\,
  \sum_i D_i^{(k)}(m_b,\mu)\,\widetilde{J}_i^{(k)}(m_c,\mu) \,.
\end{equation}
Since the velocity of the charm quark is still a dynamical degree of
freedom, in this intermediate effective theory there are only two
dimension-three operators, namely
\begin{equation}
  \widetilde{J}_1^{(0)} = \bar c\,\Gamma_1 h_v^b \,, \qquad
  \widetilde{J}_2^{(0)} = \bar c\,\Gamma_2 h_v^b \,.
\end{equation}
A major complication arises from the fact that matrix elements of the
higher-dimension operators $\widetilde{J}_i^{(k)}$ will, in general,
scale like $m_c^k$. This compensates the prefactor $1/m_b^k$.
Consequently, these operators cannot be neglected even at leading
order in the heavy quark expansion. One would thus have to deal with
an infinite number of operators in order to keep track of the full
dependence on the heavy quark masses. Ignoring this difficulty for
the moment, one may use (\ref{interm}) to scale the currents from
$\mu=m_b$ down to $\mu=m_c$, where the matching is done onto the
final effective theory with two heavy quarks. In this step, the
operator basis collapses considerably. When terms of order
$\Lambda_{\rm QCD}/m_Q$ (here I use $m_Q$ generically for $m_c$ or
$m_b$) are neglected on the level of matrix elements, only the three
operators $J_j(\mu)$ in (\ref{CiJi}) remain. Each of the operators of
the intermediate theory has an expansion in terms of these operators,
with coefficients $E_{ij}$:
\begin{equation}
  \widetilde{J}_i^{(k)}(m_c,m_c) \cong \sum_{j=1}^3 m_c^k\,\Big\{
   E_{ij}^{(k)}(m_c,w,\mu)\,J_j(\mu) + O(1/m_c) \Big\} \,.
\end{equation}
Combining this with (\ref{interm}), one obtains
\begin{equation}\label{CjJj}
  J^\mu(m_b,m_c) \cong \sum_{j=1}^3 C_j(m_b,m_c,w,\mu)\,J_j(\mu)
  + O(1/m_Q) \,,
\end{equation}
with evolution coefficients\cite{Neu6}
\begin{equation}\label{CDEful}
  C_j(m_b,m_c,w,\mu) = \sum_{k=0}^\infty \Big({m_c\over m_b}\Big)^k
   \sum_i D_i^{(k)}(m_b,m_c)\,E_{ij}^{(k)}(m_c,w,\mu) \,.
\end{equation}
In this expression, the matching scale $m$ of the first approach does
not appear. The coefficients depend either on $\alpha_s(m_b)$ or
$\alpha_s(m_c)$, i.e.\ the scaling in the intermediate region
$m_b>\mu>m_c$ is properly taken into account. To achieve this,
however, it would be necessary to consider an infinite number of
operators in the intermediate effective theory. This is, of course,
not manageable. It is important to realize, however, that the
short-distance coefficients in (\ref{CiJi}) and (\ref{CjJj}) must
agree, and that this equality must hold order by order in an
expansion in the mass ratio $m_c/m_b$. Using this fact, one can
combine the two approaches into a consistent next-to-leading order
calculation\cite{Neu6}. For details of the very tedious calculation,
the reader is referred to the
literature\cite{Falk,Neu6}$^-$\cite{Kili}. Below, I will only give
numerical results.

The short-distance coefficients can be written in the factorized form
\begin{equation}\label{CiKfac}
   C_i^{(5)}(m_b,m_c,w,\mu) = \widehat{C}_i^{(5)}(m_b,m_c,w)\,
   K_{\rm hh}(w,\mu) \,.
\end{equation}
I denote the coefficients for the vector current by $\widehat{C}_i$,
and those for the axial vector current by $\widehat{C}_i^5$. The
$\mu$-dependent function $K_{\rm hh}(w,\mu)$ is universal and
normalized to unity at zero recoil: $K_{\rm hh}(1,\mu)=1$. The
renormalization-group invariant coefficients $\widehat{C}_i^{(5)}$
contain all dependence on the heavy quark masses. In
Table~\ref{tab:3.1}, I give the numerical values of these
coefficients, which are obtained using $m_b=4.80$ GeV and $m_c=1.45$
GeV for the heavy quark masses, as well as $\Lambda_{\overline{\rm
MS}}=0.25$ GeV (for $n_f=4$) in the two-loop expression for
$\alpha_s(\mu)$. The corresponding coupling constants are
$\alpha_s(m_b)\simeq 0.20$ and $\alpha_s(m_c)\simeq 0.32$.

\begin{table}[htb]
\centerline{\parbox{11cm}{\caption{\label{tab:3.1}
Short-distance coefficients for $b\to c$ transitions.}}}
\vspace{0.5cm}
\centerline{\begin{tabular}{|c|ccc|ccc|}
\hline
\phantom{ \Big[ } $w$ \phantom{ \Big[ }
& $\widehat{C}_1$ & $\widehat{C}_2$ & $\widehat{C}_3$ &
$\widehat{C}_1^5$ & $\widehat{C}_2^5$ & $\widehat{C}_3^5$ \\
\hline
1.0 & 1.136 & $-0.085$ & $-0.021$ & 0.985 & $-0.122$ & 0.042 \\
1.1 & 1.107 & $-0.080$ & $-0.021$ & 0.965 & $-0.115$ & 0.040 \\
1.2 & 1.081 & $-0.077$ & $-0.020$ & 0.946 & $-0.109$ & 0.038 \\
1.3 & 1.056 & $-0.073$ & $-0.019$ & 0.927 & $-0.103$ & 0.036 \\
1.4 & 1.033 & $-0.070$ & $-0.018$ & 0.910 & $-0.098$ & 0.035 \\
1.5 & 1.011 & $-0.067$ & $-0.018$ & 0.894 & $-0.094$ & 0.033 \\
1.6 & 0.991 & $-0.064$ & $-0.017$ & 0.878 & $-0.089$ & 0.032 \\
1.7 & 0.972 & $-0.062$ & $-0.017$ & 0.864 & $-0.086$ & 0.031 \\
1.8 & 0.953 & $-0.059$ & $-0.016$ & 0.850 & $-0.082$ & 0.030 \\
\hline
\end{tabular}}
\vspace{0.5cm}
\end{table}

Of particular interest for the model-independent determination of
$|\,V_{cb}|$ to be discussed below is the value of the short-distance
coefficient $\widehat C_1^5(w)$ at zero recoil. One finds\cite{Neu6}
\begin{eqnarray}\label{etaa}
   \eta_A \equiv \widehat C_1^5(1) = x^{6/25}\,\bigg\{ 1
   &+& 1.561\,{\alpha_s(m_c)-\alpha_s(m_b)\over\pi}
    - {\alpha_s(m_c)\over\pi}\,\bigg( {8\over 3}
    + {2 z^2\over 1-z}\,\ln z \bigg) \nonumber\\
   &+& z\,\bigg( {25\over 54} - {14\over 27}\,x^{-9/25}
    + {1\over 18}\,x^{-12/25} + {8\over 25}\,\ln x \bigg)
    \bigg\} \,,
\end{eqnarray}
where $x=\alpha_s(m_c)/\alpha_s(m_b)$, and $z=m_c/m_b$. For
$\Lambda_{\overline{\rm MS}}=0.25\pm 0.05$ GeV and $z=0.30\pm 0.05$,
this yields $\eta_A=0.986\pm 0.006$.

Let me stop, at this point, the discussion of short-distance effects.
There have been important calculations that I have no time to mention
here, such as the renormalization of the higher-dimension operators
in the effective Lagrangian\cite{EiH1,FGL,LuMa}, and the
short-distance expansion of currents to next-to-leading order
in\cite{mepower} $1/m_Q$. A discussion of these calculations can be
found in my review article\cite{review}.

\subsection{Covariant Representation of States}

The purpose of the OPE discussed in the previous sections was to
disentangle the short-distance physics related to length scales set
by the Compton wavelengths of the heavy quarks from confinement
effects relevant at large distances. This procedure makes explicit
the $m_Q$ dependence of any Green's function of the full theory,
which contains one or more heavy quark fields. Hadronic matrix
elements of heavy quark currents have a $1/m_Q$ expansion as shown in
(\ref{OPEex}). The HQET matrix elements on the right-hand side of
this equation contain the long-distance physics associated with the
interactions of the cloud of light degrees of freedom among
themselves and with the background colour-field provided by the heavy
quarks. These hadronic quantities depend in a most complicated way on
the ``brown muck'' quantum numbers of the external states, the
quantum numbers of the current, and on the heavy quark velocities.
They are related to matrix elements of effective current operators in
HQET. Recall that these matrix elements are independent of the heavy
quark masses, if the states of the effective theory are taken to be
the eigenstates of the leading-order Lagrangian ${\cal L}_\infty$ in
(\ref{Leff}). Matrix elements evaluated using these states have a
well-defined behaviour under spin-flavour symmetry transformations.
When combined with the requirements of Lorentz covariance,
restrictive constraints on their structure can be derived. I will now
introduce a very elegant formalism due to Bjorken\cite{Bjor} and Falk
et al.\cite{Falk,AdamF}, which allows one to derive the general form
of such matrix elements in a straightforward manner. The clue is to
work with a covariant tensor representation of states with definite
transformation properties under the Lorentz group and the heavy quark
spin-flavour symmetry.

The eigenstates of HQET can be thought of as the ``would-be hadrons''
built from an infinitely heavy quark dressed with light quarks,
antiquarks and gluons. In such a state, both the heavy quark and the
cloud of light degrees of freedom have well-defined transformation
properties under the Lorentz group. The heavy quark can be
represented by a spinor $u_h(v,s)$ satisfying $\rlap/v\,u_h(v,s) =
u_h(v,s)$, where $v$ is the velocity of the hadron. Because of heavy
quark symmetry, the wave function of the state (when properly
normalized) is independent of the flavour and spin of the heavy
quark, and the states can be characterized by the quantum numbers of
the ``brown muck''. In particular, for each configuration of light
degrees of freedom with total angular momentum $j\ge 0$ and parity
$P$, there is a degenerate doublet of states with spin-parity $J^P =
(j\pm\frac{1}{2})^P$. Following Falk\cite{AdamF}, I discuss the cases
of integral and half integral $j$ separately. Hadronic states with
integral $j$ have odd fermion number and correspond to baryons;
states with half-integral $j$ have even fermion number and correspond
to mesons.

First consider the heavy baryons. In this case, the ``brown muck'' is
an object with spin-parity $j^P$ that can be represented by a totally
symmetric, traceless tensor $A^{\mu_1\ldots\mu_j}$ subject to the
transversality condition $v_\mu\,A^{\mu_1\ldots\mu_j}=0$. States are
said to have ``natural'' parity if $P=(-1)^j$, and ``unnatural''
parity otherwise. The composite heavy baryon can be represented by
the tensor wave function
\begin{equation}
   \psi^{\mu_1\ldots\mu_j} = u_h\,A^{\mu_1\ldots\mu_j} \,.
\end{equation}
Under a connected Lorentz transformation $\Lambda$, this object
transforms as a spinor-tensor field
\begin{equation}
   \psi^{\mu_1\ldots\mu_j} \to
   \Lambda_{\nu_1}^{\mu_1}\ldots\Lambda_{\nu_j}^{\mu_j}\,
   D(\Lambda)\,\psi^{\nu_1\ldots\nu_j} \,,
\end{equation}
where $D(\Lambda) = \exp(-\frac{i}{4}\omega_{\mu\nu}\sigma^{\mu\nu})$
is the usual spinor representation of $\Lambda$. A heavy quark spin
rotation $\widetilde {\Lambda}$, on the other hand, acts only on
$u_h$; hence
\begin{equation}
   \psi^{\mu_1\ldots\mu_j} \to
   D(\widetilde{\Lambda})\,\psi^{\mu_1\ldots\mu_j} \,.
\end{equation}
Here $\widetilde {\Lambda}$ is restricted to spatial rotations (in
the rest frame). The infinitesimal form of $D(\widetilde{\Lambda})$
was considered in (\ref{SU2tr}). The simplest but important case
$j^P=0^+$ corresponds to the ground-state $\Lambda_Q$-baryon with
total spin-parity $J^P = \frac{1}{2}^+$. It can be represented by a
spinor $u_\Lambda$. Since the light degrees of freedom are in a
configuration of total spin zero, the spin of the baryon is carried
by the heavy quark, and the spinor $u_\Lambda$ coincides with the
heavy quark spinor. Hence
\begin{equation}\label{uLambda}
   \psi_\Lambda = u_\Lambda(v,s) = u_h(v,s) \,.
\end{equation}
For $j\ge 0$, the object $\psi^{\mu_1\ldots\mu_j}$ does not transform
irreducibly under the Lorentz group, but is a linear combination of
two components with total spin $j\pm\frac{1}{2}$. These correspond to
a degenerate doublet of physical states, which only differ in the
orientation of the heavy quark spin relative to the angular momentum
of the light degrees of freedom. The nonrelativistic quark model
suggests that one should identify the states with $j^P=1^+$ with the
$\Sigma_Q$ ($J^P=\frac{1}{2}^+$) and $\Sigma_Q^*$ ($J^P=
\frac{3}{2}^+$) baryons. In the quark model, these states contain a
heavy quark and a light vector diquark with no orbital angular
momentum. Unlike the $\Lambda_Q$-baryons, they have unnatural parity.
This implies that decays between $\Lambda_Q$ and $\Sigma_Q^{(*)}$
must be described by parity-odd form factors\cite{Pol1,MRR1}.

Next consider the heavy mesons. I shall only discuss the case
$j^P=\frac{1}{2}^-$ in detail. Since quarks and antiquarks have
opposite intrinsic parity, the corresponding physical states with
``natural'' parity are the ground-state pseudoscalar ($J^P=0^-$) and
vector ($J^P=1^-$) mesons. As before, the heavy quark is represented
by a spinor $u_h(v,s)$ subject to the condition $\rlap/v\,u_h(v,s) =
u_h(v,s)$. The light degrees of freedom as a whole transform under
the Lorentz group as an antiquark moving at velocity $v$. They are
described by an antifermion spinor $\bar v_\ell(v,s')$ satisfying
$\bar v_\ell(v,s')\,\rlap/v = - \bar v_\ell(v,s')$. The ground-state
mesons can be represented by the composite object $\psi=u_h\,\bar
v_\ell$, which is a $4\times 4$ Dirac matrix with two spinor indices,
one for the heavy quark and one for the ``brown muck''. Under a
connected Lorentz transformation $\Lambda$, the meson wave function
$\psi$ transforms as
\begin{equation}
   \psi \to  D(\Lambda)\,\psi\,D^{-1}(\Lambda) \,,
\end{equation}
whereas under a heavy quark spin rotation $\tilde{\Lambda}$
\begin{equation}
   \psi \to D(\tilde{\Lambda})\,\psi \,.
\end{equation}
The composite $\psi$ represents a linear combination of the physical
pseudoscalar and vector meson states. It is easiest to identify these
states in the rest frame, where $u_h$ has only upper components,
whereas $\bar v_\ell$ has only lower components. The nonvanishing
components of $\psi$ are thus contained in a $2\times 2$ matrix,
which can be written as a linear combination of the identity $I$ and
the Pauli matrices $\sigma^i$. Let me choose the quantization axis in
3-direction and work with the rest-frame spinor basis
\begin{equation}
   u_h(\Uparrow) = \left(\begin{array}{c} 1\\0\\0\\0 \end{array}
    \right) \,, \quad
   u_h(\Downarrow) = \left(\begin{array}{c} 0\\1\\0\\0 \end{array}
    \right) \,, \quad
   v_\ell(\uparrow) = \left(\begin{array}{c} 0\\0\\0\\1 \end{array}
    \right) \,, \quad
   v_\ell(\downarrow) = \left(\begin{array}{c} 0\\0\\1\\0 \end{array}
    \right) \,.
\end{equation}
Then a basis of states is:
\begin{eqnarray}\label{basis}
   \big( \Uparrow\downarrow + \Downarrow\uparrow \big)
   &=& - \left( \begin{array}{cc} 0 ~&~ I \\ 0 ~&~ 0 \end{array}
    \right) \,, \nonumber\\
   \big( \Uparrow\downarrow - \Downarrow\uparrow \big)
   &=& - \left( \begin{array}{cc} 0 ~&~ \sigma^3 \\ 0 ~&~ 0
    \end{array} \right) \,, \nonumber\\
   \sqrt{2}\,\big( \Uparrow\uparrow \big) &=& - {1\over\sqrt{2}}
    \left( \begin{array}{cc} 0 & \sigma^1\!+\!i\sigma ^2 \\
    0 & 0 \end{array} \right) \,, \nonumber\\
   \sqrt{2}\,\big( \Downarrow\downarrow \big) &=& - {1\over\sqrt{2}}
    \left( \begin{array}{cc} 0 & \sigma^1\!-\!i\sigma^2 \\
    0 & 0 \end{array} \right) \,.
\end{eqnarray}
Let me furthermore define two transverse polarization vectors
$\epsilon_\pm$ and a longitudinal polarization vector $\epsilon_3$ by
\begin{equation}
   \epsilon_\pm^\mu = {1\over\sqrt{2}}\,\big( 0,1,\pm i,0 \big)
   \,,\qquad \epsilon_3^\mu = (0,0,0,1) \,.
\end{equation}
It is then obvious to identify the pseudoscalar ($P$) and vector
($V$) meson states as\cite{Falk,Bjor,AdamF}
\begin{eqnarray}\label{PVrep1}
   P(\vec{v}=0) &=& - {1 + \gamma^0\over 2}\,
    \gamma_5 \,, \nonumber\\
   V(\vec{v}=0,\epsilon) &=& \phantom{ - }
    {1+\gamma^0\over 2}\,\rlap/\epsilon \,.
\end{eqnarray}
The second state in (\ref{basis}) has longitudinal polarization,
whereas the last two states have transverse polarization.

To get familiar with this representation, consider the action of the
spin operator $\vec{\bf\Sigma}$ on $\psi$. A matrix representation of
the components ${\bf\Sigma}^i$ in the rest frame is $\Sigma^i =
\frac{1}{2}\gamma_5\gamma^0\gamma^i$, and the action of the operator
${\bf\Sigma}^i$ on the meson wave function is ${\bf\Sigma}^i\,\psi =
[\Sigma^i,\psi]$. Using this, one finds
\begin{eqnarray}
    \vec{\bf\Sigma}^2 P &=& {\bf\Sigma}^3 P = 0 \,, \nonumber\\
    \vec{\bf\Sigma}^2\,V(\epsilon) &=& 2\,V(\epsilon) \,,
     \nonumber\\
    {\bf\Sigma}^3\,V(\epsilon_\pm) &=& \pm V(\epsilon_\pm) \,,
     \nonumber\\
    {\bf\Sigma}^3\,V(\epsilon_3) &=& 0 \,,
\end{eqnarray}
which shows that $P$ has total spin zero, and $V$ has total spin one.
Next consider the action of the heavy quark spin operator $\vec{\bf
S}$. It has the same matrix representation as $\vec{\bf\Sigma}$, but
only acts on the heavy quark spinor in $\psi$: ${\bf S}^i\,\psi =
S^i\,\psi$, with $S^i=\Sigma^i$. It follows that
\begin{eqnarray}
   {\bf S}^3 P &=& {1\over 2}\,V(\epsilon_3) \,, \nonumber\\
   {\bf S}^3\,V(\epsilon_3) &=& {1\over 2}\,P \,, \nonumber\\
   {\bf S}^3\,V(\epsilon_\pm) &=& \pm {1\over 2}\,
    V(\epsilon_\pm) \,,
\end{eqnarray}
in accordance with the spin assignments for the heavy quark in
(\ref{basis}).

In a general frame, the tensor wave functions in (\ref{PVrep1}) can
be readily generalized in a Lorentz-covariant way by replacing
$\gamma^0$ with $\rlap/v$. The covariant representation of states can
be used to determine in a very efficient way the structure of
hadronic matrix elements in the effective theory. The goal is to find
a minimal form factor decomposition consistent with Lorentz
covariance, parity invariance of the strong interactions, and heavy
quark symmetry. The flavour symmetry is manifest when one uses
mass-independent wave functions. The correct transformation
properties under the spin symmetry are guaranteed when one collects a
spin-doublet of states into a single object. In case of the
ground-state pseudoscalar and vector mesons, for instance, one
introduces a covariant tensor wave function ${\cal M}(v)$ that
represents both $P(v)$ and $V(v,\epsilon)$ by
\begin{equation}\label{PVrep}
   {\cal M}(v) = {1+\rlap/v\over 2}\,
   \cases{ -\gamma_5 \,; &pseudoscalar meson, \cr
           \rlap/\epsilon \,; &vector meson. \cr}
\end{equation}
It has the important property ${\cal M}(v) = P_+\,{\cal M}(v)\,P_-$,
where $P_\pm=\frac{1}{2} (1\pm\rlap/v)$. This is often used to
simplify expressions.

Consider now a transition between two heavy mesons, $M(v)\to M'(v')$,
mediated by a renormalized effective current operator $\bar h'_{v'}\,
\Gamma\,h_v$, which changes a heavy quark $Q$ into another heavy
quark $Q'$. According to the Feynman rules of HQET, the ``heavy quark
part'' of the decay amplitude is simply proportional to $\bar
u'_h\,\Gamma\,u_h$; interactions of the heavy quarks with gluons do
not modify the Dirac structure of $\Gamma$. Since the heavy quark
spinors are part of the tensor wave functions associated with the
hadron states, it follows that the amplitude must be proportional to
$\overline{\cal M}'(v')\,\Gamma\,{\cal M}(v)$. This is a Dirac matrix
with two indices representing the light degrees of freedom. Since the
total matrix element is a scalar, these indices must be contracted
with those of a matrix $\Xi$. Hence one may write
\begin{equation}\label{Ximatr}
   \langle M'(v')|\,\bar h'_{v'}\,\Gamma\,h_v\,|M(v)\rangle
   = {\rm Tr}\Big\{\, \Xi(v,v',\mu)\,\overline{\cal M}'(v')\,
   \Gamma\,{\cal M}(v) \Big\} \,.
\end{equation}
The matrix $\Xi$ contains all long-distance dynamics. It is a most
complicated object, only constrained by the symmetries of the
effective theory. Heavy quark symmetry requires that it be
independent of the spins and masses of the heavy quarks, as well as
of the Dirac structure of the current. Hence, $\Xi$ can only be a
function of the meson velocities and of the renormalization scale
$\mu$. Lorentz covariance and parity invariance imply that $\Xi$ must
transform as a scalar with even parity. This allows the decomposition
\begin{equation}
   \Xi(v,v',\mu) = \Xi_1 + \Xi_2\,\rlap/v + \Xi_3\,\rlap/v'
   + \Xi_4\,\rlap/v\,\rlap/v' \,,
\end{equation}
with coefficients $\Xi_i=\Xi_i(v\cdot v',\mu)$. But using the
projection properties of the tensor wave functions, one finds that
under the trace
\begin{equation}
   \Xi(v,v',\mu) \to \Xi_1 - \Xi_2 - \Xi_3 + \Xi_4 \equiv
   - \xi(v\cdot v',\mu) \,.
\end{equation}
Therefore\cite{Falk},
\begin{equation}\label{IWmaster}
   \langle M'(v')|\,\bar h'_{v'}\,\Gamma\,h_v\,|M(v)\rangle =
   - \xi(v\cdot v',\mu)\,{\rm Tr}\Big\{\,
   \overline{\cal M}'(v')\,\Gamma\,{\cal M}(v) \Big\} \,.
\end{equation}
The sign is chosen such that the universal form factor $\xi(v\cdot
v',\mu)$ coincides with the Isgur--Wise function, which is the single
form factor that describes semileptonic weak decay processes of heavy
mesons in the infinite quark mass limit. Equation (\ref{IWmaster})
summarizes in a compact way the results derived in Section 3.4. Using
the explicit form of the meson wave functions, one can readily
recover the relations (\ref{elast}), (\ref{inelast}), and
(\ref{PVff}). The only new feature is that the Isgur--Wise function
depends on the renormalization scale $\mu$. This is necessary to
compensate the scale dependence of the Wilson coefficients, which
multiply the renormalized current operators in the short-distance
expansion.

\newpage
\subsection{Meson Decay Form Factors}

One of the most important applications of heavy quark symmetry is to
derive relations between the form factors parametrizing the exclusive
weak decays $B\to D\,\ell\,\bar\nu$ and $B\to D^*\ell\, \bar\nu$. A
detailed theoretical understanding of these processes is necessary
for a reliable determination of the element $|\,V_{cb}|$ of the
Kobayashi--Maskawa matrix. Let me start by introducing a set of six
hadronic form factors $h_i(w)$, which parametrize the relevant meson
matrix elements of the flavour-changing vector and axial vector
currents $V^\mu=\bar c\, \gamma^\mu b$ and $A^\mu=\bar
c\,\gamma^\mu\gamma_5\,b$:
\begin{eqnarray}\label{hiwdef}
   \phantom{ \Big[ }
   \langle D(v') |\,V^\mu\,| B(v) \rangle &=& h_+(w)\,(v+v')^\mu
    + h_-(w)\,(v-v')^\mu \,, \nonumber\\
   \phantom{ \bigg[ }
   \langle D^*(v',\epsilon) |\,V^\mu\,| B(v) \rangle &=&
    i\,h_V(w)\,\epsilon^{\mu\nu\alpha\beta}\,\epsilon_\nu^*\,
    v'_\alpha\,v_\beta \,, \\
   \langle D^*(v',\epsilon) |\,A^\mu\,| B(v) \rangle &=&
    h_{A_1}(w)\,(w+1)\,\epsilon^{\ast\mu} \!-\! \Big[
    h_{A_2}(w)\,v^\mu + h_{A_3}(w)\,v'^\mu \Big]
    \epsilon^*\!\cdot\!v \,. \nonumber
\end{eqnarray}
Here $w=v\cdot v'$ is the velocity transfer of the mesons, and
$\epsilon$ denotes the polarization vector of the $D^*$-meson. At
leading order in the $1/m_Q$ expansion, one can derive expressions
for these form factors by using the short-distance expansion
(\ref{CiJi}) of the flavour-changing currents, as well as the tensor
formalism outlined above. According to (\ref{CiKfac}), the
$\mu$ dependence of the Wilson coefficients of any bilinear heavy
quark current can be factorized into a universal function $K_{\rm
hh}(w,\mu)$, which is normalized at zero recoil. The $\mu$ dependence
of this function has to cancel against that of the Isgur--Wise
function. One can use this fact to define a renormalization-group
invariant Isgur--Wise form factor as
\begin{equation}\label{IWFren}
   \xi_{\rm ren}(w) \equiv \xi(w,\mu)\,K_{\rm hh}(w,\mu) \,,
   \qquad \xi_{\rm ren}(1) = 1 \,.
\end{equation}
Neglecting terms of order $1/m_Q$, one then obtains\cite{Neu6}
\begin{equation}\label{Neurel}
   \langle M'(v')|\,J^\mu\,|M(v)\rangle = - \xi_{\rm ren}(w)
    \sum_{i=1}^3 \widehat{C}_i^{(5)}(w)\,{\rm Tr}\Big\{\,
    \overline{\cal M}'(v')\,\Gamma_i\,{\cal M}(v) \Big\} \,,
\end{equation}
where the Dirac matrices $\Gamma_i$ have been defined in
(\ref{vecmatr}) and (\ref{aximatr}). It is now straightforward to
evaluate the traces to find
\begin{eqnarray}\label{hilead}
   h_+(w) &=& \bigg[ \widehat{C}_1(w) + {w+1\over 2}\,
    \Big( \widehat{C}_2(w) + \widehat{C}_3(w) \Big) \bigg]\,
    \xi_{\rm ren}(w) \,, \nonumber\\
   h_-(w) &=& {w+1\over 2}\,\Big[ \widehat{C}_2(w)
    - \widehat{C}_3(w) \Big]\,\xi_{\rm ren}(w) \,, \nonumber\\
   h_V(w) &=& \phantom{ \bigg[ }
    \widehat{C}_1(w)\,\xi_{\rm ren}(w) \,, \nonumber\\
   \phantom{ \bigg[ } h_{A_1}(w) &=&
    \widehat{C}_1^5(w)\,\xi_{\rm ren}(w) \,, \nonumber\\
   \phantom{ \bigg[ } h_{A_2}(w) &=&
    \widehat{C}_2^5(w)\,\xi_{\rm ren}(w) \,, \nonumber\\
   \phantom{ \bigg[ } h_{A_3}(w) &=&
    \Big[ \widehat{C}_1^5(w) + \widehat{C}_3^5(w) \Big]\,
    \xi_{\rm ren}(w) \,.
\end{eqnarray}
This is the correct generalization of (\ref{inelast}) and
(\ref{PVff}) in the presence of short-distance corrections. The fact
that, to leading order in $1/m_Q$, the meson form
factors are given in terms of a single universal function $\xi_{\rm
ren}(w)$ was the discovery of Isgur and Wise\cite{Isgu}.
Short-distance QCD corrections affect the form factors in a
calculable way. Their effects are contained in the various
combinations of short-distance coefficients, which can be evaluated
using the numerical results given in Table~\ref{tab:3.1}.

\subsection{Power Corrections and Luke's Theorem}

Using the covariant tensor formalism and the short-distance
expansions of the effective Lagrangian and currents beyond the
leading order in $1/m_Q$, one can investigate in a systematic way the
structure of power corrections to the relations derived in the
previous section. I have given a simple (tree-level) example of the
structure of power corrections in (\ref{toypower}). For the more
complicated case of meson weak decay form factors, the analysis at
order $1/m_Q$ was performed by Luke\cite{Luke}. Later, short-distance
corrections were included to all orders in perturbation
theory\cite{mepower,Neu7}. Falk and myself have also analysed the
structure of $1/m_Q^2$ corrections for both meson and baryon weak
decay form factors\cite{FaNe}. I shall not discuss these rather
technical issues in detail, but nevertheless give you the main
results.

Luke has shown that, for transitions between two heavy ground-state
(pseudoscalar or vector) mesons, the $1/m_Q$ corrections can be
parametrized by a set of four additional universal functions of the
velocity transfer $w$. The most important outcome of his analysis
concerns the zero recoil limit, where an analogue of the
Ademollo--Gatto theorem\cite{AGTh} can be proved. This is Luke's
theorem\cite{Luke}, which states that the matrix elements describing
the leading $1/m_Q$ corrections to meson decay amplitudes vanish at
zero recoil. As a consequence, in the limit $v=v'$ there are no terms
of order $1/m_Q$ in the hadronic matrix elements in (\ref{hiwdef}).
This theorem is independent of the structure of the Wilson
coefficients and thus valid to all orders in perturbation
theory\cite{FaNe,Neu7,ChGr}.

There is considerable confusion in the literature about the
implications of this result. It is often claimed that the theorem
would protect any meson decay rate, or even all form factors that are
normalized in the spin-flavour symmetry limit, from first-order power
corrections at zero recoil. However, these claims are erroneous. The
reason is simple but somewhat subtle. Luke's theorem only protects
the form factors $h_+$ and $h_{A_1}$ in (\ref{hiwdef}), since all the
others are multiplied by kinematic factors which vanish for $v=v'$.
In fact, an explicit calculation shows that the $1/m_Q$ corrections
to $h_-$, $h_V$, $h_{A_2}$, and $h_{A_3}$ do not vanish at zero
recoil. The fact that these functions are kinematically suppressed
does not imply that they could not contribute to physical decay
rates. This is often overlooked. Consider, as an example, the process
$B\to D\,\ell\,\bar\nu$ in the limit of vanishing lepton mass. By
angular momentum conservation, the two pseudoscalar mesons must be in
a relative $p$-wave in order to match the helicities of the lepton
pair. The amplitude is proportional to the velocity $|\vec v_D|$ of
the $D$-meson in the $B$-meson rest frame, which leads to a factor
$(w^2-1)$ in the decay rate. In such a situation, form factors that
are kinematically suppressed can contribute\cite{Neu1}. Indeed, the
$B\to D\,\ell\,\bar\nu$ decay rate is proportional to
\begin{equation}
   (w^2-1)^{3/2}\,\bigg| h_+(w) - {m_B-m_D\over m_B+m_D}\,h_-(w)
   \bigg|^2 \,.
\end{equation}
Both form factors, $h_+$ and $h_-$, contribute with similar strength
to the rate near $w=1$, although $h_-$ is kinematically suppressed in
(\ref{hiwdef}). Consequently, the decay rate at zero recoil does
receive corrections of order $1/m_Q$. The situation is different for
$B\to D^*\ell\,\bar\nu$ transitions\cite{Neu9}. Because the vector
meson has spin one, the decay can proceed in an $s$-wave, and there
is no helicity suppression near zero recoil. One finds that close to
$w=1$ the decay rate is proportional to $(w^2-1)^{1/2}\,
|h_{A_1}(w)|^2$. Since the form factor $h_{A_1}$ is protected by
Luke's theorem, these transitions are ideally suited for a precision
measurement of $|\,V_{cb}|$ from an extrapolation of the momentum
spectrum of the $D^*$-meson to zero recoil\cite{Vol2,Neu9}. There,
the normalization of the decay rate is known in a model-independent
way up to corrections of order $1/m_Q^2$:
\begin{equation}\label{Lukeha1}
   h_{A_1}(1) = \eta_A + \delta_{1/m^2} + \ldots \,,
\end{equation}
where $\eta_A\simeq 0.99$ contains the short-distance corrections
[cf.~(\ref{etaa})].

One expects higher-order power corrections to be of order
$\delta_{1/m^2}\sim (\Lambda_{\rm QCD}/m_c)^2\sim 3\%$, but of course
such a na\"\i ve estimate could be too optimistic. For a precision
measurement of $|\,V_{cb}|$, it is important to know the structure of
such corrections in more detail. Although in principle
straightforward, the analysis of $1/m_Q^2$ corrections in HQET is a
tedious enterprise\cite{FaNe}. Three classes of corrections have to
be distinguished: matrix elements of local dimension-five current
operators, ``mixed'' corrections resulting from the combination of
corrections to the current and to the Lagrangian, and corrections
from one or two insertions of operators from the effective Lagrangian
into matrix elements of the leading-order currents. Within these
classes, one can distinguish corrections proportional to $1/m_b^2$,
$1/m_c^2$, or $1/m_b m_c$. More than thirty new universal functions
are necessary to parametrize the second-order power corrections to
meson form factors. When radiative corrections are neglected, eleven
combinations of these functions contribute to the hadronic form
factors $h_i(w)$. The general results greatly simplify at zero
recoil, however. There the equation of motion and the Ward identities
of the effective theory can be used to prove that matrix elements of
local dimension-five current operators, as well as matrix elements of
time-ordered products containing a dimension-four current and an
insertion from the effective Lagrangian, can be expressed in terms of
only two parameters $\lambda_1$ and $\lambda_2$, which are related to
the $1/m_Q$ corrections to the physical meson masses. Moreover, the
conservation of the flavour-diagonal vector current in the full
theory forces certain combinations of the universal functions to
vanish at zero recoil. The consequence is that whenever a form factor
is protected by Luke's theorem, the structure of second-order power
corrections at zero recoil becomes rather simple. The quantity
$\delta_{1/m^2}$ in (\ref{Lukeha1}) is of this type. A careful
estimate based on some mild model assumptions gives\cite{FaNe,ThMa}
$-3\% < \delta_{1/m^2} < -1\%$. One can combine this result with
(\ref{Lukeha1}) to obtain one of the most important, and certainly
most precise predictions of HQET:
\begin{equation}\label{glory}
   h_{A_1}(1) = 0.97 \pm 0.04 \,.
\end{equation}

\subsection{Properties of the Isgur--Wise Function}

The establishment of heavy quark symmetry as an exact limit of the
strong interactions enables one to derive approximate relations
between decay amplitudes, and normalization conditions for certain
form factors, which are similar to the relations and normalization
conditions that can be derived for Goldstone-boson scattering
amplitudes from the low-energy theorems of current algebra. HQET
provides the framework for a systematic investigation of the
corrections to the limit of an exact spin-flavour symmetry. The
output of such a model-independent analysis is a short-distance
expansion of decay amplitudes, in which the dependence on the heavy
quark masses is explicit. At each order in the $1/m_Q$ expansion, the
long-distance physics is parametrized by a minimal set of universal
form factors, which are independent of the heavy quark masses. As
presented so far, the analysis is completely model-independent. Since
hadronic decay processes are of a genuine nonperturbative nature,
however, it is clear that predictions that can be made based on
symmetries only are limited. In particular, very little can be said
on general grounds about the properties of the form factors of the
effective theory. But there is a lot of information contained in
these functions. Much like the hadron structure functions probed in
deep-inelastic scattering, they are fundamental nonperturbative
quantities in QCD, which describe the properties of the light degrees
of freedom in the background of the colour field provided by the
heavy quarks. Since a static colour source is the most direct way to
probe the strong interactions of quarks and gluons at large
distances, a theoretical understanding of the universal form factors
would not only enlarge the predictive power of HQET, but would also
teach us in a very direct way about the nonperturbative nature of the
strong interactions.

The leading-order Isgur--Wise function $\xi(w)$ plays a central role
in the description of the weak decays of heavy mesons. It contains
the long-distance physics associated with the strong interactions of
the light degrees of freedom and cannot be calculated from first
principles. Nevertheless, some important properties of this function
can be derived on general grounds, such as its normalization at zero
recoil, which is a consequence of current conservation. According to
(\ref{elast}), the Isgur--Wise function is the elastic form factor of
a ground-state heavy meson in the limit where power corrections are
negligible. As such, $\xi(w)$ must be a monotonically decreasing
function of the velocity transfer $w=v\cdot v'$, which is analytic in
the cut $w$-plane with a branch point at $w=-1$, corresponding to the
threshold $q^2=4 m_Q^2$ for heavy quark pair production. However,
being obtained from a limiting procedure, the Isgur--Wise function
can have stronger singularities than the physical elastic form
factor. In fact, the short-distance corrections contained in the
function $K_{\rm hh}(w,\mu)$ lead to an essential singularity at
$w=-1$ in the renormalized Isgur--Wise function defined in
(\ref{IWFren}).

When using a phenomenological parametrization of the universal form
factor, one should incorporate the above properties. Some legitimate
forms suggested in the literature are:
\begin{eqnarray}\label{IWFtoy}
   \xi_{\rm BSW}(w) &=& {2\over w+1}\,\exp\bigg\{ - (2\varrho^2-1)\,
    {w-1\over w+1} \bigg\} \,, \nonumber\\
   \xi_{\rm ISGW}(w) &=& \exp\Big\{ -\varrho^2\,(w-1) \Big\} \,,
    \phantom{ \bigg[ } \nonumber\\
   \xi_{\rm pole}(w) &=& \bigg({2\over w+1}\bigg)^{2\varrho^2} .
\end{eqnarray}
The first function is the form factor derived by Rieckert and
myself\cite{Neu1} from an analysis of the BSW model\cite{Wirb}, the
second one corresponds to the ISGW model\cite{ISGW}, and the third
one is a pole-type ansatz. Of particular interest is the behaviour of
the Isgur--Wise function close to zero recoil, which is determined by
the slope parameter $\varrho^2>0$ defined by $\xi'(1)=-\varrho^2$, so
that
\begin{equation}\label{rhodef}
   \xi(w) = 1 - \varrho^2\,(w-1) + O\big[(w-1)^2\big] \,.
\end{equation}
It is important to realize that the kinematic region accessible in
semileptonic decays is small ($1<w<1.6$). As long as $\varrho^2$ is
the same, different functional forms of $\xi(w)$ will give similar
results. A precise knowledge of the slope parameter would thus
basically determine the Isgur--Wise function in the physical region.

Bjorken has shown that $\varrho^2$ is related to the form factors of
transitions of a ground-state heavy meson into excited states, in
which the light degrees of freedom carry quantum numbers
$j^P=\frac{1}{2}^+$ or $\frac{3}{2}^+$, by a sum rule which is an
expression of quark--hadron duality: In the infinite mass limit, the
inclusive sum of the probabilities for decays into hadronic states is
equal to the probability for the free quark transition. If one
normalizes the latter probability to unity, the sum rule has the
form\cite{Bjor,IsgW,BjDT}
\begin{eqnarray}\label{inclsum}
   1 &=& {w+1\over 2}\,\bigg\{ |\xi(w)|^2 + \sum_l |\xi^{(l)}(w)|^2
    \bigg\} \nonumber\\
   &&\mbox{}+ (w-1)\,\bigg\{ 2\sum_m |\tau_{1/2}^{(m)}(w)|^2
    + (w+1)^2 \sum_n |\tau_{3/2}^{(n)}(w)|^2 \bigg\} \nonumber\\
   \phantom{ \bigg[ }
   &&\mbox{}+ O\big[(w-1)^2\big] \,,
\end{eqnarray}
where $l,m,n$ label the radial excitations of states with the same
spin-parity quantum numbers. The sums are understood in a generalized
sense as sums over discrete states and integrals over continuum
states. The terms on the right-hand side of the sum rule in the first
line correspond to transitions into states with ``brown muck''
quantum numbers $j^P=\frac{1}{2}^-$. The ground state gives a
contribution proportional to the Isgur--Wise function, and excited
states contribute proportionally to analogous functions
$\xi^{(l)}(w)$. Because at zero recoil these states must be
orthogonal to the ground state, it follows that $\xi^{(l)}(1)=0$, and
one can conclude that the corresponding contributions to
(\ref{inclsum}) are of order $(w-1)^2$. The contributions in the
second line correspond to transitions into states with
$j^P=\frac{1}{2}^+$ or $\frac{3}{2}^+$. Because of the change in
parity, these are $p$-wave transitions. The amplitudes are
proportional to the velocity $|\vec v_f|= (w^2-1)^{1/2}$ of the final
state in the rest frame of the initial state, which explains the
suppression factor $(w-1)$ in the decay probabilities. The functions
$\tau_j(w)$ are the analogues of the Isgur--Wise function for these
transitions. Transitions into excited states with quantum numbers
other than the above proceed via higher partial waves and are
suppressed by at least a factor $(w-1)^2$.

For $w=1$, eq.~(\ref{inclsum}) reduces to the normalization condition
for the Isgur--Wise function. The Bjorken sum rule is obtained by
expanding in powers of $(w-1)$ and keeping terms of first order.
Taking into account the definition of the slope parameter $\varrho^2$
in (\ref{rhodef}), one finds that\cite{Bjor,IsgW}
\begin{equation}\label{Bjsr}
   \varrho^2 = {1\over 4} + \sum_m |\tau_{1/2}^{(m)}(w)|^2
   + 2 \sum_n |\tau_{3/2}^{(n)}(w)|^2 > {1\over 4} \,.
\end{equation}
Notice that the lower bound is due to the prefactor
$\frac{1}{2}(w+1)$ of the first term in (\ref{inclsum}) and is of
purely kinematic origin. In the analogous sum rule for
$\Lambda_Q$-baryons, this factor is absent, and consequently the
slope parameter of the baryon Isgur--Wise function is only subject to
the trivial constraint\cite{Neu1,IWYo} $\varrho^2>0$.

Based on various model calculations, there is a general belief that
the contributions of excited states in the Bjorken sum rule are
sizeable, and that $\varrho^2$ is substantially larger than $1/4$.
For instance, Blok and Shifman have estimated the contributions of
the lowest-lying excited states to (\ref{Bjsr}) using QCD sum rules
and find that\cite{BlokS} $0.35<\varrho^2<1.15$. The experimental
observation that semileptonic $B$-decays into excited $D^{**}$-mesons
have a large branching ratio of about\cite{Dstar} 2.5\% gives further
support to the importance of such contributions.

Voloshin has derived another sum rule involving the form factors for
transitions into excited states, which is the analogue of the
``optical sum rule'' for the dipole scattering of light in atomic
physics. It reads\cite{Volo}
\begin{equation}\label{Volsr}
   {m_M-m_Q\over 2} = \sum_m E_{1/2}^{(m)}\,|\tau_{1/2}^{(m)}(w)|^2
   + 2 \sum_n  E_{3/2}^{(n)}\,|\tau_{3/2}^{(n)}(w)|^2 \,,
\end{equation}
where $E_j$ are the excitation energies relative to the mass $m_M$ of
the ground-state heavy meson. The important point is that one can
combine this relation with the Bjorken sum rule to obtain an upper
bound for the slope parameter $\varrho^2$:
\begin{equation}
   \varrho^2 < {1\over 4} + {m_M-m_Q\over 2 E_{\rm min}} \,,
\end{equation}
where $E_{\rm min}$ denotes the minimum excitation energy. In the
quark model, one ex\-pects\footnote{Strictly speaking, the lowest
excited ``state'' contributing to the sum rule is $D+\pi$, which has
an excitation energy spectrum with a threshold at $m_\pi$. However,
one expects that this spectrum is broad, so that this contribution
will not invalidate the upper bound for $\varrho^2$ derived here.}
that $E_{\rm min}\simeq m_M-m_Q$, and one may use this as an estimate
to obtain $\varrho^2<0.75$.

The above discussion of the sum rules ignores renormalization
effects. However, both the slope parameter $\varrho^2$ in
(\ref{Bjsr}) and the heavy quark mass $m_Q$ in (\ref{Volsr}) are
renormalization-scheme dependent quantities. Although there exist
some qualitative ideas of how to account for the $\mu$ dependence of
the various parameters\cite{IsgW,Volo}, there is currently no known
way to include renormalization effects quantitatively. One should
therefore consider the bounds on $\varrho^2$ as somewhat uncertain.
To account for this, I relax the upper bound derived from the
Voloshin sum rule and conclude that
\begin{equation}\label{rhorange}
   0.25 < \varrho^2 < 1.0 \,,
\end{equation}
where it is expected that the actual value is close to the upper
bound. Recently, de~Rafael and Taron claimed to have derived an upper
bound $\varrho^2<0.48$ from general analyticity arguments\cite{deRa}.
If true, this had severely constrained the form of the Isgur--Wise
function near zero recoil. It took quite some time to become clear
what went wrong with their derivation: The effects of resonances
below the threshold for heavy meson pair production invalidate the
argument\cite{KPDo}$^-$\cite{Carl}. It is possible to derive a new
bound, which takes into account the known properties of the
$\Upsilon$-states\cite{Tarnew}. However, it is too loose to be of any
phenomenological relevance, and one is thus left with the sum rule
result (\ref{rhorange}).

\begin{figure}[htb]
   \vspace{0.5cm}
   \epsfxsize=8cm
   \centerline{\epsffile{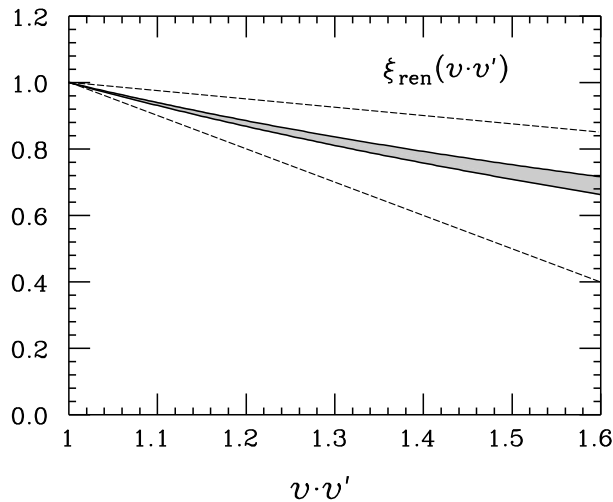}}
   \centerline{\parbox{11cm}{\caption{\label{fig:3.6}
QCD sum rule prediction for the Isgur--Wise function in the kinematic
region accessible in semileptonic decays\protect\cite{Neu10}. The
dashed lines indicate the bounds on the slope at $v\cdot v'=1$
derived by Bjorken and Voloshin [see (\protect\ref{rhorange})].
   }}}
\end{figure}

Any sensible calculation of the Isgur--Wise function has to respect
these bounds. As an example, I show in Fig.~\ref{fig:3.6} the result
of a QCD sum rule analysis of the renormalized universal form
factor\cite{Neu10}$^-$\cite{Rady}. Similar predictions have been
obtained for the functions that describe the $1/m_Q$ corrections to
the meson form factors\cite{Lige}.

\subsection{Model-Independent Determination of $|\,V_{cb}|$}

One of the most important results of HQET is the prediction
(\ref{glory}) of the normalization of the hadronic form factor
$h_{A_1}$ at zero recoil. It can be used to obtain a
model-independent measurement of the element $|\,V_{cb}|$ of the
Kobayashi--Maskawa matrix. The semileptonic decay $B\to
D^*\ell\,\bar\nu$ is ideally suited for this purpose\cite{Neu9}.
Experimentally, this is a particularly clean mode, since the
reconstruction of the $D^*$-meson mass provides a powerful rejection
against background. From the theoretical point of view, it is ideal
since the decay rate at zero recoil is protected by Luke's theorem
against first-order power corrections in $1/m_Q$. In terms of the
hadronic form factors defined in (\ref{hiwdef}), one finds
\begin{equation}
   \lim_{w\to 1} {1\over\sqrt{w^2-1}}\,
   {{\rm d}\Gamma(B\to D^*\ell\,\bar\nu)\over{\rm d}w}
   = {G_F^2\,|\,V_{cb}|^2\over 4\pi^3}\,(m_B-m_{D^*})^2\,m_{D^*}^3\,
   |h_{A_1}(1)|^2 \,.
\end{equation}
At zero recoil, the normalization of $h_{A_1}$ is known in a
model-independent way with an accuracy of 4\%. Ideally, then, one can
extract $|\,V_{cb}|$ with a theoretical uncertainty of well below
10\% from an extrapolation of the spectrum to $w=1$.

Presently, the proposal to measure $|\,V_{cb}|$ close to zero recoil
poses quite a challenge to the experimentalists. First, there is the
fact that the decay rate vanishes at zero recoil because of phase
space. Therefore the statistics gets worse as one tries to measure
close to $w=1$. However, I do not believe that this will be an
important limitation of the method. The phase-space suppression is
proportional to $\sqrt{w^2-1}$ and is in fact a rather mild one. When
going from the endpoint $w_{\rm max}\simeq 1.5$ down to $w=1.05$,
the change in the statistical error in $|\,V_{cb}|^2$ due to the
variation of the phase-space factor is not even a factor of 2. A
more serious problem is related to the fact that, for experiments
working on the $\Upsilon(4s)$ resonance, the zero-recoil limit
corresponds to a situation where both the $B$- and the $D^*$-mesons
are approximately at rest in the laboratory. Then the pion in the
subsequent decay $D^*\to D\,\pi$ is very soft and can hardly be
detected. Thus, the present experiments have to make cuts which
disfavour the zero recoil region, leading to large systematic
uncertainties for values of $w$ smaller than about 1.15. This second
problem would be absent at an asymmetric $B$-factory, where the rest
frame of the parent $B$-meson is boosted relative to the laboratory
frame.

In view of these difficulties, one presently has to rely on an
extrapolation over a wide range in $w$ to obtain a measurement of
$|\,V_{cb}|$. In general, the differential decay rate can be written
as\cite{review,Neu9}
\begin{eqnarray}\label{BDnew}
   {{\rm d}\Gamma(B\to D^*\ell\,\bar\nu)\over{\rm d}w}
   &=& {G_F^2\over 48\pi^3}\,(m_B-m_{D^*})^2\,m_{D^*}^3\,\eta_A^2\,
    \sqrt{w^2-1}\,(w+1)^2 \nonumber\\
   \phantom{ \bigg[ }
   &&\times \bigg[ 1 + {4w\over w+1}\,{1-2wr+r^2\over(1-r)^2}
    \bigg]\,|\,V_{cb}|^2\,\widehat{\xi}^{\,2}(w) \,,
\end{eqnarray}
where $\eta_A=0.99$ is the short-distance correction to the form
factor $h_{A_1}(w)$ at zero recoil, and $r=m_{D^*}/m_B$. Equation
(\ref{BDnew}) is written in such a way that the deviations from the
heavy quark symmetry limit are absorbed into the form factor
$\widehat{\xi}(w)$, which in the absence of symmetry-breaking
corrections would be the Isgur--Wise function. Since everything
except $|\,V_{cb}|$ and $\widehat{\xi}(w)$ is known, a measurement of
the differential decay rate is equivalent to a measurement of the
product $|\,V_{cb}|\,\widehat{\xi}(w)$. However, theory predicts the
normalization of $\widehat{\xi}(w)$ at zero recoil:
\begin{equation}\label{xihatnorm}
   \widehat{\xi}(1) = \eta_A^{-1}\,h_{A_1}(1) = 1 + \delta_{1/m^2}
   = 0.98\pm 0.04 \,,
\end{equation}
where the uncertainty comes from power corrections of order
$1/m_Q^2$. Using this information, $|\,V_{cb}|$ and
$\widehat{\xi}(w)$ can be obtained separately from a measurement of
the differential decay rate.

I have applied this strategy for the first time\cite{Neu9} to the
combined sample\cite{Stone} of the data on $B^0\to
D^{*+}\ell\,\bar\nu$ decays collected until 1989 by the ARGUS and
CLEO collaborations. For the extrapolation to zero recoil, I used the
parametrizations given in (\ref{IWFtoy}) for the function
$\widehat{\xi}(w)$, treating its slope at zero recoil as a free
parameter. The result obtained for $|\,V_{cb}|$ was
\begin{equation}\label{Vcbold}
   |\,V_{cb}|\,\bigg({\tau_{B^0}\over 1.5~{\rm ps}}\bigg)^{1/2}
   = 0.039\pm 0.006 \,.
\end{equation}
Since this original analysis the data have changed. In particular,
the branching ratio for $D^{*+}\to D^0\,\pi^+$ has increased
from\cite{PDG92} 55\% to\cite{Sheld} 68\%. This lowers the decay rate
for $B^0\to D^{*+}\ell\,\bar\nu$, and correspondingly decreases
$|\,V_{cb}|$ by 10\%. However, the new data recently reported by the
ARGUS\cite{Dstar} and CLEO\cite{CLEOVcb} collaborations give a larger
branching ratio than the old data, indicating that further changes in
the analysis must have taken place. It is thus not possible to simply
rescale the result for $|\,V_{cb}|$ given above.

\begin{figure}[htb]
   \vspace{0.5cm}
   \epsfxsize=10cm
   \centerline{\epsffile{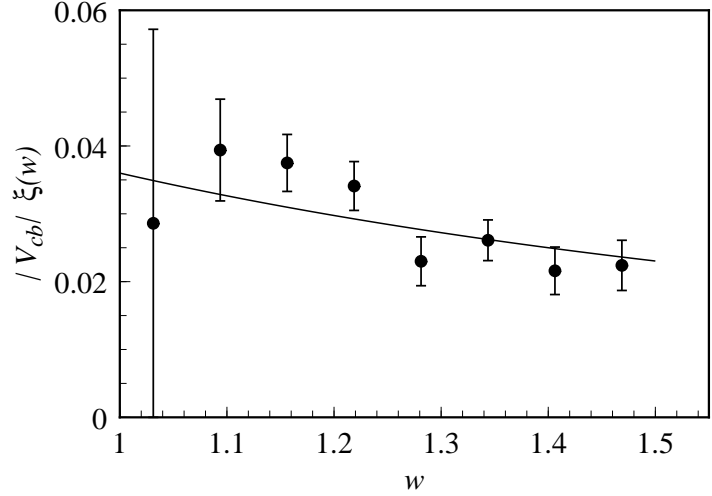}}
   \vspace{-0.5cm}
   \centerline{\parbox{11cm}{\caption{\label{fig:3.7}
ARGUS data\protect\cite{Dstar} for the product $|\,V_{cb}|\,
\widehat{\xi}(w)$ as a function of the recoil $w=v\cdot v'$, assuming
$\tau_{B^0}=1.5$ ps. Using the normalization condition
(\protect\ref{xihatnorm}), $|\,V_{cb}|$ is obtained from an
extrapolation to $w=1$. The fit curve is explained in the text.
   }}}
\end{figure}

In Fig.~\ref{fig:3.7}, I show the new ARGUS data\cite{Dstar} for the
product $|\,V_{cb}|\,\widehat{\xi}(w)$. From an unconstrained fit
using again the parametrizations in (\ref{IWFtoy}), the following
value is obtained:
\begin{equation}\label{ARGUSVcb}
   |\,V_{cb}|\,\bigg({\tau_{B^0}\over 1.5~{\rm ps}}\bigg)^{1/2}
   = 0.049\pm 0.008 \,.
\end{equation}
However, the fit gives very large values for the slope parameter
$\varrho^2$, between 1.9 and 2.3. Since the slope of
$\widehat{\xi}(w)$ agrees with the slope of the Isgur--Wise function
up to power corrections of order $\Lambda_{\rm QCD}/m_c$, I believe
that such large values cannot be tolerated in view of the Voloshin
sum rule. In Fig.~\ref{fig:3.7}, I therefore show a fit to the data
which uses the pole ansatz in (\ref{IWFtoy}) with the constraint that
$\varrho^2\le 1$. This leads to a significantly smaller value:
\begin{equation}
   |\,V_{cb}|\,\bigg({\tau_{B^0}\over 1.5~{\rm ps}}\bigg)^{1/2}
   = 0.037\pm 0.006 \,.
\end{equation}

\begin{figure}[htb]
   \vspace{0.5cm}
   \epsfxsize=10cm
   \centerline{\epsffile{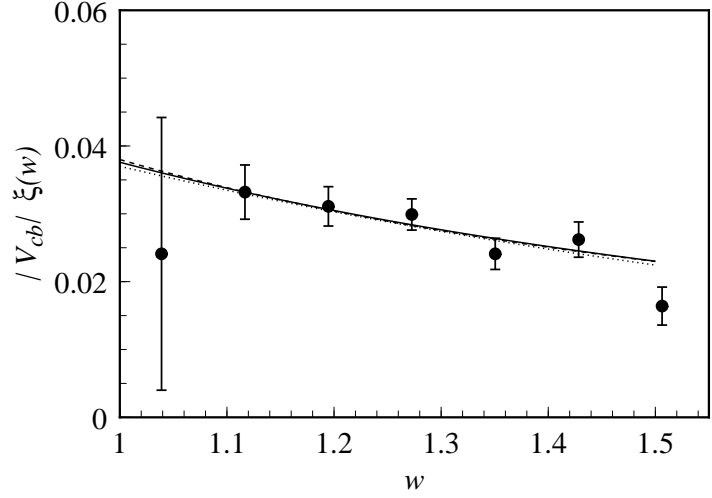}}
   \vspace{-0.5cm}
   \centerline{\parbox{11cm}{\caption{\label{fig:3.8}
CLEO data\protect\cite{CLEOVcb} for the product $|\,V_{cb}|\,
\widehat{\xi}(w)$ as a function of the recoil $w$.
   }}}
\end{figure}

Very recently, the CLEO collaboration has reported new results for
the recoil spectrum with higher statistics\cite{CLEOVcb}. They have
applied tight cuts in order to reduce the systematic errors. These
data are shown in Fig.~\ref{fig:3.8}. The curves correspond to the
results of a fit using the various functions in (\ref{IWFtoy}).
Without any constraint on the slope at zero recoil, one obtains
\begin{equation}
   |\,V_{cb}|\,\bigg({\tau_{B^0}\over 1.5~{\rm ps}}\bigg)^{1/2}
   = 0.039\pm 0.006 \,,
\end{equation}
together with values $1.0\pm 0.4<\varrho^2<1.2\pm 0.7$. It is
reassuring that the result for the slope parameter is in agreement
with the bound derived from the Voloshin sum rule. Because of the
careful analysis of systematic uncertainties, I consider these
results to be the best numbers for $|\,V_{cb}|$ and $\varrho^2$ that
are currently available.

The model-independent determination of $|\,V_{cb}|$ is one of the
many applications of heavy quark symmetry that have been explored
over the last years. I hope that the above presentation gives you a
flavour of the potential of the HQET formalism. For a more
comprehensive discussion, I refer to my review paper\cite{review}.

\subsection{Exercises}

\begin{itemize}
\item
Go through the derivation of the effective Lagrangian of HQET and
prove eqs.~(\ref{Lhchi})--(\ref{Lsubl}).
\item
Derive the Feynman rules depicted in Fig.~\ref{fig:3.2} from the
effective Lagrangian (\ref{Leff}).
\item
Convince yourself that (\ref{PVff}) is the correct covariant
expression for the pseudo\-scalar-to-vector transition matrix
element.
\item
Derive the one-loop expression given in (\ref{ZfacMS}) for the
wave-function renormalization constant $Z_h$.
\item
By performing the appropriate Dirac traces in (\ref{Neurel}), derive
the expressions given in (\ref{hilead}) for the meson form factors
$h_i(w)$.
\item
Derive the Bjorken sum rule (\ref{Bjsr}) by taking the limit $w\to 1$
in (\ref{inclsum}).
\end{itemize}

\newpage
\noindent
{\bf Concluding Remarks}
\vspace{0.4cm}

In these lectures, I have presented an introduction to the theory and
phenomenology of heavy quark masses and mixing angles, as well as to
the exciting new field of heavy quark spin-flavour symmetry. The
synthesis of these topics is interesting in itself, in that it
combines rather mature subjects with one of the most active fields of
research in particle physics. As far as quark masses and mixing
angles are concerned, the theoretical concepts have been set many
years ago. Yet new theoretical tools have to be developed to
determine these parameters more and more accurately. Heavy quark
effective field theory, on the other hand, has received broad
attention in recent years only, although some ideas of a spin-flavour
symmetry for heavy quarks had been around for much longer. These new
developments start to have an impact on the precision with which one
can determine some elements of the Kobayashi--Maskawa matrix. I have
discussed this in detail for the extraction of $|\,V_{cb}|$ from the
semileptonic decays $B\to D^*\ell\,\bar\nu$.

Besides reviewing the status of the theoretical developments, my
purpose was to convince you that $B$-physics is a rich and diverse
area of research. Presently, this field is characterized by a very
fruitful interplay between theory and experiments, which has led to
many significant discoveries and developments on both sides. Heavy
quark physics has the potential to determine important parameters of
the flavour sector of the electroweak theory, and at the same time
provides stringent tests of the standard model. On the other hand,
weak decays of hadrons containing a heavy quark are an ideal
laboratory to study the nature of nonperturbative phenomena in QCD.

When I presented these lectures in the summer of 1993, the future of
$B$-physics was uncertain. Today, the prospects for further
significant developments in this field look rather promising. With
the approval of the first asymmetric $B$-factory at SLAC, and with
ongoing $B$-physics programs at the existing facilities at Cornell,
Fermilab, and LEP, there are clearly $B$eautiful times ahead of us!

\vspace{1cm}
{\it Acknowledgements:\/}
It is a great pleasure to thank the organizers of TASI-93 for the
invitation to present these lectures and for providing a stimulating
and relaxing atmosphere, which helped to initiate many physics
discussions. Coming to TASI-93 also gave me the opportunity to meet,
for the first time, my collaborator Zoltan Ligeti, and to complete a
piece of work with him. As far as fun and sports are concerned, I
will remember an honest attempt to climb Longs Peak with Zoltan, and
a most pleasant day of rock climbing in the Boulder area with Tom
DeGrand. Writing lecture notes, on the other hand, is a painful and
time-consuming experience. It is only because of the continuous
support and queries of Stuart Raby that I finally finished this
manuscript. I very much appreciated his patience with me.

Many discussions with colleagues helped me to prepare the lectures
and had an impact on the form of these notes. I am particularly
grateful to my collaborators Adam Falk, Zoltan Ligeti, Yossi Nir,
Michael Luke, and Thomas Mannel. During the school, I have enjoyed
discussions with many of the students, as well as with Andy Cohen,
Chris Hill, and Peter Lepage. Before coming to Boulder, talking to
Michael Peskin was an invaluable asset in preparing the lectures. In
addition, I would like to thank David Cassel and Persis Drell for
keeping me updated about the CLEO analysis of semileptonic
$B$-decays. Finally, it is my great pleasure to thank Suzy Vascotto
for her careful reading of the manuscript, and Maria Girone for
teaching me {\tt PAW}.


\newpage
\vspace{0.9cm}
\noindent
{\bf References}

\begin{thebibliography}{999}


%%%   Lecture 1   %%%%%%%%%%%%%%%%%%%%%%%%%%%%%%%%%%%%%%%%%%%%%%%%%%%

\bibitem {BBbar1}
H. Albrecht et al.\ (ARGUS collaboration), Phys.\ Lett.\ B {\bf 192},
245 (1987).

\bibitem {BBbar2}
C. Albajar et al.\ (UA1 collaboration), Phys.\ Lett.\ B {\bf 186},
247 (1987).

\bibitem {btou1}
H. Albrecht et al.\ (ARGUS collaboration), Phys.\ Lett.\ B {\bf 234},
409 (1990).

\bibitem {btou2}
R. Fulton et al.\ (CLEO collaboration), Phys.\ Rev.\ Lett.\ {\bf 64},
16 (1990).

\bibitem {btou3}
H. Albrecht et al.\ (ARGUS collaboration), Phys.\ Lett.\ B {\bf 255},
297 (1991).

\bibitem {btou4}
J. Bartelt et al.\ (CLEO collaboration), Phys.\ Rev.\ Lett.\ {\bf
71}, 4111 (1993).

\bibitem {BKstar}
R. Ammar et al.\ (CLEO collaboration), Phys.\ Rev.\ Lett.\ {\bf 71},
674 (1993).

\bibitem {Shu1}
E.V. Shuryak, Phys.\ Lett.\ B {\bf 93}, 134 (1980); Nucl.\ Phys.\ B
{\bf 198}, 83 (1982).

\bibitem {Clos}
J.E. Paschalis and G.J. Gounaris, Nucl.\ Phys.\ B {\bf 222}, 473
(1983); F.E. Close, G.J. Gounaris, and J.E. Paschalis, Phys.\ Lett.\
B {\bf 149}, 209 (1984).

\bibitem {Nuss}
S. Nussinov and W. Wetzel, Phys.\ Rev.\ D {\bf 36}, 130 (1987).

\bibitem {Vol1}
M.B. Voloshin and M.A. Shifman, Yad.\ Fiz.\ {\bf 45}, 463 (1987)
[Sov.\ J.\ Nucl.\ Phys.\ {\bf 45}, 292 (1987)].

\bibitem {Vol2}
M.B. Voloshin and M.A. Shifman, Yad.\ Fiz.\ {\bf 47}, 801 (1988)
[Sov.\ J.\ Nucl.\ Phys.\ {\bf 47}, 511 (1988)].

\bibitem {Isgu}
N. Isgur and M.B. Wise, Phys.\ Lett.\ B {\bf 232}, 113 (1989);
{\bf 237}, 527 (1990).

\bibitem {EiFe}
E. Eichten and F. Feinberg, Phys.\ Rev.\ D {\bf 23}, 2724 (1981).

\bibitem {CasL}
W.E. Caswell and G.P. Lepage, Phys.\ Lett.\ B {\bf 167}, 437 (1986).

\bibitem {Eich}
E. Eichten, in: Field Theory on the Lattice, edited by A. Billoire et
al., Nucl.\ Phys.\ B (Proc.\ Suppl.) {\bf 4}, 170 (1988).

\bibitem {Thac}
G.P. Lepage and B.A. Thacker, in: Field Theory on the Lattice, edited
by A. Billoire et al., Nucl.\ Phys.\ B (Proc.\ Suppl.) {\bf 4}, 199
(1988).

\bibitem {PoWi}
H.D. Politzer and M.B. Wise, Phys.\ Lett.\ B {\bf 206}, 681 (1988);
{\bf 208}, 504 (1988).

\bibitem {EiH1}
E. Eichten and B. Hill, Phys.\ Lett.\ B {\bf 234}, 511 (1990); {\bf
243}, 427 (1990).

\bibitem {Grin}
B. Grinstein, Nucl.\ Phys.\ B {\bf 339}, 253 (1990).

\bibitem {Geor}
H. Georgi, Phys.\ Lett.\ B {\bf 240}, 447 (1990).

\bibitem {Falk}
A.F. Falk, H. Georgi, B. Grinstein, and M.B. Wise, Nucl.\ Phys.\ B
{\bf 343}, 1 (1990).

\bibitem {FGL}
A.F. Falk, B. Grinstein, and M.E. Luke, Nucl.\ Phys.\ B {\bf 357},
185 (1991).

\bibitem {Luke}
M.E. Luke, Phys.\ Lett.\ B {\bf 252}, 447 (1990).

\bibitem {review}
M. Neubert, SLAC preprint SLAC-PUB-6263 (1993), to appear in Phys.\
Rep.

\bibitem {Tarr}
R. Tarrach, Nucl.\ Phys.\ B {\bf 183}, 384 (1981).

\bibitem {Bura}
W.A. Bardeen, A.J. Buras, D.W. Duke, and T. Muta, Phys.\ Rev.\ D
{\bf 18}, 3998 (1978); A.J. Buras, Rev.\ Mod.\ Phys.\ {\bf 52}, 199
(1980).

\bibitem {tHo2}
G. 't Hooft, Nucl.\ Phys.\ B {\bf 61}, 455 (1973).

\bibitem {SVZ}
M.A. Shifman, A.I. Vainshtein, and V.I. Zakharov, Nucl.\ Phys.\ B
{\bf 147}, 385 (1979); {\bf 147}, 448 (1979).

\bibitem {FNL}
A.F. Falk, M. Neubert, and M.E. Luke, Nucl.\ Phys.\ B {\bf 388}, 363
(1992).

\bibitem {Neu3}
M. Neubert, Phys.\ Rev.\ D {\bf 46}, 3914 (1992).

\bibitem {tHo1}
G. 't Hooft and M. Veltman, Nucl.\ Phys.\ B {\bf 44}, 189 (1972).

\bibitem {Boll}
C.G. Bollini and J.J. Giambiagi, Phys.\ Lett.\ B {\bf 40}, 566
(1972); Nuovo Cim.\ B {\bf 12}, 20 (1972).

\bibitem {DGro}
D.J. Gross, in: Methods in Field Theory, edited by C.R. Balian and
J. Zinn-Justin (North-Holland, Amsterdam, 1976), p.~141.

\bibitem {Gray}
N. Gray, D.J. Broadhurst, W. Grafe, and K. Schilcher, Z.\ Phys.\ C
{\bf 48}, 673 (1990).

\bibitem {Tara}
O.V. Tarasov, JINR preprint P2-82-900 (1982), unpublished.

\bibitem {Gori}
S.G. Gorishny, A.L. Kataev, and S.A. Larin, Phys.\ Lett.\ B {\bf
135}, 457 (1984).

\bibitem {Gros}
D.J. Gross and F. Wilczek, Phys.\ Rev.\ Lett.\ {\bf 30}, 1343 (1973).

\bibitem {Poli}
H.D. Politzer, Phys.\ Rev.\ Lett.\ {\bf 30}, 1346 (1973).

\bibitem {Casw}
W.E. Caswell, Phys.\ Rev.\ Lett.\ {\bf 33}, 244 (1974).

\bibitem {Jone}
D.R.T. Jones, Nucl.\ Phys.\ B {\bf 75}, 531 (1974).

\bibitem {Bela}
A.A. Belavin and A.A. Migdal, JETP Lett.\ {\bf 19}, 181 (1974) [ZhETF
Pis.\ Red.\ {\bf 19}, 317 (1974)].

\bibitem {Flor}
E.G. Floratos, D.A. Ross, and C.T. Sachrajda, Nucl.\ Phys.\ B {\bf
129}, 66 (1977) [E: {\bf 139}, 545 (1978)].

\bibitem {BJLW}
A.J. Buras, M. Jamin, M.E. Lautenbacher, and P.H. Weisz, Nucl.\
Phys.\ B {\bf 370}, 69 (1992).

\bibitem {ApCa}
T. Appelquist and J. Carazzone, Phys.\ Rev.\ D {\bf 11}, 2856 (1975).

\bibitem {PeRo}
B. Pendleton and G.G. Ross, Phys.\ Lett.\ B {\bf 98}, 291 (1981).

\bibitem {CHil}
C.T. Hill, Phys.\ Rev.\ D {\bf 24}, 691 (1981); C.T.~Hill,
C.N.~Leung, and S.~Rao, Nucl.\ Phys.\ B {\bf 262}, 517 (1985).

\bibitem {yyy1}
J. Bagger, S. Dimopoulos, and E. Mass\'o, Phys.\ Rev.\ Lett.\ {\bf
55}, 920 (1985).

\bibitem {yyy2}
M. Carena et al., Nucl.\ Phys.\ B {\bf 369}, 33 (1992); M.~Carena,
S.~Pokorski, and C.E.M.~Wagner, Nucl.\ Phys.\ B {\bf 406}, 59 (1993).

\bibitem {yyy3}
V. Barger, M.S. Berger, and P. Ohmann, Phys.\ Rev.\ D {\bf 47}, 1093
(1993).

\bibitem {BHL}
W.A. Bardeen, C.T. Hill, and M. Lindner, Phys.\ Rev.\ D {\bf 41},
1647 (1990).

\bibitem {PDG92}
K. Hikasa et al., Review of Particle Properties, Phys.\ Rev.\ D {\bf
45}, Part II (1992).

\bibitem {Gube}
B. Guberina, R. Meckbach, R.D. Peccei, and R. R\"uckl, Nucl.\ Phys.\
B {\bf 184}, 476 (1981).

\bibitem {Rein}
L.J. Reinders, Phys.\ Rev.\ D {\bf 38}, 947 (1988); L.J. Reinders,
H.R. Rubinstein, and S. Yazaki, Phys.\ Rep.\ {\bf 127}, 1 (1985).

\bibitem {Novi}
V.A. Novikov et al., Phys.\ Rep.\ {\bf 41}, 1 (1978).

\bibitem {VoSR}
M.B. Voloshin, Yad.\ Fiz.\ {\bf 29}, 1368 (1979) [Sov.\ J.\ Nucl.\
Phys.\ {\bf 29}, 703 (1979)].

\bibitem {VoZa} M.B. Voloshin and Yu.M. Zaitsev, Usp.\ Fiz.\ Nauk
{\bf 152}, 361 (1987) [Sov.\ Phys.\ Usp.\ {\bf 30}, 553 (1987)].

\bibitem {Nari}
S. Narison, Phys.\ Lett.\ B {\bf 210}, 238 (1988); {\bf 308}, 305
(1993).

\bibitem {Baga}
E. Bagan, P. Ball, V.M. Braun, and H.G. Dosch, Phys.\ Lett.\ B {\bf
278}, 457 (1992).

\bibitem {Neu2}
M. Neubert, Phys.\ Rev.\ D {\bf 46}, 1076 (1992).

\bibitem {renom1}
I.I. Bigi, M.A. Shifman, N.G. Uraltsev, and A.I. Vainshtein, CERN
preprint CERN-TH.7171/94 (1994).

\bibitem {renom2}
M. Beneke and V.M. Braun, Munich preprint MPI-PhT/94-9 (1994).

\bibitem {reno1}
G. 't Hooft, in: The Whys of Subnuclear Physics, Proceedings of the
15th International School on Subnuclear Physics, Erice, Sicily, 1977,
edited by A.~Zichichi (Plenum Press, New York, 1979), p.~943.

\bibitem {reno2}
B. Lautrup, Phys.\ Lett.\ B {\bf 69}, 438 (1977).

\bibitem {reno3}
G. Parisi, Phys.\ Lett.\ B {\bf 76}, 65 (1978); Nucl.\ Phys.\ B
{\bf 150}, 163 (1979).

\bibitem {reno4}
F. David, Nucl.\ Phys.\ B {\bf 234}, 237 (1984); {\bf 263}, 637
(1986).

\bibitem {reno5}
A.H. Mueller, Nucl.\ Phys.\ B {\bf 250}, 327 (1985); Phys.\ Lett.\ B
{\bf 308}, 355 (1993).


%%%   Lecture 2   %%%%%%%%%%%%%%%%%%%%%%%%%%%%%%%%%%%%%%%%%%%%%%%%%%%

\bibitem {BuHa}
A.J. Buras and M.K. Harlander, in: Heavy Flavours, edited by A.J.
Buras and M. Lindner (World Scientific, Singapore,1992), p.~58.

\bibitem {Yosi}
Y. Nir, in: The Third Family and the Physics of Flavour, Proceedings
of the 20th SLAC Summer Institute on Particle Physics, Stanford,
California, 1992, edited by L. Vassilian (SLAC Report No.\ 412,
Stanford, 1993), p.~81; Y. Nir and H.R. Quinn, Annu.\ Rev.\ Nucl.\
Part.\ Sci.\ {\bf 42}, 221 (1992).

\bibitem {Cabi}
N. Cabibbo, Phys.\ Rev.\ Lett.\ {\bf 10}, 531 (1963).

\bibitem {KoMa}
M. Kobayashi and K. Maskawa, Prog.\ Theor.\ Phys.\ {\bf 49}, 652
(1973).

\bibitem {Jarl}
C. Jarlskog, in: CP Violation, edited by C. Jarlskog (World
Scientific, Singapore, 1989), p.~3.

\bibitem {Jar1}
C. Jarlskog, Phys.\ Rev.\ Lett.\ {\bf 55}, 1039 (1985); Z.\ Phys.\ C
{\bf 29}, 491 (1985).

\bibitem {Chau}
L.L. Chau and W.-Y. Keung, Phys.\ Rev.\ Lett.\ {\bf 53}, 1802 (1984).

\bibitem {Wolf}
L. Wolfenstein, Phys.\ Rev.\ Lett.\ {\bf 51}, 1945 (1983).

\bibitem {Sirl}
A. Sirlin and R. Zucchini, Phys.\ Rev.\ Lett.\ {\bf 57}, 1994 (1986);
A. Sirlin, Phys.\ Rev.\ D {\bf 35}, 3423 (1987).

\bibitem {Jaus}
W. Jaus and G. Rasche, Phys.\ Rev.\ D {\bf 35}, 3420 (1987).

\bibitem {LeRo}
H. Leutwyler and M. Roos, Z.\ Phys.\ C {\bf 25}, 91 (1984).

\bibitem {DHKl}
J.F. Donoghue, B.R. Holstein, and S.W. Klimt, Phys.\ Rev.\ D {\bf
35}, 934 (1987).

\bibitem {AGTh}
M. Ademollo and R. Gatto, Phys.\ Rev.\ Lett.\ {\bf 13}, 264 (1964).

\bibitem {Wirb}
M. Wirbel, B. Stech and M. Bauer, Z.\ Phys.\ C {\bf 29}, 637 (1985);
M. Bauer, B.~Stech, and M. Wirbel, Z.\ Phys.\ C {\bf 34}, 103 (1987).

\bibitem {GWI}
B. Grinstein, M.B. Wise, and N. Isgur, Phys.\ Rev.\ Lett.\ {\bf 56},
298 (1986).

\bibitem {ISGW}
N. Isgur, D. Scora, B. Grinstein, and M.B. Wise, Phys.\ Rev.\ D {\bf
39}, 799 (1989).

\bibitem {KS}
J.G. K\"orner and G.A. Schuler, Z.\ Phys.\ C {\bf 38}, 511 (1988) [E:
{\bf 41}, 690 (1989)]; {\bf 46}, 93 (1990).

\bibitem {Brod}
S.J. Brodsky and G.P. Lepage, Phys.\ Rev.\ D {\bf 22}, 2157 (1980).

\bibitem {Suzu}
M. Suzuki, Nucl.\ Phys.\ B {\bf 258}, 553 (1985).

\bibitem {Alto}
T. Altomari and L. Wolfenstein, Phys.\ Rev.\ Lett.\ {\bf 58},
1583 (1987).

\bibitem {Ali2}
T.M. Aliev and V.L. Eletskij, Yad.\ Fiz.\ {\bf 38}, 1537 (1983)
[Sov.\ J.\ Nucl.\ Phys.\ {\bf 38}, 936 (1983)]; T.M. Aliev, V.L.
Eletskij, and Y.I. Kogan, Yad.\ Fiz.\ {\bf 40}, 823 (1984) [Sov.\ J.\
Nucl.\ Phys.\ {\bf 40}, 527 (1984)].

\bibitem {Ovch}
A.A. Ovchinnikov, V.A. Slobodenyuk, Z.\ Phys.\ C {\bf 44}, 433
(1989); A.A.~Ovchinnikov, Yad.\ Fiz.\ {\bf 50}, 1433 (1989) [Sov.\
J.\ Nucl.\ Phys.\ {\bf 50}, 891 (1989)].

\bibitem {Baie}
V.N. Baier and A.G. Grozin, Z.\ Phys.\ C {\bf 47}, 669 (1990).

\bibitem {BBDN}
P. Ball, V.M. Braun, H.G. Dosch, and M. Neubert, Phys.\ Lett.\ B
{\bf 259}, 481 (1991); P. Ball, V.M. Braun, and H.G. Dosch, Phys.\
Rev.\ D {\bf 44}, 3567 (1991); Phys.\ Lett.\ B {\bf 273}, 316 (1991).

\bibitem {PBal}
P. Ball, Phys.\ Lett.\ B {\bf 281}, 133 (1992); in: Perturbative QCD
and Hadronic Interactions, Proceedings of the 27th Rencontres de
Moriond, edited by J.~Tran Thanh Van, Gif-sur-Yvette, France, 1992
(Editions Fronti\`eres, Gif-sur-Yvette, 1992), p.~343.

\bibitem {Neu10}
M. Neubert, Phys.\ Rev.\ D {\bf 45}, 2451 (1992); {\bf 47}, 4063
(1993).

\bibitem {Buch}
M. Neubert, V. Rieckert, B. Stech, and Q.P. Xu, in: Heavy Flavours,
edited by A.J. Buras and M. Lindner (World Scientific, Singapore,
1992), p.~286.

\bibitem {Rady}
A.V. Radyushkin, Phys.\ Lett.\ B {\bf 271}, 218 (1991).

\bibitem {Lige} M. Neubert, Z. Ligeti, and Y. Nir, Phys.\ Lett.\ B
{\bf 301}, 101 (1993); Phys.\ Rev.\ D {\bf 47}, 5060 (1993); {\bf
49}, 1302 (1994).

\bibitem {Chri}
J.H. Christenson, J.W. Cronin, V.L. Fitch, and R. Turlay, Phys.\
Rev.\ Lett.\ {\bf 13}, 138 (1964); Phys.\ Rev.\ B {\bf 140}, 74
(1965).

\bibitem {IsWeVub}
N. Isgur and M.B. Wise, Phys.\ Rev.\ D {\bf 42}, 2388 (1990).

\bibitem {Dib}
C.O. Dib and F. Vera, Phys.\ Rev.\ D {\bf 47}, 3938 (1993).

\bibitem {BLNN}
G. Burdman, Z. Ligeti, M. Neubert, and Y. Nir, Phys.\ Rev.\ D {\bf
49}, 2331 (1994).

\bibitem {AlPi}
A. Ali and E. Pietarinen, Nucl.\ Phys.\ B {\bf 154}, 519 (1979).

\bibitem {CCM}
G. Corb\`o, Nucl.\ Phys.\ B {\bf 212}, 99 (1983); N. Cabibbo, G.
Corb\`o, and L.~Maiani, Nucl.\ Phys.\ B {\bf 155}, 93 (1979).

\bibitem {JeKu}
M. Jezabek and J.H. K\"uhn, Nucl.\ Phys.\ B {\bf 320}, 20 (1989).

\bibitem {ACM}
G. Altarelli et al., Nucl.\ Phys.\ B {\bf 208}, 365 (1982).

\bibitem {Pasc}
A. Bareiss and E.A. Paschos, Nucl.\ Phys.\ B {\bf 327}, 353 (1989);
C.H. Jin, W.F.~Palmer, and E.A. Paschos, Dortmund preprint
DO-TH-93/21 (1993).

\bibitem {shape}
M. Neubert, Phys.\ Rev.\ D {\bf 49}, 3392 (1994).

\bibitem {photon}
M. Neubert, CERN preprint CERN-TH.7113/93 (1993), to appear in
Phys.\ Rev.\ D {\bf 49}, no.~9 (1994).

\bibitem {Fermi}
I.I. Bigi, M.A. Shifman, N.G. Uraltsev, and A.I. Vainshtein, CERN
preprint CERN-TH.7129/93 (1993).

\bibitem {FJMW}
A.F. Falk, E. Jenkins, A.V. Manohar, and M.B. Wise, San Diego
preprint UCSD/PTH-93-38 (1993).

\bibitem {bcshap}
T. Mannel and M. Neubert, CERN preprint CERN-TH.7156/94 (1994).

\bibitem {Chay}
J. Chay, H. Georgi, and B. Grinstein, Phys.\ Lett.\ B {\bf 247}, 399
(1990).

\bibitem {Bigi}
I.I. Bigi, N.G. Uraltsev, and A.I. Vainshtein, Phys.\ Lett.\ B {\bf
293}, 430 (1992); I.I. Bigi, M.A. Shifman, N.G. Uraltsev, and A.I.
Vainshtein, Phys.\ Rev.\ Lett.\ {\bf 71}, 496 (1993); I.I. Bigi et
al., in: Proceedings of the Annual Meeting of the Division of
Particles and Fields of the American Physical Society, Batavia,
Illinois, 1992, edited by C. Albright et al.\ (World Scientific,
Singapore, 1993), p.~610.

\bibitem {Blok}
B. Blok, L. Koyrakh, M.A. Shifman, and A.I. Vainshtein, Phys.\ Rev.\
D {\bf 49}, 3356 (1994).

\bibitem {MaWe}
A.V. Manohar and M.B. Wise, Phys.\ Rev.\ D {\bf 49}, 1310 (1994).

\bibitem {Adam}
A.F. Falk, M. Luke, and M.J. Savage, Phys.\ Rev.\ D {\bf 49}, 3367
(1994).

\bibitem {Thom}
T. Mannel, Nucl.\ Phys.\ B {\bf 413}, 396 (1994).

\bibitem {btau1}
L. Koyrakh, Minnesota preprint TPI-MINN-93/47-T (1993).

\bibitem {btau2}
A.F. Falk, Z. Ligeti, M. Neubert, and Y. Nir, CERN preprint
CERN-TH.7124/93 (1993), to appear in Phys.\ Lett.\ B.

\bibitem {btau3}
S. Balk, J.G. K\"orner, D. Pirjol, and K. Schilcher, Mainz preprint
MZ-TH/93-32 (1993).

\bibitem {LukSav}
M. Luke and M.J. Savage, Phys.\ Lett.\ B {\bf 321}, 88 (1994).

\bibitem {PDG88}
G.P. Yost et al., Review of Particle Properties, Phys.\ Lett.\ B {\bf
204} (1988).

\bibitem {Sheldon}
S. Stone, to appear in the 2nd edition of: $B$ Decays, edited by S.
Stone (World Scientific, Singapore, 1991).

\bibitem {tauBav}
M. Danilov, to appear in: Proceedings of the International
Europhysics Conference on High Energy Physics, Marseille, France,
July 1993, edited by J.~Carr and M. Perrottet (Editions Fronti\`eres,
Gif-sur-Yvettes).

\bibitem {tauB}
T. Hessing, to appear in: Electroweak Interactions and Unified
Theories, Proceedings of the 29th Rencontres de Moriond, M\'eribel,
France, March 1993.

\bibitem {Neu9}
M. Neubert, Phys.\ Lett.\ B {\bf 264}, 455 (1991).

\bibitem {CLEOVcb}
G. Crawford et al.\ (CLEO collaboration), to appear in: Proceedings
of the 16th International Symposium on Lepton and Photon Interaction,
Ithaca, New York, Aug.\ 1993.

\bibitem {Bjtr}
J.D. Bjorken, presented in discussions at the Workshop on
Experiments, Detectors and Experimental Areas for the Supercollider,
Berkeley, California, July 1987 (Proceedings edited by R. Donaldson
and M.G.D. Gilchriese, World Scientific, Singapore, 1988).

\bibitem {JaSt}
C. Jarlskog and R. Stora, Phys.\ Lett.\ B {\bf 208}, 268 (1988).

\bibitem {Khoz}
I.I. Bigi, V.A. Khoze, N.G. Uraltsev, and A.I. Sanda, in: CP
Violation, edited by C. Jarlskog (World Scientific, Singapore, 1989),
p.~175.

\bibitem {DDGN}
C.O. Dib, I. Dunietz, F.J. Gilman, and Y. Nir, Phys.\ Rev.\ D {\bf
41}, 1522 (1990); F.J. Gilman and Y. Nir, Annu.\ Rev.\ Nucl.\ Part.\
Sci.\ {\bf 40}, 213 (1990).

\bibitem {KRYu}
C.S. Kim, J.L. Rosner, and S.-P. Yuan, Phys.\ Rev.\ D {\bf 42}, 96
(1990).

\bibitem {ScSc}
M. Schmidtler and K.R. Schubert, Z.\ Phys.\ C {\bf 53}, 347 (1992).

\bibitem {LMMR}
M. Lusignoli, L. Maiani, G. Martinelli, and L. Reina, Nucl.\ Phys.\ B
{\bf 369}, 139 (1992).

\bibitem {HaRo}
G.R. Harris and J.L. Rosner, Phys.\ Rev.\ D {\bf 45}, 946 (1992).

\bibitem {BBARG}
H. Albrecht et al.\ (ARGUS collaboration), Z.\ Phys.\ C {\bf 55}, 357
(1992).

\bibitem {BBCLEO}
J. Bartelt et al.\ (CLEO collaboration), Phys.\ Rev.\ Lett.\ {\bf
71}, 1680 (1993).

\bibitem {dmMoriond}
D. Abbaneo, to appear in: Electroweak Interactions and Unified
Theories, Proceedings of the 29th Rencontres de Moriond, M\'eribel,
France, March 1993.

\bibitem {BJWe}
A.J. Buras, M. Jamin, and P.H. Weisz, Nucl.\ Phys.\ B {\bf 347}, 491
(1990).

\bibitem {InLi}
T. Inami and C.S. Lim, Progr.\ Theor.\ Phys.\ {\bf 65}, 297 (1981);
{\bf 65}, 1772 (1981).

\bibitem {Bur1}
A.J. Buras, Phys.\ Rev.\ Lett.\ {\bf 46}, 1354 (1981).

\bibitem {Penbox}
G. Buchalla, A.J. Buras, and M.K. Harlander, Nucl.\ Phys.\ B {\bf
349}, 1 (1991).

\bibitem {DGHo}
J.F. Donoghue, E. Golowich, and B.R. Holstein, Phys.\ Lett.\ B {\bf
119}, 412 (1982).

\bibitem {PdeR}
A. Pich and E. de Rafael, Phys.\ Lett.\ B {\bf 158}, 477 (1985);
Nucl.\ Phys.\ B {\bf 358}, 311 (1991).

\bibitem {RDec}
R. Decker, in: Hadronic Matrix Elements and Weak Decays, edited by
A.J.~Buras, J.-M. G\'erard, and W. Huber, Nucl.\ Phys.\ B (Proc.\
Suppl.) {\bf 7a}, 180 (1989), and references therein.

\bibitem {BBGe}
W.A. Bardeen, A.J. Buras, and J.-M. G\'erard, Phys.\ Lett.\ B {\bf
211}, 343 (1988).

\bibitem {Gave}
M.B. Gavela et al., Phys.\ Lett.\ B {\bf 206}, 113 (1988); Nucl.\
Phys.\ B {\bf 306}, 677 (1988).

\bibitem {KSGP}
G.W. Kilcup, S.R. Sharpe, R. Gupta, and A. Patel, Phys.\ Rev.\ Lett.\
{\bf 64}, 25 (1990).

\bibitem {BSon}
C. Bernard and A. Soni, in: Lattice 89, edited by N. Cabibbo et
al., Nucl.\ Phys.\ B (Proc.\ Suppl.) {\bf 17}, 495 (1990).

\bibitem {CrisS}
C. Sachrajda, in: QCD-20 Years Later, Proceedings of the Workshop on
QCD, Aachen, Germany, 1992, edited by P.M. Zerwas and H.A.~Kastrup
(World Scientific, Singapore, 1993), p.~668, and references therein.

\bibitem {SShar}
S.R. Sharpe and A. Patel, Washington preprint UW-PT-93-1 (1993);
S.R.~Sharpe, to appear in: Proceedings of the 11th Symposium on
Lattice Field Theory (Lattice 93), Dallas, Texas, Oct.~1993.

\bibitem {NA31}
H. Burckhardt et al.\ (NA31 collaboration), Phys.\ Lett.\ B {\bf
206}, 169 (1988).

\bibitem {E731}
J.R. Patterson et al.\ (E731 collaboration), Phys.\ Rev.\ Lett.\ {\bf
64}, 1491 (1990).

\bibitem {Burs}
G. Buchalla, A.J. Buras, and M.K. Harlander, Nucl.\ Phys.\ B {\bf
337}, 313 (1990).

\bibitem {BJLa}
A.J. Buras, M. Jamin, and M.E. Lautenbacher, Nucl.\ Phys.\ B {\bf
408}, 209 (1993).

\bibitem {Marti}
M. Ciuchini, E. Franco, G. Martinelli, and L. Reina, Phys.\ Lett.\ B
{\bf 301}, 263 (1993); Rome preprint 93/913 (1993).

\bibitem {Berto}
S. Bertolini, in: Higgs Particle(s): Physics Issues and Searches in
High-Energy Collisions, Proceedings of the 8th INFN Eloisatron
Project Workshop, Erice, Sicily, 1989, edited by A. Ali (Plenum
Press, New York, 1990), p.~243.

\bibitem {Ali}
A. Ali, to appear in the 2nd edition of: $B$ Decays, edited by S.
Stone (World Scientific, Singapore, 1991).

\bibitem {Bert}
S. Bertolini, F. Borzumati, and A. Masiero, Phys.\ Rev.\ Lett.\ {\bf
59}, 180 (1987).

\bibitem {Desh}
N.G. Deshpande et al., Phys.\ Rev.\ Lett.\ {\bf 59}, 183 (1987).

\bibitem {Grin}
B. Grinstein, R. Springer, and M.B. Wise, Phys.\ Lett.\ B {\bf 202},
138 (1988); Nucl.\ Phys.\ B {\bf 339}, 269 (1990).

\bibitem {Grig}
R. Grigjanis, P.J. O'Donnel, M. Sutherland, and H. Navelet, Phys.\
Lett.\ B {\bf 213}, 355 (1988); {\bf 223}, 239 (1989); {\bf 237}, 252
(1990).

\bibitem {Cell}
G. Cella, G. Curci, G. Ricciardi, and A. Vicere, Phys.\ Lett.\ B {\bf
248}, 181 (1990).

\bibitem {Misi}
M. Misiak, Phys.\ Lett.\ B {\bf 269}, 161 (1991); Nucl.\ Phys.\ B
{\bf 393}, 23 (1993).

\bibitem {Ciuc}
M. Ciuchini et al., Phys.\ Lett.\ B {\bf 316}, 127 (1993).

\bibitem {AlGr}
A. Ali and C. Greub, Z.\ Phys.\ C {\bf 49}, 431 (1991); Phys.\ Lett.\
B {\bf 259}, 182 (1991); {\bf 287}, 191 (1992).

\bibitem {AMan}
A. Ali and T. Mannel, Phys.\ Lett.\ B {\bf 264}, 447 (1991) [E: {\bf
274}, 526 (1992)].

\bibitem {Ball}
P. Ball, Munich preprint TUM-T31-43/93 (1993).

\bibitem {SNar}
S. Narison, CERN preprint CERN-TH.7166/94 (1994).


%%%   Lecture 3   %%%%%%%%%%%%%%%%%%%%%%%%%%%%%%%%%%%%%%%%%%%%%%%%%%%

\bibitem {GeRev}
H. Georgi, in: Perspectives in the Standard Model, Proceedings of the
Theoretical Advanced Study Institute in Elementary Particle Physics
(TASI-91), Boulder, Colorado, 1991, edited by R.K.~Ellis, C.T.~Hill,
and J.D.~Lykken (World Scientific, Singapore, 1992), p.~589.

\bibitem {GrRev}
B. Grinstein, in: High Energy Phenomenology, Proceedings of the
Workshop on High Energy Phenomenology, Mexico City, Mexico, 1991,
edited by M.A.~P\'eres and R.~Huerta (World Scientific, Singapore,
1992), p.~161.

\bibitem {IWRev}
N. Isgur and M.B. Wise, in: Heavy Flavours, edited by A.J. Buras and
M.~Lindner (World Scientific, Singapore, 1992), p.~234.

\bibitem {MaRev}
T. Mannel, in: QCD-20 Years Later, Proceedings of the Workshop on
QCD, Aachen, Germany, 1992, edited by P.M.~Zerwas and H.A.~Kastrup
(World Scientific, Singapore, 1993), p.~634.

\bibitem {Appe}
T. Appelquist and H.D. Politzer, Phys.\ Rev.\ Lett.\ {\bf 34}, 43
(1975).

\bibitem {SVZ1}
M.A. Shifman, A.I. Vainshtein, and V.I. Zakharov, Nucl.\ Phys.\ B
{\bf 120}, 316 (1977).

\bibitem {Witt}
E. Witten, Nucl.\ Phys.\ B {\bf 122}, 109 (1977).

\bibitem {Polc}
J. Polchinski, Nucl.\ Phys.\ B {\bf 231}, 269 (1984).

\bibitem {Wils}
K. Wilson, Phys.\ Rev.\ {\bf 179}, 1499 (1969); Phys.\ Rev.\ D
{\bf 3}, 1818 (1971).

\bibitem {Zimm}
W. Zimmermann, Ann.\ of Phys.\ {\bf 77}, 536 (1973); {\bf 77}, 570
(1973).

\bibitem {Alta}
G. Altarelli and L. Maiani, Phys.\ Lett.\ B {\bf 52}, 351 (1974).

\bibitem {Gail}
M.K. Gaillard and B.W. Lee, Phys.\ Rev.\ Lett.\ {\bf 33}, 108 (1974).

\bibitem {Gilm}
F.J. Gilman and M.B. Wise,  Phys.\ Rev.\ D {\bf 27}, 1128 (1983).

\bibitem {Mann}
T. Mannel, W. Roberts and Z. Ryzak, Nucl.\ Phys.\ B {\bf 368}, 204
(1992).

\bibitem {Soto}
J. Soto and R. Tzani, Phys.\ Lett.\ B {\bf 297}, 358 (1992).

\bibitem {GGW}
H. Georgi, B. Grinstein, and M.B. Wise, Phys.\ Lett.\ B {\bf 252},
456
(1990).

\bibitem {IsWi}
N. Isgur and M.B. Wise, Phys.\ Rev.\ Lett.\ {\bf 66}, 1130 (1991).

\bibitem {LEPBs}
D. Buskulic et al.\ (ALEPH collaboration), Phys.\ Lett.\ B {\bf 311},
425 (1993).

\bibitem {FaNe}
A.F. Falk and M. Neubert, Phys.\ Rev.\ D {\bf 47}, 2965 (1993);
{\bf 47}, 2982 (1993).

\bibitem {Prep}
G. Preparata and W.I. Weisberger, Phys.\ Rev.\ {\bf 175}, 1965
(1968).

\bibitem {JiMu}
X. Ji and M.J. Musolf, Phys.\ Lett.\ B {\bf 257}, 409 (1991).

\bibitem {BrGr}
D.J. Broadhurst and A.G. Grozin, Phys.\ Lett.\ B {\bf 267}, 105
(1991).

\bibitem {BGSc}
D.J. Broadhurst, N. Gray, and K. Schilcher, Z.\ Phys.\ C {\bf 52},
111 (1991).

\bibitem {DeWi}
B.S. DeWitt, Phys.\ Rev.\ {\bf 162}, 1195 (1967).

\bibitem {tHo3}
G. 't Hooft, in: Functional and Probabilistic Methods in Quantum
Field Theory, Proceedings of the 12th Winter School of Theoretical
Physics, Karpacz, Poland, Acta Univ.\ Wratisl.\ {\bf 38}, Vol.~1
(1975).

\bibitem {Boul}
D. Boulware, Phys.\ Rev.\ D {\bf 23}, 389 (1981).

\bibitem {Abbo}
L.F. Abbott, Nucl.\ Phys.\ B {\bf 185}, 189 (1981); Acta Phys.\ Pol.\
B {\bf 13}, 33 (1982).

\bibitem {Klug}
H. Kluberg-Stern and J.B. Zuber, Phys.\ Rev.\ D {\bf 12}, 3159
(1975).

\bibitem {Neu6}
M. Neubert, Phys.\ Rev.\ D {\bf 46}, 2212 (1992).

\bibitem {KoRa}
G.P. Korchemsky and A.V. Radyushkin, Nucl.\ Phys.\ B {\bf 283}, 342
(1987); Phys.\ Lett.\ B {\bf 279}, 359 (1992); G.P. Korchemsky, Mod.\
Phys.\ Lett.\ A {\bf 4}, 1257 (1989).

\bibitem {FaGr}
A.F. Falk and B. Grinstein, Phys.\ Lett.\ B {\bf 247}, 406 (1990);
{\bf 249}, 314 (1990).

\bibitem {Neu4}
M. Neubert, Nucl.\ Phys.\ B {\bf 371}, 149 (1992).

\bibitem {Kili}
W. Kilian, P. Manakos, and T. Mannel, Phys.\ Rev.\ D {\bf 48}, 1321
(1993).

\bibitem {LuMa}
M. Luke and A.V. Manohar, Phys.\ Lett.\ B {\bf 286}, 348 (1992).

\bibitem {mepower}
M. Neubert, Phys.\ Lett.\ B {\bf 306}, 357 (1993); Phys.\ Rev.\ D
{\bf 49}, 1542 (1994).

\bibitem {Bjor}
J.D. Bjorken, in: Results and Perspectives in Particle Physics,
Proceedings of the 4th Rencontres de Physique de la Vall\'e d'Aoste,
La Thuile, Italy, 1990, edited by M. Greco (Editions Fronti\`eres,
Gif-sur-Yvette, 1990), p.~583; in: Gauge Bosons and Heavy Quarks,
Proceedings of the 18th SLAC Summer Institute on Particle Physics,
Stanford, California, 1990, edited by J.F.~Hawthorne (SLAC Report
No.\ 378, Stanford, 1991), p.~167.

\bibitem {AdamF}
A.F. Falk, Nucl.\ Phys.\ B {\bf 378}, 79 (1992).

\bibitem {Pol1}
H.D. Politzer, Phys.\ Lett.\ B {\bf 250}, 128 (1990).

\bibitem {MRR1}
T. Mannel, W. Roberts, and Z. Ryzak, Phys.\ Lett.\ B {\bf 271}, 421
(1991).

\bibitem {Neu7}
M. Neubert, Nucl.\ Phys.\ B {\bf 416}, 786 (1994).

\bibitem {ChGr}
P. Cho and B. Grinstein, Phys.\ Lett.\ B {\bf 285}, 153 (1992).

\bibitem {Neu1}
M. Neubert and V. Rieckert, Nucl.\ Phys.\ B {\bf 382}, 97 (1992).

\bibitem {ThMa}
T. Mannel, CERN preprint CERN-TH.7162/94 (1994).

\bibitem {IsgW}
N. Isgur and M.B. Wise, Phys.\ Rev.\ D {\bf 43}, 819 (1991).

\bibitem {BjDT}
J.D. Bjorken, I. Dunietz, and J. Taron, Nucl.\ Phys.\ B {\bf 371},
111 (1992).

\bibitem {IWYo}
N. Isgur, M.B. Wise, and M. Youssefmir, Phys.\ Lett.\ B {\bf 254},
215
(1991).

\bibitem {BlokS}
B. Blok and M. Shifman, Phys.\ Rev.\ D {\bf 47}, 2949 (1993).

\bibitem {Dstar}
H. Albrecht et al.\ (ARGUS collaboration), Z.\ Phys.\ C {\bf 57}, 533
(1993).

\bibitem {Volo}
M.B. Voloshin, Phys.\ Rev.\ D {\bf 46}, 3062 (1992).

\bibitem {deRa}
E. de Rafael and J. Taron, Phys.\ Lett.\ B {\bf 282}, 215 (1992).

\bibitem {KPDo}
C. Dominguez, J.G. K\"orner, and D. Pirjol, Phys.\ Lett.\ B {\bf
301}, 257 (1993).

\bibitem {ALWi}
A.F. Falk, M. Luke, and M.B. Wise, Phys.\ Lett.\ B {\bf 299}, 123
(1993).

\bibitem {GrMe}
B. Grinstein and P.F. Mende, Phys.\ Lett.\ B {\bf 299}, 127
(1993).

\bibitem {Carl}
C.E. Carlson et al., Phys.\ Lett.\ B {\bf 299}, 133 (1993).

\bibitem {Tarnew}
E. de Rafael and J. Taron, Marseille preprint CPT-93/P.2908 (1993).

\bibitem {Stone}
D. Bortoletto and S. Stone, Phys.\ Rev.\ Lett.\ {\bf 65}, 2951
(1990).

\bibitem {Sheld}
S. Stone, in: Heavy Flavours, edited by A.J. Buras and M. Lindner
(World Scientific, Singapore, 1992), p.~334.


\end{thebibliography}
\end{document}